\documentclass[preprint,3p,12pt]{elsarticle}
\usepackage{lineno,hyperref}
\usepackage{amsmath,bm,geometry}
\usepackage{color}
\modulolinenumbers[5]

\begin{document}

\begin{frontmatter}

\title{Limitation Principle for Computational Fluid Dynamics}

\author[mysecondaryaddress]{Chang Liu}
\ead{cliuaa@connect.ust.hk}

\author[mymainaddress]{Guangzhao Zhou}
\ead{zgz@pku.edu.cn}

\author[mysecondaryaddress1]{Wei Shyy}
\ead{weishyy@ust.hk}

\author[mysecondaryaddress,mysecondaryaddress1,mythirdaddress]{Kun Xu\corref{cor1}}
\ead{makxu@ust.hk}

\address[mysecondaryaddress]{Department of Mathematics, Hong Kong University of Science and Technology, Kowloon, Hong Kong, China}
\address[mysecondaryaddress1]{Department of Mechanical and Aerospace Engineering, Hong Kong University of Science and Technology, Kowloon, Hong Kong, China}
\address[mythirdaddress]{Shenzhen Research Institute, Hong Kong University of Science and Technology, Shenzhen 518057, China}
\address[mymainaddress]{College of Engineering, Peking University, Beijing 100871, China}
\cortext[cor1]{Corresponding author}


\begin{abstract}

For theoretical gas dynamics, the flow regimes are classified according to the physical Knudsen number $K\!n_s$ which is defined as the ratio of
particle mean free path over the characteristic length scale.
The Navier-Stokes and Boltzmann equations are the corresponding modeling eqautions
 in the continuum $K\!n_s \leq 10^{-3}$ and rarefied  $K\!n_s \sim 1$ regimes.
For computational fluid dynamics (CFD), the numerical solution is the projection of physical flow field onto a discretized space,
and the cell Knudsen number $K\!n_c$, defined as the ratio of particle mean free path over the cell size, will effect the solution as well.
The final CFD result is controlled by a numerical Knudsen number $K\!n_n$,
which is a function of $K\!n_s$ and $K\!n_c$.
The limitation principle is about their connections among $K\!n_s$, $K\!n_c$, and $K\!n_n$, and
states the best result a numerical scheme can provide.
This principle is the uncertainty theory for the capturing of a physical solution with a limited mesh resolution,
{which means the numerical flow regime changes with different numerical resolution.}
In order to give the best result, a numerical scheme must be a multiscale method which
correctly identifies flow physics in the corresponding cell Knudsen number $K\!n_c$, such as
capturing the hydrodynamic scale  wave propagation in the coarse mesh case,
to the kinetic scale particle transport in the fine mesh case.
The unified gas-kinetic scheme (UGKS) is such a multiscale method which connects
the NS to Boltzmann solutions  seamlessly through the continuum,
near-continuum, and the non-equilibrium solutions with a variation of cell Knudsen number.
Even from a multiscale method, the CFD solution may not be the same as the physical solution once
the mesh resolution isn't enough to resolve the physical flow structure.
In this paper, the mesh size effect on the representation of physical solution is explicitly analyzed through examples.
The current CFD practice, which targets on the cell converged solution of a fixed governing equation, such as NS,
is inadequate to recover a truly multiple scale nature of gas dynamics.
On the other hand, a scheme cannot target on the Boltzmann solution alone due to the limitation of computational resources.
For an efficient CFD simulation, the representable flow physics should depend on the cell resolution
under the constraint of limitation principle.

\end{abstract}

\end{frontmatter}

\section{Introduction}
The computational fluid dynamics (CFD) is more complicated than the theoretical one.
For a computation, three Knudsen numbers, {namely} the physical, cell, and numerical ones, will effect the numerical
solution and determine the quality of the scheme in the representation of a physical solution.
According to physical Knudsen number, the flow regimes are classified into rarefied ($K\!n\ge1$),
transitional ($0.1<K\!n\le1$), near-continuum  ($0.001<K\!n\le0.1$), and continuum one ($K\!n\le0.001$).
In the continuum regime, the Euler and Navier-Stokes (NS) equations are widely used as model equations.
For the Euler equations, there is no dissipative length and relaxation time scales.
The instinct achievement of local equilibrium state for the Euler system is equivalent to the zero particle collision time.
As a result, in the process of mesh refinement, there is always infinite number of pseudo-particle collisions within any small numerical time step to form  shock, contact, and rarefaction waves, such as the solution in the Riemann solver \cite{toro}, and this practice
will never get mesh converged solution in any complicated flow simulation, especially for the flow with shock and shear instability.
For the NS equations, instead of the equilibrium assumption,
the non-equilibrium physics will take effect to introduce a dissipative mechanism in the gas system.
The viscosity and heat conduction coefficients are associated with a limited particle mean free path and particle collision time.
The current numerical NS solutions are mostly obtained through operator splitting approach,
i.e., the Riemann solver for the inviscid flux and the central difference for the viscous terms.
The physical consistency in the operator-splitting approach for NS equations
need to be investigated \cite{li-li-xu}, especially in the process of mesh refinement.
The gas kinetic scheme (GKS) constructs the numerical flux from the kinetic equation,
and calculate the inviscid flux and viscous terms in a coupled way \cite{xu-book}.
Comparing to the splitting approach, GKS is more physically reliable,
especially under mesh refinement. As the cell size and the particle mean free path go to the same order,
there is no physical basis for the distinctive wave interaction from the macroscopic equations.
However, GKS still represents NS solution due to the adoption of the Chapman-Enskog expansion for the initial gas distribution function
at the beginning of each time step.
In reality, the flow physics modeling should be changed as well with the mesh refinement, instead of solving
purely the Euler, NS, or Boltzmann equations.

For a flow field with high frequency mode or large gradients, the local Knudsen number can become
large and non-equilibrium flow dynamics may emerge.
The NS equations cease {to appropriately} describe such a phenomenon.
Under such a circumstance, the converged NS solution is not physically valid as the mesh size goes to the particle mean free path.
With the increase of mesh resolution, the CFD as a method for capturing physically valid solution must be associated with
multiple scale modeling, such as solving different equations in different scales.
The physical solution from a CFD algorithm  should be synchronized with the cell resolution.
Unfortunately, we don't have such a traditional partial differential equation-based governing equation
which could connect flow physics in different regimes smoothly.
Numerically, as a direct modeling method, the unified gas kinetic scheme (UGKS) is constructed according to the resolvable scale of
a discretized scale, such as mesh size, and captures the solutions in all flow regimes through the variation of the cell Knudsen number \cite{xu-book}.
However, even with perfect capturing of physical solution in different mesh size scale,
there is still limitation due to the mismatch between the physical flow structure thickness and the
 and the pre-defined mesh resolution. In this paper, a limitation principle in CFD simulation will be proposed and it is
related to the differences in the physical, cell, and numerical Knudsen numbers,  to represent a physical flow field in a
discretized space.

\section{The limitation principle in  flow simulation}

Theoretically, a physical flow field is a nature existence with different structure and
dynamics in different scales. The kinetic scale description follows the particle transport and collision, such as the physical process
in the Boltzmann equation, where the separate treatment of particle transport and collision determines its applicable scale within the
particle mean free path.
The hydrodynamic equations, such as NS, are based on the macroscopic description and  closed fluid elements with gigantic amount of
particles inside are used in the modeling,
even though the scale of the fluid element has never been clearly defined \cite{xu-liu}.
Numerically, an algorithm in CFD becomes more complicated to represent a  flow field in a discretized space with
 basic scales of cell size $\Delta x$ and time step $\Delta t$, and to construct flow evolution in such a space.
The numerical flow is the projection of the physical flow field on a discretized space.
Both flow physics and the cell resolution play dynamic roles for capturing multiple scale flow evolution.
The limitation principle is about the limiting process in the representation of a physical flow field in a discretized space.

A multiscale method captures macroscopic flow behavior on a coarse mesh with the inclusion of
a large amount of particles inside each control volume, and presents non-equilibrium flow dynamics
on a fine mesh with the resolvable small kinetic scale physics.
In a mesh refinement process, an idealized numerical method should provide a changeable dynamics in different scales.
A multiscale method is the one to capture the corresponding flow physics  on different mesh resolution, such as the UGKS.
Still, due to the possible mismatch between the cell resolution and the physical flow structure thickness,
there is a limitation principle about the best result an idealized multiscale method can provide.

The flow regime in CFD can be characterized by the numerical Knudsen number $K\!n_{n}$,
which is defined as the ratio of the molecular mean free path $\lambda$ to the resolvable solution's length scale $L_n$, i.e., $K\!n_{n}=\lambda/L_n$.
The numerical Knudsen number depends on the physical Knudsen number $K\!n_s$, and the cell Knudsen number $K\!n_c$.
The relationship among $K\!n_{n}$, $K\!n_s$, and $K\!n_c$ gives the limitation principle about the real
solution ($K\!n_s$) to be represented ($K\!n_n$) numerically due to the cell resolution ($K\!n_c$).

The physical Knudsen number is defined as the ratio of the molecular mean free path $\lambda$ to the physical characteristic length scale $L$, i.e., $K\!n_s = \lambda/L$.
The theoretical flow mechanics states that the NS equations are the modeling equations in the continuum flow regime at small Knudsen number,
with local thermodynamic equilibrium assumption.
For the compressible flow, the Knudsen number can be expressed as a function of  Mach number $M\!a$ and the Reynolds number $Re$ by
\begin{equation}
K\!n_s = \sqrt{\frac{\gamma \pi}{2}} \frac{M\!a}{Re},
\end{equation}
where $\gamma$ is the specific heat ratio.
Obviously $K\!n_s$ reflects physical description of a flow field in the characteristic length scale $L$.
Physically, it is more reasonable to define the length scale of a flow structure, such as $L_s = \rho/|\bigtriangledown \rho |$
based on density $\rho$, as the scale of a physical solution, i.e.,
$K\!n_s = \lambda/L_s$.
The NS modeling is applicable for flow structure with small Knudsen number, such as $K\!n_s \ll 1$.

For a numerical scheme with a mesh size $\Delta x$, it is important to define a cell Knudsen number:
\begin{equation} \label{eq: Kn_c}
K\!n_c = \frac{\lambda}{\Delta x},
\end{equation}
which is the ratio between the particle mean free path $\lambda$ and the local cell size $\Delta x$.
For a viscous flow, the particle mean free path, such as for a hard sphere molecule, is related to the viscosity coefficient \cite{vincenti},
\begin{equation}
\lambda = \sqrt{\frac{\pi m}{2kT}} \frac{\mu}{\rho},
\end{equation}
where $m$ is the molecular mass, $k$ is the Boltzmann constant, $T$ is the temperature,
$\mu$ is the dynamical viscosity coefficient, and $\rho$ is the local density.
Alternatively, the cell Knudsen number can be expressed as
\begin{equation} \label{eq: Kn_c2}
K\!n_c = \sqrt{\frac{\gamma \pi}{2}} \frac{M\!a}{Re_c},
\end{equation}
where $M\!a = |u|/a$ is now the local Mach number, $Re_c = |u| \Delta x/ \nu$ is the cell Reynolds number.
Here $u$ is the local flow velocity, $a$ is the speed of sound, and $\nu$ is the kinematic viscosity coefficient $\nu = \mu /\rho$.
The $K\!n_c$ is related to the particle transport mechanism in the mesh resolution, through the waves in the Riemann solution ($\lambda \ll \Delta x $), or particle free transport in the Boltzmann solver ($\lambda \sim \Delta x $).
With a mesh refinement, $K\!n_c$ continuously increases.
In the current CFD practice,
it isn't surprising to see the mesh size approaching to the particle mean free path in order to resolve the
 boundary layer solution in the hypersonic flow simulation.
For an ultimate  CFD method, the dynamics in the algorithm should change with the cell resolution $K\!n_c$.
When the mesh size $\Delta x$ gets to the particle mean free path $\Delta x \simeq \lambda$,
there are no intensive particle collisions around a cell interface to form the Riemann solution.
The condition for the formation of distinct shock, contact, and rarefaction waves in the Riemann solver
cannot be satisfied at all in the mesh refinement process.
Therefore, to keep on using Riemann solver to get mesh converged solution is not
physically founded, even for the NS solution.

\begin{figure}
\centering
\includegraphics[width=0.45\textwidth]{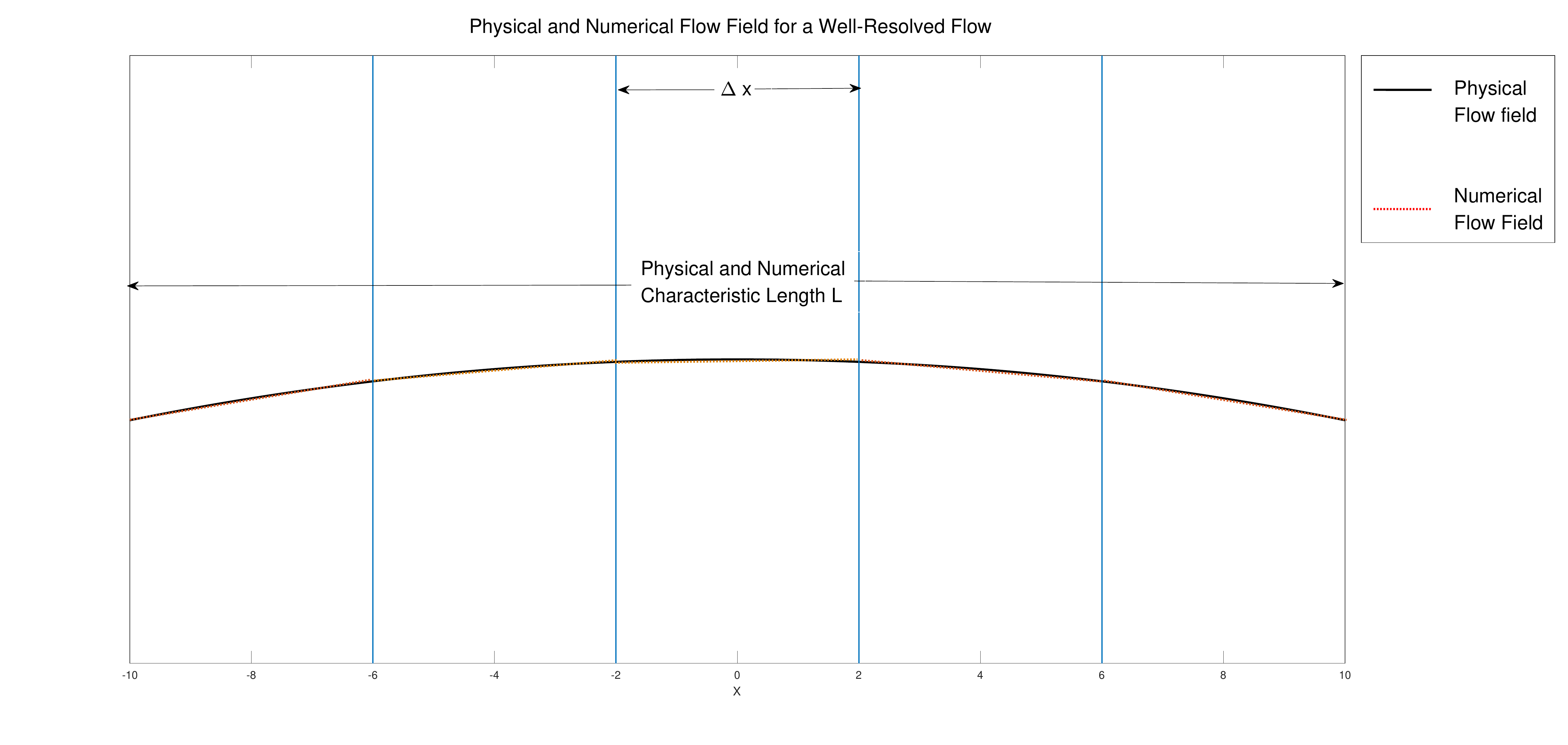}(a)\\
\includegraphics[width=0.45\textwidth]{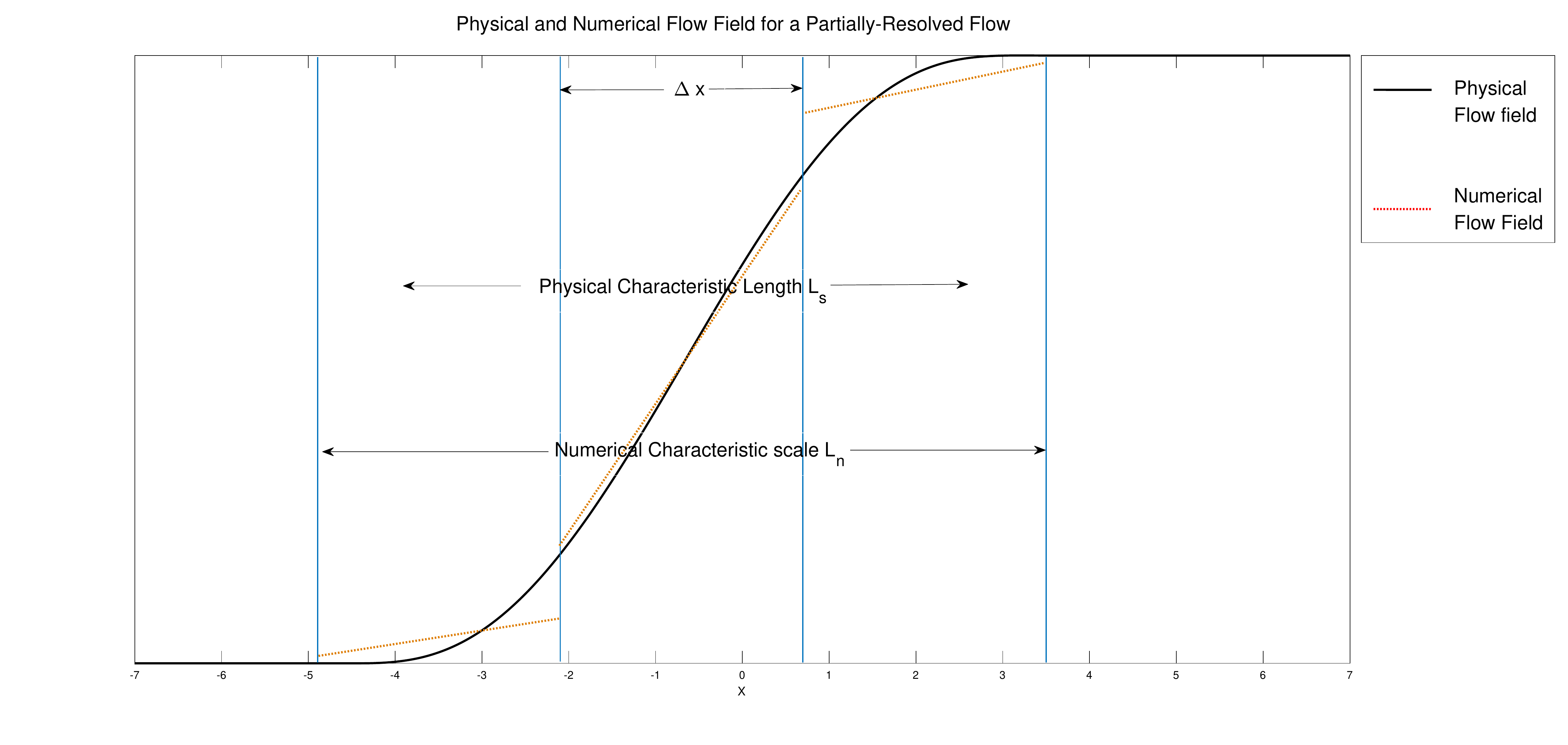}(b)
\includegraphics[width=0.45\textwidth]{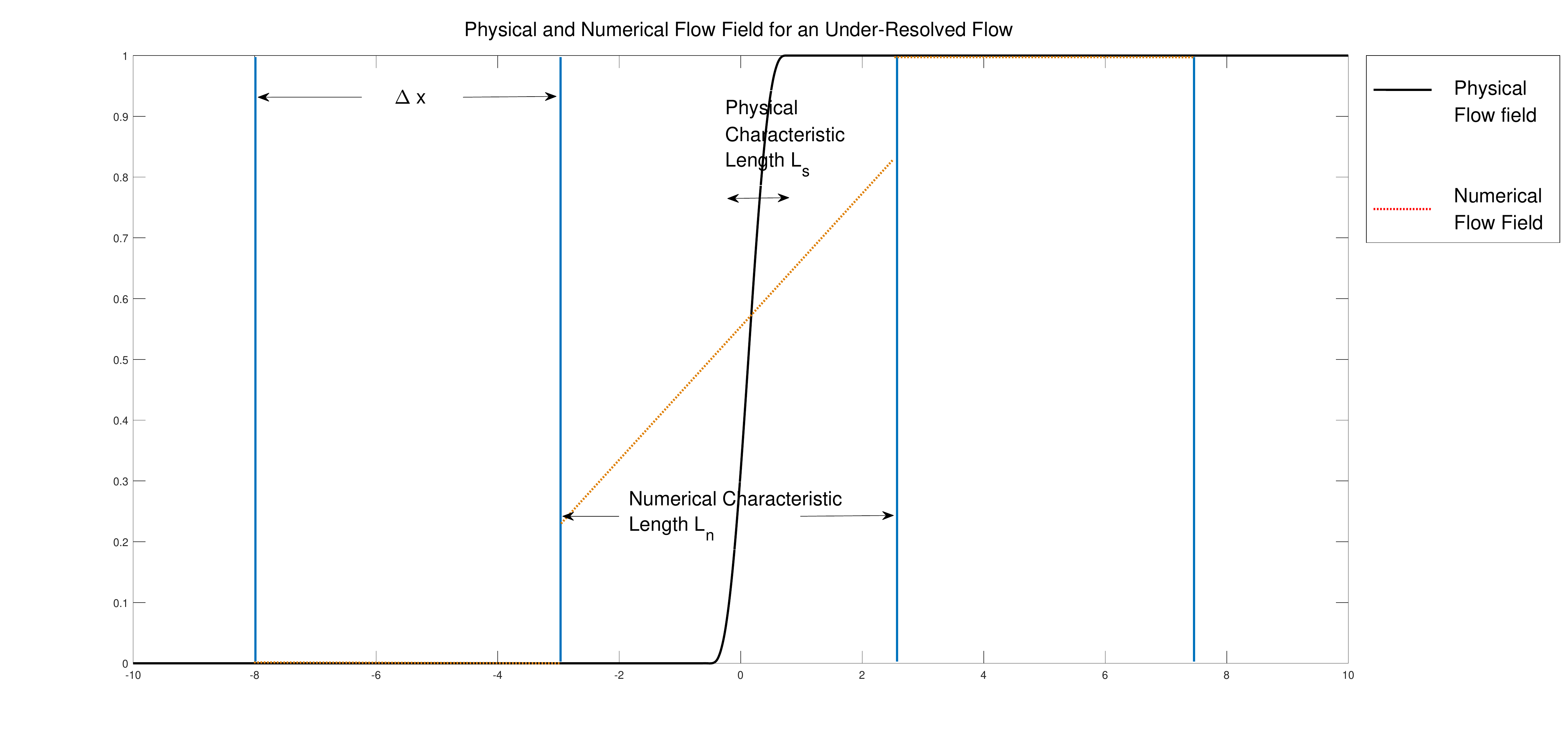}(c)
\caption{Diagram for physical flow field (solid line) and reconstructed numerical flow field (dotted line) for a well-resolved smooth flow (a), a partially-resolved flow (b), and an under-resolved flow with large gradient (c).}
\label{knn}
\end{figure}

The CFD provides a physical evolution solution in a discretized space.
The resolvable flow regimes are controlled by the numerical Knudsen number $K\!n_n$, instead of the physical and cell Knudsen numbers. The numerical Knudsen number is a function of physical and cell Knudsen numbers due to the
 representation of the flow physics in a cell size scale.
As show in Fig. \ref{knn}, for a well-resolved flow, the numerical characteristic length becomes the same as the physical characteristic length.
In such a case, the numerical Knudsen number goes to the physical Knudsen number $K\!n_n\rightarrow K\!n_s$.
For a partially-solved flow, the numerical Knudsen number is in-between the physical Knudsen number and the cell Knudsen number.
For an un-resolved flow, the structure of a numerical solution is a purely numerical one with cell size thickness,
and the characteristic length $L_n\sim\Delta x$.
In such a case, the numerical Knudsen number goes to the cell Knudsen number $K\!n_n\rightarrow K\!n_c$.
Therefore, the numerical Knudsen number is a function of physical and cell Knudsen numbers by some soft minimum functions
\begin{equation}\label{kn-relation}
  K\!n_{n}=\text{softmin}(K\!n_s,K\!n_c).
\end{equation}
The solution obtained with the corresponding $K\!n_n$ is the best result a best multiscale method can provide.
The multiscale method is the one to capture
flow dynamics in the mesh size scale.
Here the multiscale method can smoothly connect the NS and Boltzmann solutions.
With a continuous variation of cell resolution, the multiscale method recovers the dynamics from the rarefied particle transport and collision
to the hydrodynamic wave propagation seamlessly.

\section{Multiscale method: unified gas kinetic scheme}

It is shown in Section 2 that the numerical solution depends on the numerical Knudsen number instead of the physical and cell Knudsen numbers, but they are closely connected.
Here we present a multiscale method, which could give the corresponding flow physics in the cell size scale, but this cell size may not be able to
resolve the physical flow structure.
 The final solution from the multiscale method in any realistic computation
 will depend on the numerical Knudsen number, instead of {the physical and} cell Knudsen number.

 The multiscale method is different from the flow solvers targeting on the single scale NS and Boltzmann equations.
Macroscopic equations such as the Euler and NS systems are valid in hydrodynamic scale, so called continuum flow regime.
The NS equations may fail to provide a physically consistent solution in a highly non-equilibrium flow regime \cite{xu-liu}.
In order to capture the physical solution under cell resolution, a multiscale method is needed.
Following the direct modeling methodology \cite{xu-book,liu2016unified}, the unified gas kinetic scheme (UGKS) has been
constructed with the inclusion of cell size and time step effect, and becomes a multiscale method.
In the modeling of UGKS, the coupling of the particle transport and collision is based on the local flow physics on the time step scale, which is crucial for the multiscale nature of UGKS.
The UGKS is constructed starting from the kinetic equation
\begin{equation}\label{kinetic-eqn}
\frac{\partial f}{\partial t}+ \mathbf{v}\cdot\nabla_\mathbf{x} f = Q,
\end{equation}
where $f(\mathbf{x},t,\mathbf{v})$ is the velocity distribution function and $Q$ is collision term.
For a discrete phase space
$$\mathbf{X}\times \mathbf{V}=\sum_{i,j}\Omega_{ij}=\sum_{i,j} \Omega_{x_i}\times\Omega_{v_j},$$
the evolution of cell averaged distribution function
\begin{equation}\label{f-bar}
  f_{ij}=\frac{1}{|\Omega_{ij}|} \int_{\Omega_{ij}} f(\mathbf{x},t,\mathbf{v}) d\mathbf{v}d\mathbf{x},
\end{equation}
is coupled with the evolution of cell averaged macroscopic conservative variables
\begin{equation}\label{w-bar}
  \mathbf{W}_i=\frac{1}{|\Omega_i|} \int_{\Omega_{i}} \left(\begin{array}{c}\rho \\
  \rho \mathbf{U} \\ \rho E \\ \end{array} \right) d\mathbf{x}.
\end{equation}
Note the above $f_{ij}$ and $\mathbf{W}_i$ are cell averaged variables with a variable cell size.
The governing equation for $f_{ij}$ is different from the Boltzmann equation, which is valid in the kinetic scale only, such as the scale less
than the particle mean free path.

The evolution equation for the cell-averaged velocity distribution function is
\begin{equation}\label{update-f}
  f_{ij}^{n+1}=f_{ij}^{n}-\frac{1}{|\Omega_i|}\int_{t^n}^{t^{n+1}}\oint_{\partial \Omega_i}
  \mathbf{v}\cdot \mathbf{n} f_{\partial \Omega_i}(t,\mathbf{v}_j) ds dt
  +\beta^n Q^n+(\Delta t-\beta^n)Q^{n+1},
\end{equation}
and the evolution equation for the conservative variable is
\begin{equation}\label{update-W}
  W_i^{n+1}=W_i^{n}-\frac{1}{|\Omega_i|}\int^{t^{n+1}}_{t^n}\oint_{\partial \Omega_i}
  \psi \mathbf{v}\cdot \mathbf{n} f_{\partial \Omega_i}(t,\mathbf{v}) ds dt.
\end{equation}
The above two equations are based on the direct modeling of conservation laws in a physical scale of cell size and time step.
In the following formulation, we choose $Q^n$ to be the Boltzmann collision term and $Q^{n+1}$ to be the Shakhov model.
The explicit Boltzmann collision term is calculated by the fast spectral method \cite{wu}, and the evolution equation for the distribution function then becomes
\begin{equation}\footnotesize
  f_{ij}^{n+1}=\left(1+\frac{\Delta t-\beta^n}{\tau^{n+1}}\right)^{-1}
  \left[f_{ij}^n-\frac{1}{|\Omega_i|}\int_{t^n}^{t^{n+1}}\oint_{\partial \Omega_i}
  \mathbf{v}\cdot \mathbf{n} f_{\partial \Omega_i}(t,\mathbf{v}_j) ds dt
  +\beta^nQ(f^n,f^n)+\frac{\Delta t-\beta^n}{\tau^{n+1}}f_{ij}^{+(n+1)}\right],
\end{equation}
where the post collision distribution function $f^+$ is defined as
\begin{equation}\label{shakhov-g}
  f^+=g\left(1+(1-\text{Pr})\mathbf{c}\cdot
   \mathbf{q}\left(\mathbf{c}^2\frac{m}{k_BT}-5\right)\frac{m}{5pk_BT}\right),
\end{equation}
with the Prandtl number Pr,  the peculiar velocity $\mathbf{c}$, and  the heat flux $\mathbf{q}$.
The detail formulation of $\beta^n$ can be found in \cite{liu2016unified}.
The local Maxwellian distribution $g(\mathbf{x},t,\mathbf{v})$ is obtained from the local conservative flow variables.
The numerical fluxes for the distribution function and conservative variables are calculated from the integral solution $f_{\partial \Omega_i}(t,\mathbf{v})$ of the kinetic equation Eq.\eqref{kinetic-eqn} with Shakhov collision term,
\begin{equation}\label{integral-solution}
f_{\partial \Omega_i}(t,\mathbf{v})=\frac{1}{\tau}\int_{t^n}^{t^n+ t} f^+(\mathbf{x}^\prime,t^\prime,\mathbf{v})\mathrm{e}^{-(t-t^\prime)/\tau}dt^\prime+
\mathrm{e}^{-t/\tau}f_0(\mathbf{x}_{\partial \Omega_i}-\mathbf{v}t,\mathbf{v}),
\end{equation}
where $\mathbf{x}'=\mathbf{x}_{\partial \Omega_i}-\mathbf{u}(t-t')$ is the particle trajectory and $f_0$ is the distribution function at time $t^n$.
Based on the integral solution Eq.\eqref{integral-solution}, the UGKS couples two effects in the particle transport for the numerical flux evaluation.
The particle free transport and collision are connected based on the ratio of local time step over the particle collision time,
which is a function of local cell Knudsen number $K\!n_c$.
The UGKS provides dynamics in different flow regime from the particle free transport to the NS and Euler wave
interactions \cite{xu-book}.
The multiscale UGKS can only provide a solution corresponding to the numerical Knudsen number.
In different flow regimes, in order to resolve the physical solution the cell size can be varied significantly in comparison with the local
particle mean free path.
The possible discrepancy between the cell size and the dissipative flow structure thickness is the root of the limitation principle.
Theoretically, the UGKS can always capture the physical solution in the mesh refinement, even in the highly non-equilibrium region.
But, sometimes we don't need such a refined mesh to resolve all dissipative wave structures, such as the shock structure.
Therefore, the limitation principle still applies to UGKS once there is discrepancy between the cell size resolution and
physical structure thickness. As shown next in the numerical examples, the multiscale method provides the best result
a numerical scheme can give efficiently. For a single scale flow solver, such as NS, the mesh converged solution may not be the physical one at all.
For the Boltzmann solver, even though it can provide accurate cell converged solution, but it always requires the cell size on the scale
of particle mean free path, which is too expensive to be used in many applications.

\section{Numerical Experiments}

In this section, six numerical experiments will be used to demonstrate the multiscale solutions under different mesh resolutions.
The examples include the viscous shock tube, density sine wave propagation,
1-D shock interface interaction, 2-D shock interface interaction, shock bubble interaction, and force-driven Poiseuille flow.
In the viscous shock tube problem, the high-order gas-kinetic scheme (GKS) is used for the NS solution.
However, based on the cell Knudsen number distribution, it is realized that for a reasonable NS solution the mesh size
can easily go to the scale of the particle mean free path. The NS modeling becomes questionable in certain regions.
For the density wave propagation and the shock-interface interaction problems,
the NS solutions are compared with the multiscale results under different mesh resolution.
It can be observed that when the cell Knudsen number is small, the NS solutions agree with the multiscale results,
which means that the NS equations give reasonable solutions on large space and time scales.
Relative large differences between the NS and multiscale solutions appear as the cell Knudsen number increases in the mesh refinement process,
and the NS equations don't provide accurate physical solution in the mesh refinement process.
The shock-bubble interaction problem shows that for a fixed cell resolution, as the local Knudsen number increases, the single scale NS solutions  deviate from the multiscale solutions.
Even for small Knudsen number, there is still deviation between the NS and multiscale solutions in the high order moment,
such as the heat flux inside a shock layer.
The study of force-driven Poiseuille flow concludes that the conventional heat flux modeling, i.e. Fourier's law, is incomplete
at all for a flow under large external force field.
{The heat flux induced by the external force $q^F_x$ is on the same order as the Fourier's heat flux induced by the temperature gradient for $K\!n\ge10^{-2}$ and normalized external force term $F_x\ge10^{-2}$. With the decreasing of Knudsen number and external force, the force induced heat flux $q^F_x$ decays linearly with respect to Knudsen number and external force $q^F_x\propto K\!n\cdot F_x$.}
This means that even in the continuum regime, the Navier-Stokes-Fourier equations are not adequate to provide a valid description of physical phenomena. These example show the multiple scale nature of gas dynamics, and the necessity of a multiscale method. At the same time, the final
numerical solution is constrained by the limitation principle even from a multiscale method.

\subsection{Viscous shock tube}

For any time-marching scheme, the time step $\Delta t$ is determined by the CFL condition.
For the Euler equations with the absence of dissipative terms, the time step is merely related to the
local convective wave speed,
\begin{equation}\label{eq: CFL_Euler}
\Delta t \leq \sigma \left( \frac{|u|+a}{\Delta x} \right)^{-1},
\end{equation}
where $\sigma$ is the maximum CFL number on the order of $1$.
For the NS equations, besides the above time limitation, the viscosity coefficient will restrict the time step as well.
With the definition of kinematic viscosity coefficient $\nu$, the time step for the NS equations is determined by \cite{Blazek}:
\begin{equation}\label{eq: CFL_NS}
\Delta t \leq \sigma \left( \frac{|u|+a}{\Delta x} + C \frac{\nu}{\Delta x^2}\right)^{-1},
\end{equation}
where $C$ is a constant on the order of $1$.
The gas kinetic theory shows that the particle collision time $\tau$ is related to the dynamic viscosity coefficient $\mu ( =\nu \rho) $ and local pressure $p$,
\begin{equation}
\tau = \frac{\mu}{p} = \frac{\gamma \nu}{a^2}.
\end{equation}
Then we get
\begin{equation} \label{eq: tau_dt}
\frac{\tau}{\Delta t} \sim \frac{\gamma \nu}{\sigma a^2} \left( \frac{|u|+a}{\Delta x} + C \frac{\nu}{\Delta x^2}\right) = \frac{\gamma}{\sigma} \frac{M\!a}{Re_c} \left( C \frac{M\!a}{Re_c}+ M\!a +1\right).
\end{equation}
With the definition cell Knudsen number in Eq.~\eqref{eq: Kn_c}, Eq.~\eqref{eq: tau_dt} yields
\begin{equation}\label{eq: tau_dt2}
\frac{\tau}{\Delta t} \sim K\!n_c^2 + \frac{M\!a + 1}{C} K\!n_c.
\end{equation}
It indicates that for NS calculation $\tau / \Delta t$ is a quadratic function of the cell Knudsen number $K\!n_c$, and it is affected by the local Mach number $M\!a$ as well.

In recent years, Daru and Tenaud \cite{Daru2001} defined a viscous shock tube problem for high-speed NS solutions.
In this test case, a diaphragm is vertically located in the middle of a square 2-D shock tube with unit side length,
separating the space into two parts with different initial states. Due to the symmetry, the flow in the lower half of the tube is simulated.
The diaphragm is removed instantly at time $t=0$,
resulting in a system of waves including a right-moving shock with the Mach number $2.37$ and their interactions.
Non-slip conditions are used at all wall boundaries. Due to the boundary layer, incident shock, and reflecting shock wave interactions,
complicated flow structures emerge with a large variation of wave structures.
The flow fields at the Reynolds number 200 and 1000 and time $t=1$ are used in the following investigation.

The problem has been studied extensively  and the results presented here are from a simplified high-order gas-kinetic scheme \cite{Zhou2017a},
which provides similar converged solutions as other schemes on commonly agreed mesh size \cite{Zhou2017b}.
Though the GKS does not directly solve the NS equations, it is an accurate NS solver in the well-resolved region due to the use of Chapman-Enskog expansion for the initial gas distribution function reconstruction.
Note that GKS is for NS solution and UGKS is a multiscale method for flow simulation in all regimes.

\begin{figure}[!htb]
\centering
\begin{minipage}[t]{0.45\textwidth}
\centering
\includegraphics[width=\textwidth]{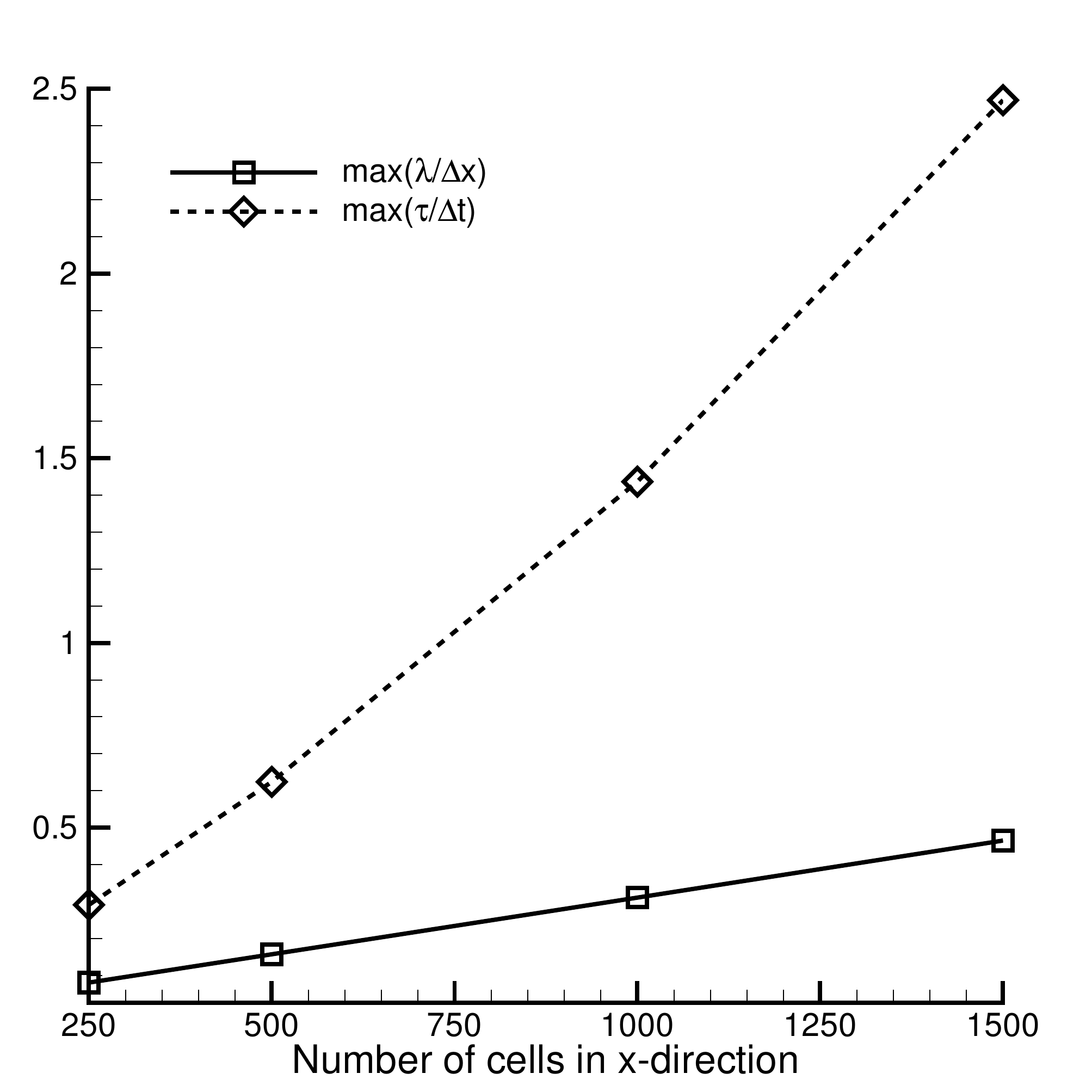}
\centerline{\footnotesize (a)}
\end{minipage}%
\begin{minipage}[t]{0.45\textwidth}
\centering
\includegraphics[width=\textwidth]{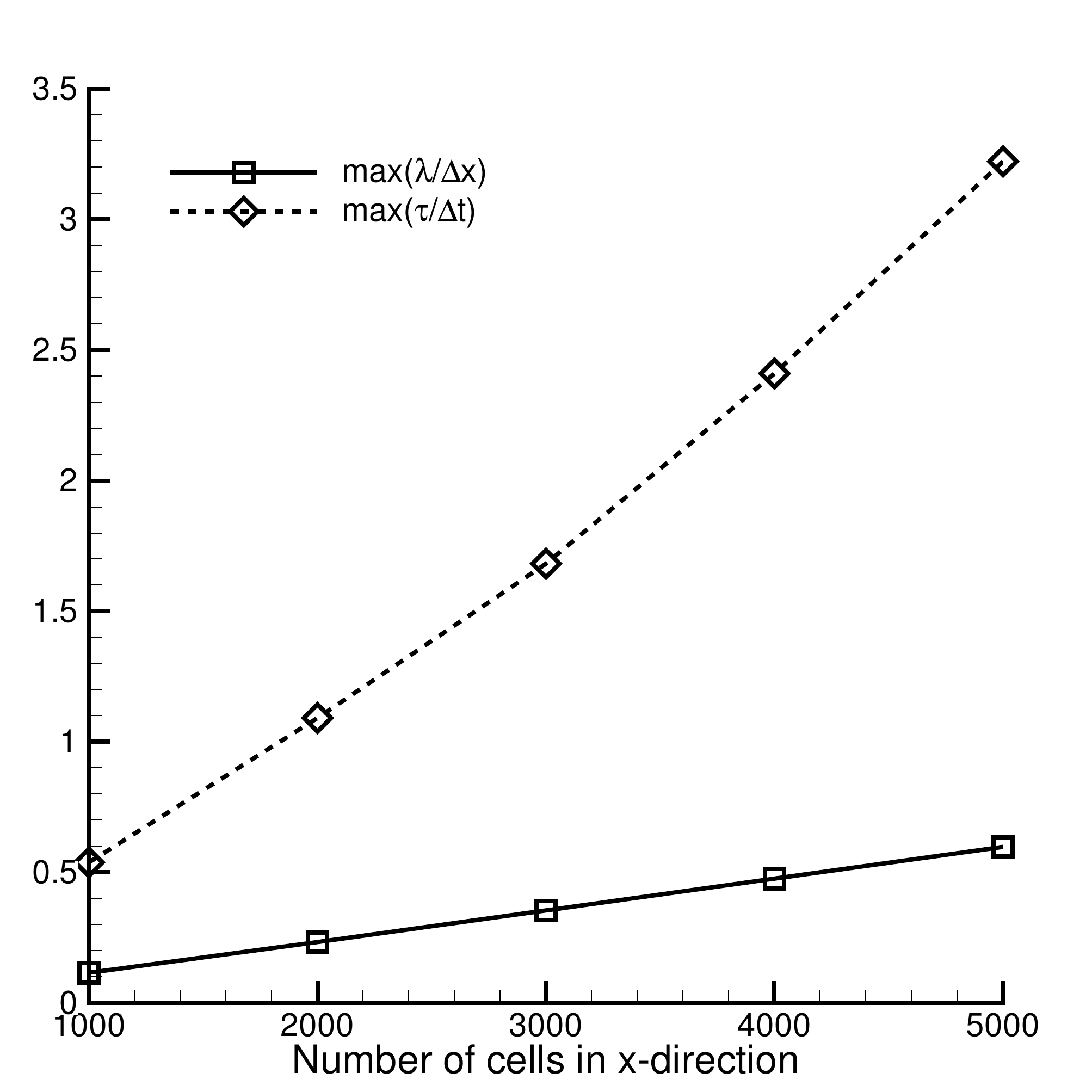}
\centerline{\footnotesize (b)}
\end{minipage}
\centering
\begin{minipage}[t]{\textwidth}
\centering
\caption{Maximum $\lambda / \Delta x$ and $\tau / \Delta t$ for different cell numbers at (a) Re = 200 and (b) Re = 1000.}\label{fig: curve}
\end{minipage}
\end{figure}

Fig.~\ref{fig: curve}(a) shows the variation of two dimensionless quantities $\lambda / \Delta x$ and $\tau / \Delta t$ in the whole computation domain at the Reynolds number 200.
It is clear that the maximum $\tau / \Delta t$ on the $1500 \times 750$ mesh is around 2.5, indicating that the computational time step is already smaller than the particle collision time. In many real engineering applications, it is common to see such a mesh in the resolved
hypersonic viscous flow computations.

It should be noted that although a smaller kinematic viscosity $\nu$ leads to a larger $\Delta t$ according to Eq.~\eqref{eq: CFL_NS}, in order to resolve the dissipative layer corresponding to smaller viscosity coefficient, a smaller cell size $\Delta x$ is also needed.
As a result, $\tau / \Delta t$ can be very large for higher Reynolds number simulations.
To make it clear, Fig.~\ref{fig: curve}(b) plots the same curves as Fig.~\ref{fig: curve}(a) but with a Reynolds number 1000. The mesh for a converged flow field at Re = 200 is $500 \times 250$,
whereas the result is not converged for Re = 1000 case until the mesh is refined to $3000 \times 1500$ \citep{Zhou2017b}.
Therefore, the value $\tau / \Delta t$ is still large on a fine mesh for Re = 1000 case.

For both cases, the physical Knudsen number for the whole problem is on the order of $0.001 \sim 0.01$, because the flow has a Mach number on the order $1$ and the Reynolds number between $100$ and $1000$.
However, with a fine mesh resolution the cell Knudsen number can easily reach the order of $1$.
As shown in the figure, the ratio $\tau / \Delta t$ can go beyond $3$. This clearly indicates the necessity to use a multiscale method for its solution at local high $K\!n_c$ region.

\begin{figure}[!htb]
\centering
\begin{minipage}[t]{0.5\textwidth}
\centering
\includegraphics[width=\textwidth]{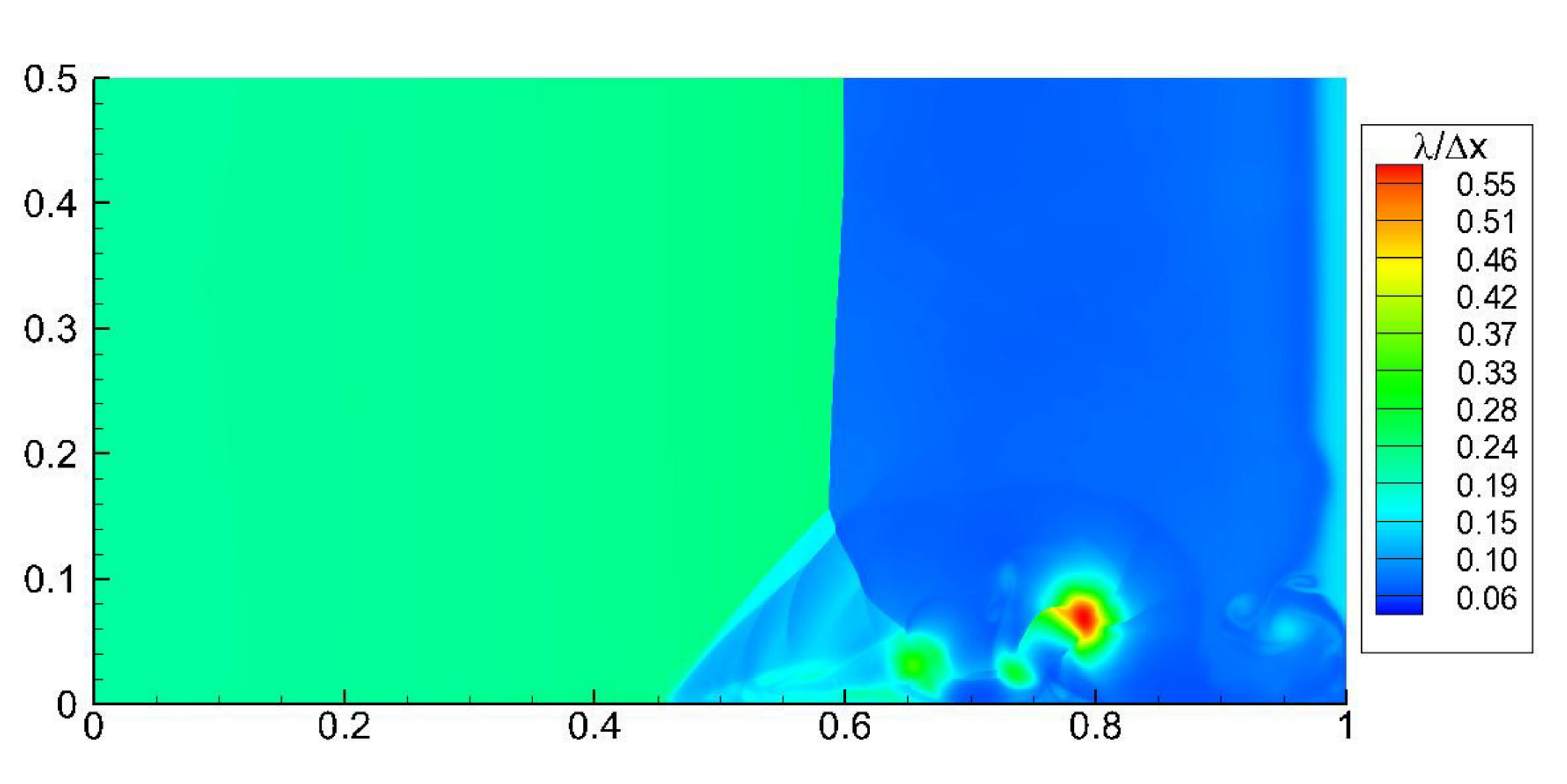}
\centerline{\footnotesize (a)}
\end{minipage}%
\begin{minipage}[t]{0.5\textwidth}
\centering
\includegraphics[width=\textwidth]{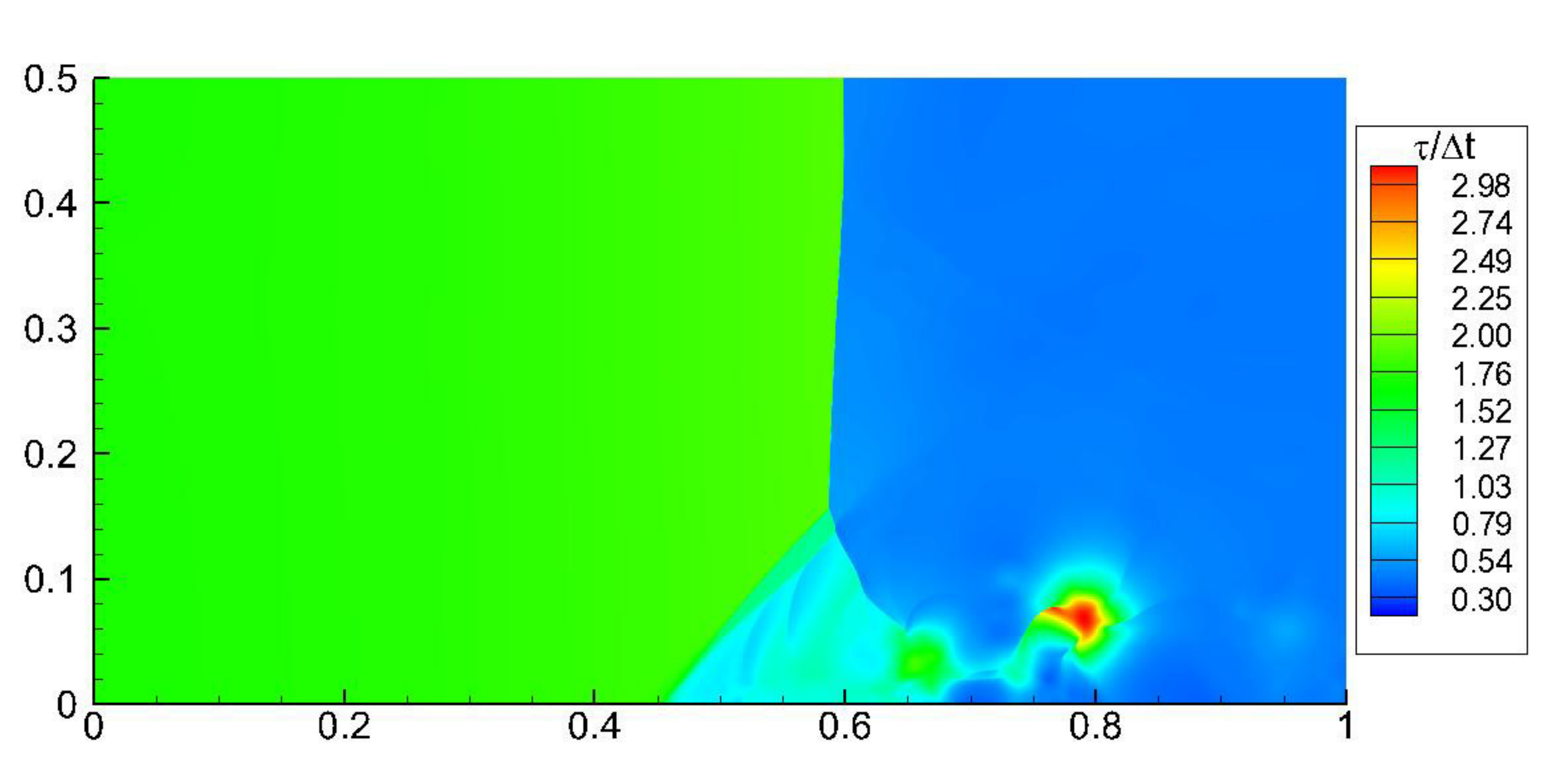}
\centerline{\footnotesize (b)}
\end{minipage}
\centering
\begin{minipage}[t]{\textwidth}
\centering
\caption{Contour map of (a) $\lambda / \Delta x$ and (b) $\tau / \Delta t$ of the viscous shock tube problem at $t=1$ with Re = 1000. $5000 \times 2500$ cells are used for computation.}\label{fig: distribution}
\end{minipage}
\end{figure}

\begin{figure}[!htb]
\centering
\begin{minipage}[t]{0.33\textwidth}
\centering
\includegraphics[width=\textwidth]{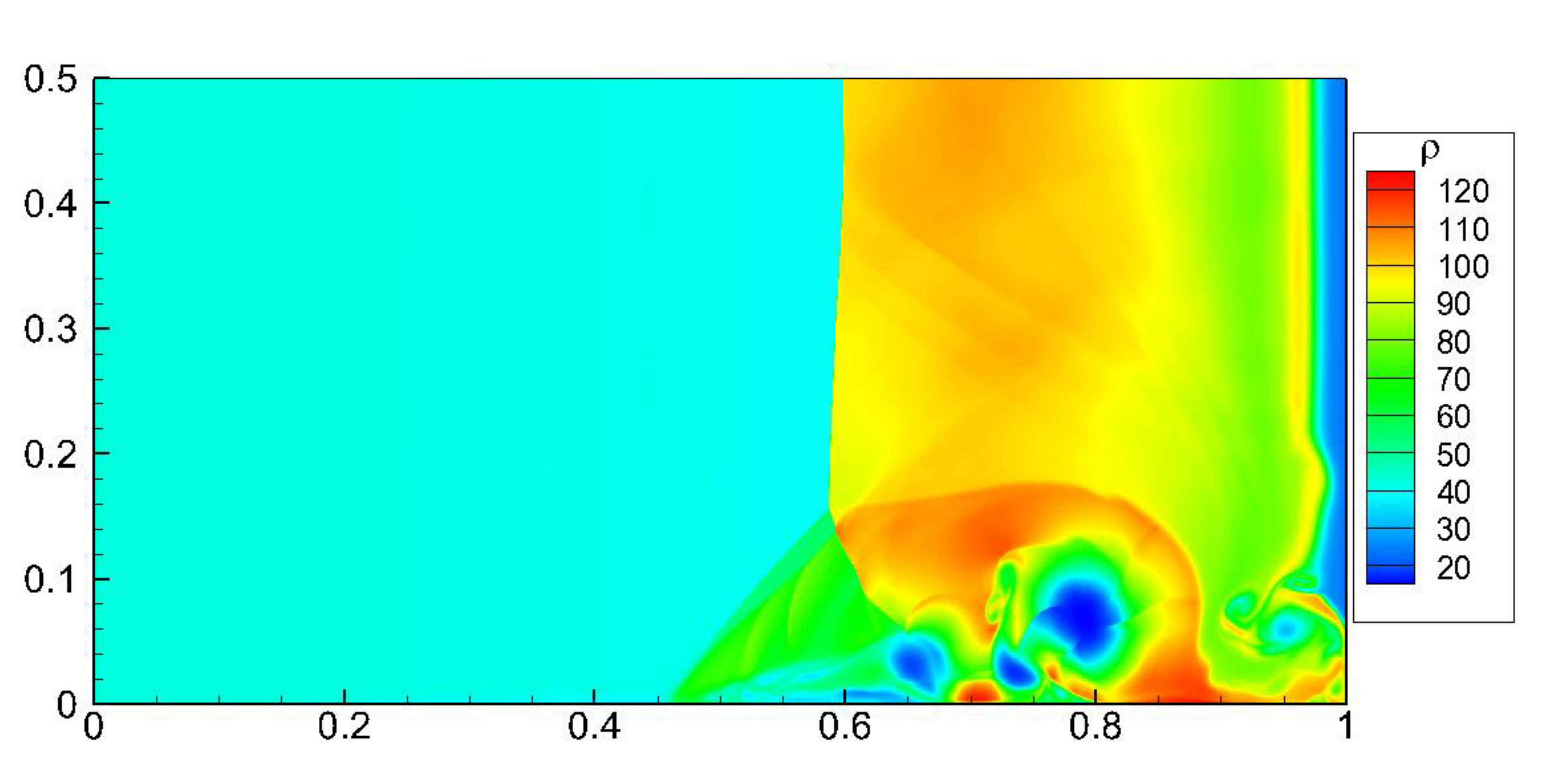}
\centerline{\footnotesize (a)}
\end{minipage}%
\begin{minipage}[t]{0.33\textwidth}
\centering
\includegraphics[width=\textwidth]{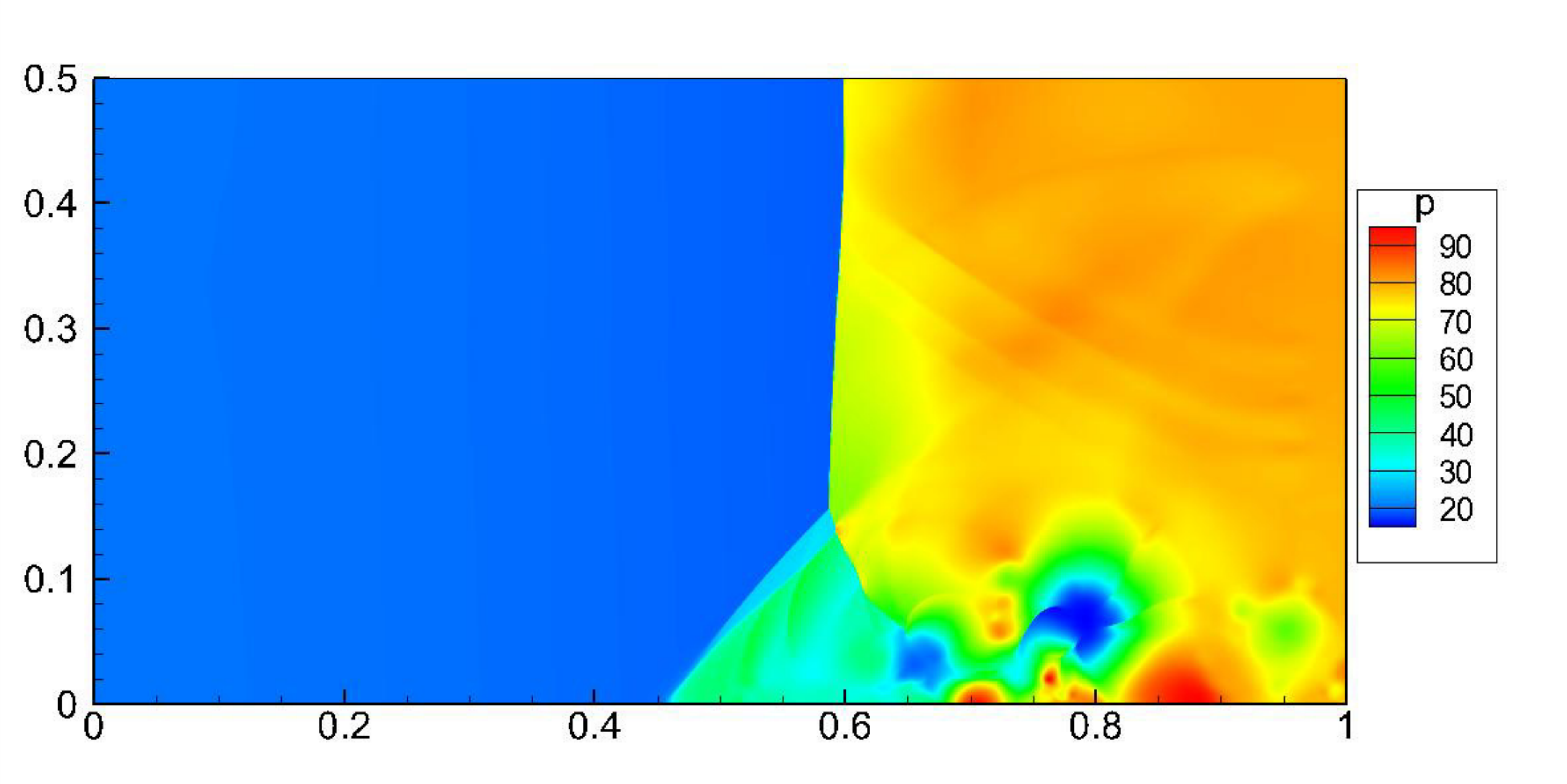}
\centerline{\footnotesize (b)}
\end{minipage}%
\begin{minipage}[t]{0.33\textwidth}
\centering
\includegraphics[width=\textwidth]{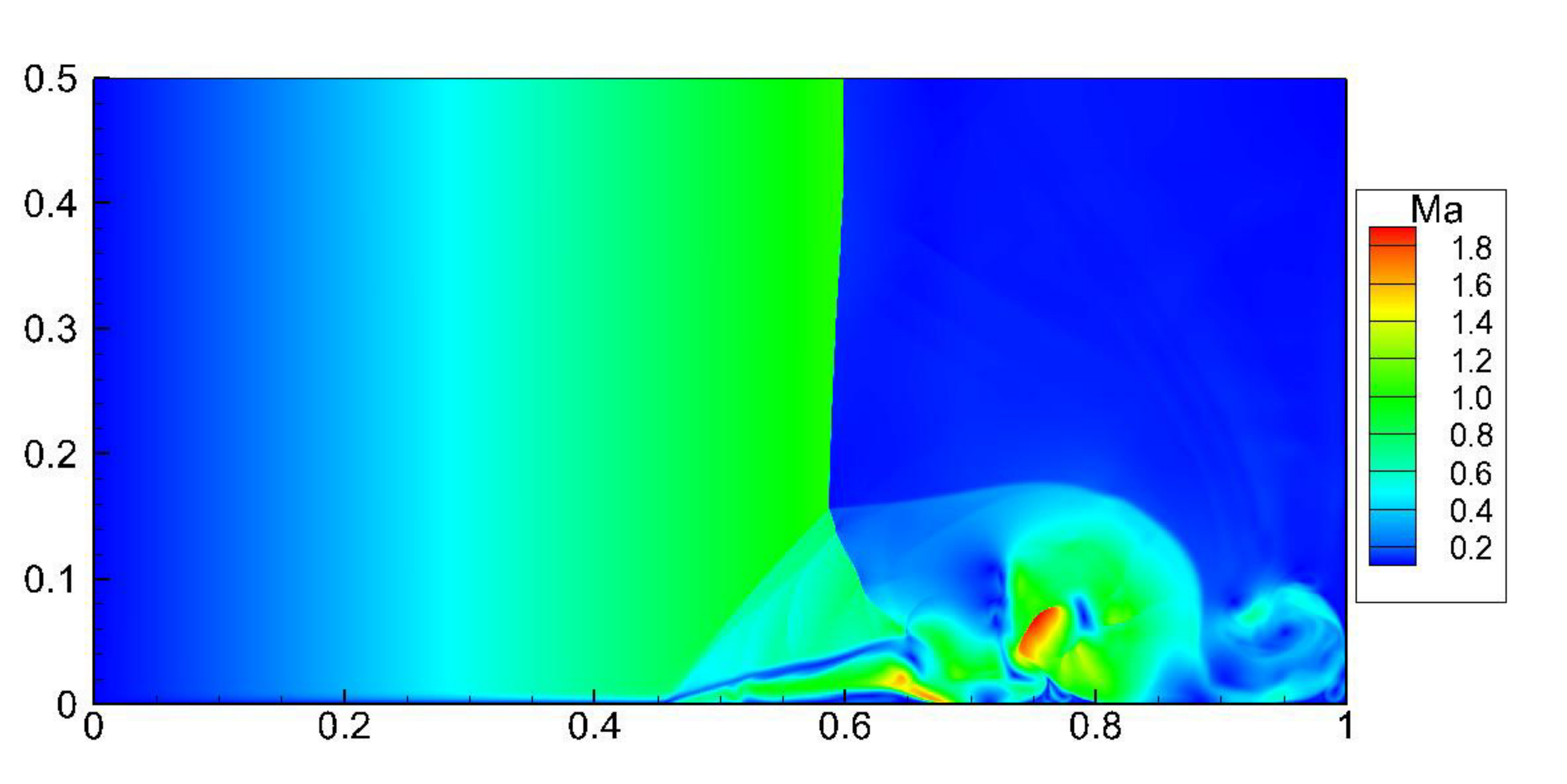}
\centerline{\footnotesize (c)}
\end{minipage}
\centering
\begin{minipage}[t]{\textwidth}
\centering
\caption{Contour map of (a) density, (b) pressure and (c) Mach number of the viscous shock tube problem at $t=1$ with Re = 1000. $5000 \times 2500$ cells are used for computation.}\label{fig: rho_u_ma}
\end{minipage}
\end{figure}

Eq.~\eqref{eq: Kn_c2} shows that $K\!n_c$ is inversely proportional to $\sqrt{\rho p} \Delta x$.
Eq.~\eqref{eq: tau_dt2} indicates that a large Mach number leads to a large $\tau / \Delta t$.
For the Re = 1000 case, the distributions of $\lambda / \Delta x$ and $\tau / \Delta t$ are presented in Fig.~\ref{fig: distribution}.
And the distributions of density, pressure and Mach number are shown in Fig.~\ref{fig: rho_u_ma}.
All of them are obtained with a $5000 \times 2500$ points mesh, which is a commonly agreed mesh resolution for a converged NS solution.

For the NS equations, the particle collision time is a true physical parameter which is related to the viscosity coefficient.
The current CFD computation seeks for mesh converged solution, and the time step can reach a very small value with mesh refinement.
As shown in the calculation, the mesh size can go to the order of particle mean free path in some regions,
and the time step is even smaller than the particle collision time.
Under such a physical situation, the particle will take a limited number of collisions within a time step.
The flux modeling in the CFD method should consider the particle transport mechanism,
where there is no corresponding macroscopic wave interactions.
However, the CFD for the NS solutions is mostly based on the the Godunov-type scheme with distinct wave interactions in the
Riemann solver for the inviscid flow and the central difference for the viscous terms.
In order to get the Riemann solution starting from a cell interface discontinuity,
such as the formation of shock, contact, and rarefaction waves,
an infinite number of particle collisions are assumed.
This is clearly inconsistent with the physical reality in this problem.
In other words, there have intrinsic modeling inaccuracy in modern CFD method under the NS framework, even for the continuum flow simulations.
The Riemann solver based CFD for high speed compressible may lose its physical foundation underlying the numerical algorithm.

\subsection{Density wave propagation}

\begin{figure}
\centering
\includegraphics[width=0.45\textwidth]{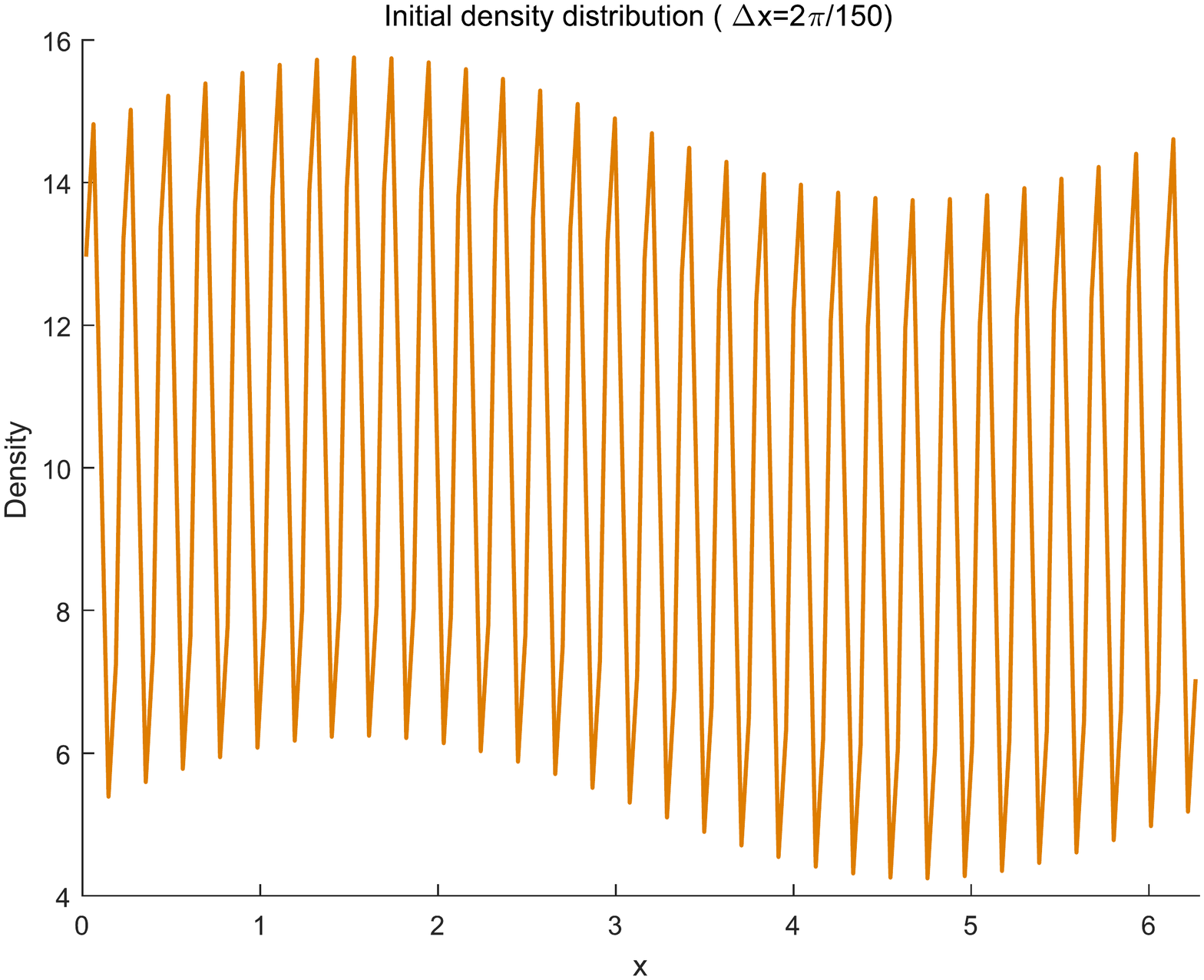}{a}
\includegraphics[width=0.45\textwidth]{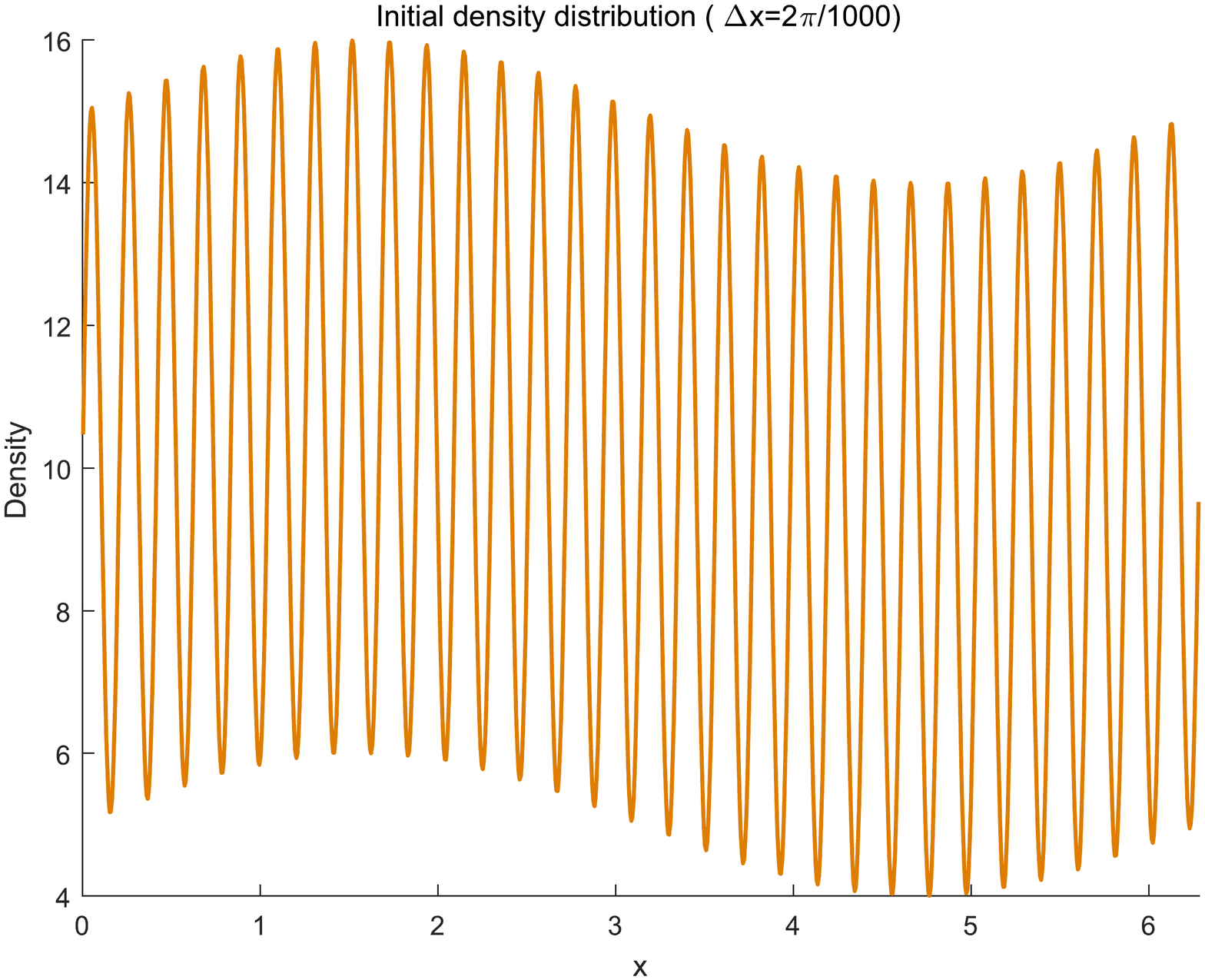}{b}
\caption{The initial density distribution of the density wave propagation on coarse mesh (a) and on fine mesh (b).}
\label{sine-initial}
\end{figure}

\begin{figure}
\centering
\includegraphics[width=0.45\textwidth]{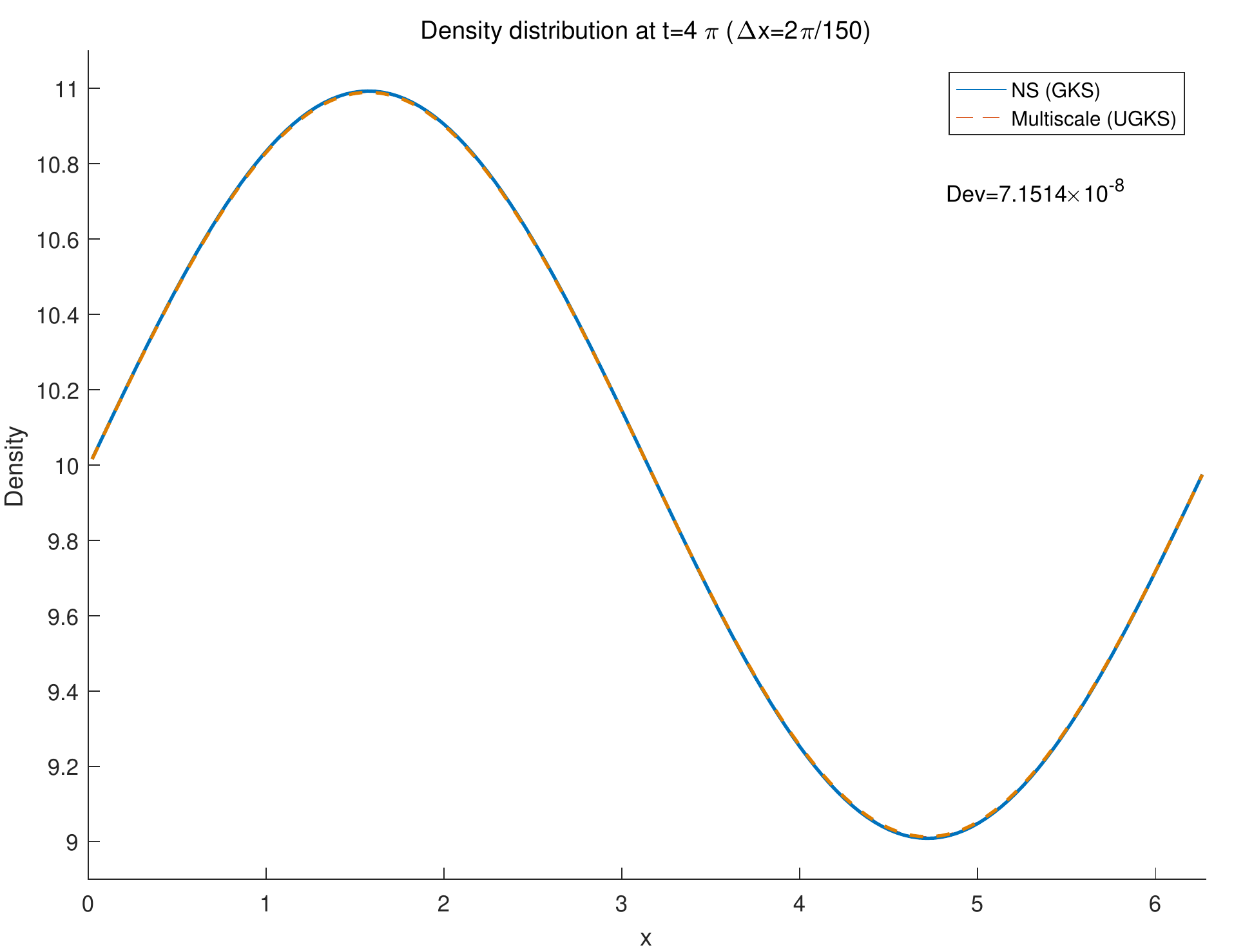}{a}
\includegraphics[width=0.45\textwidth]{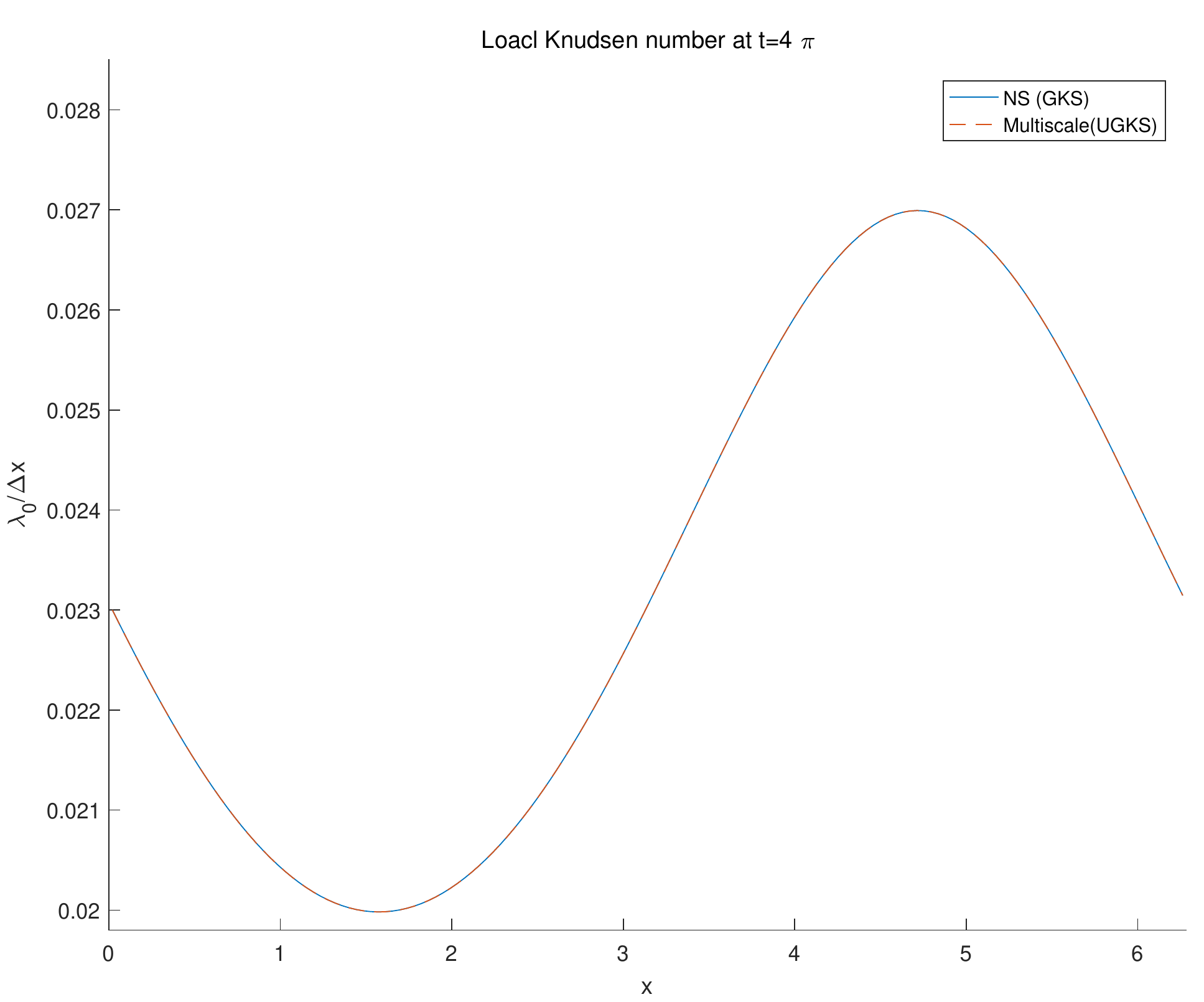}{b}
\caption{The density distribution at $t=4\pi$ on coarse mesh (a),
and the cell Knudsen number at $t=4\pi$ on coarse mesh (b).}
\label{sine-coarse}
\end{figure}

\begin{figure}
\centering
\includegraphics[width=0.45\textwidth]{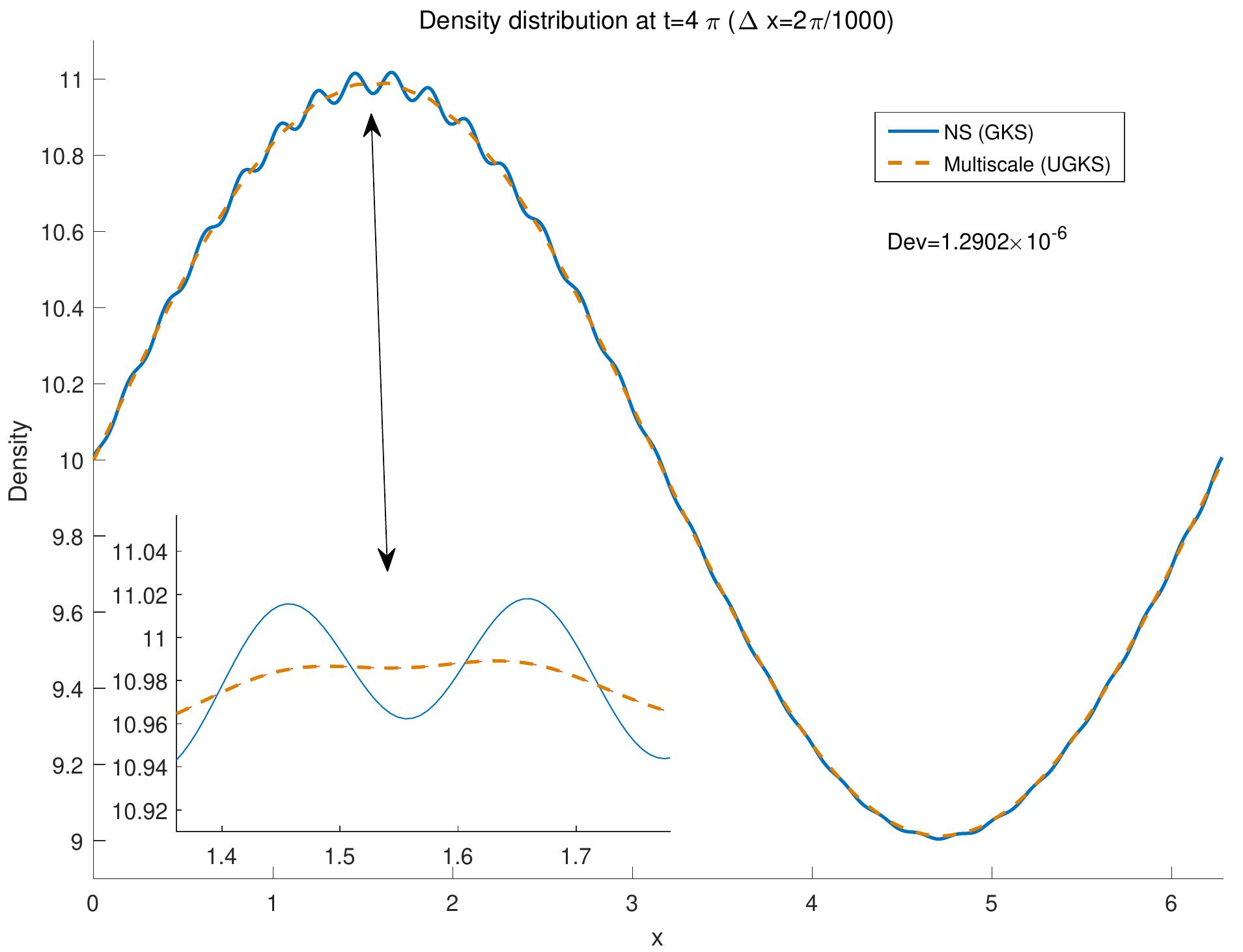}{a}
\includegraphics[width=0.45\textwidth]{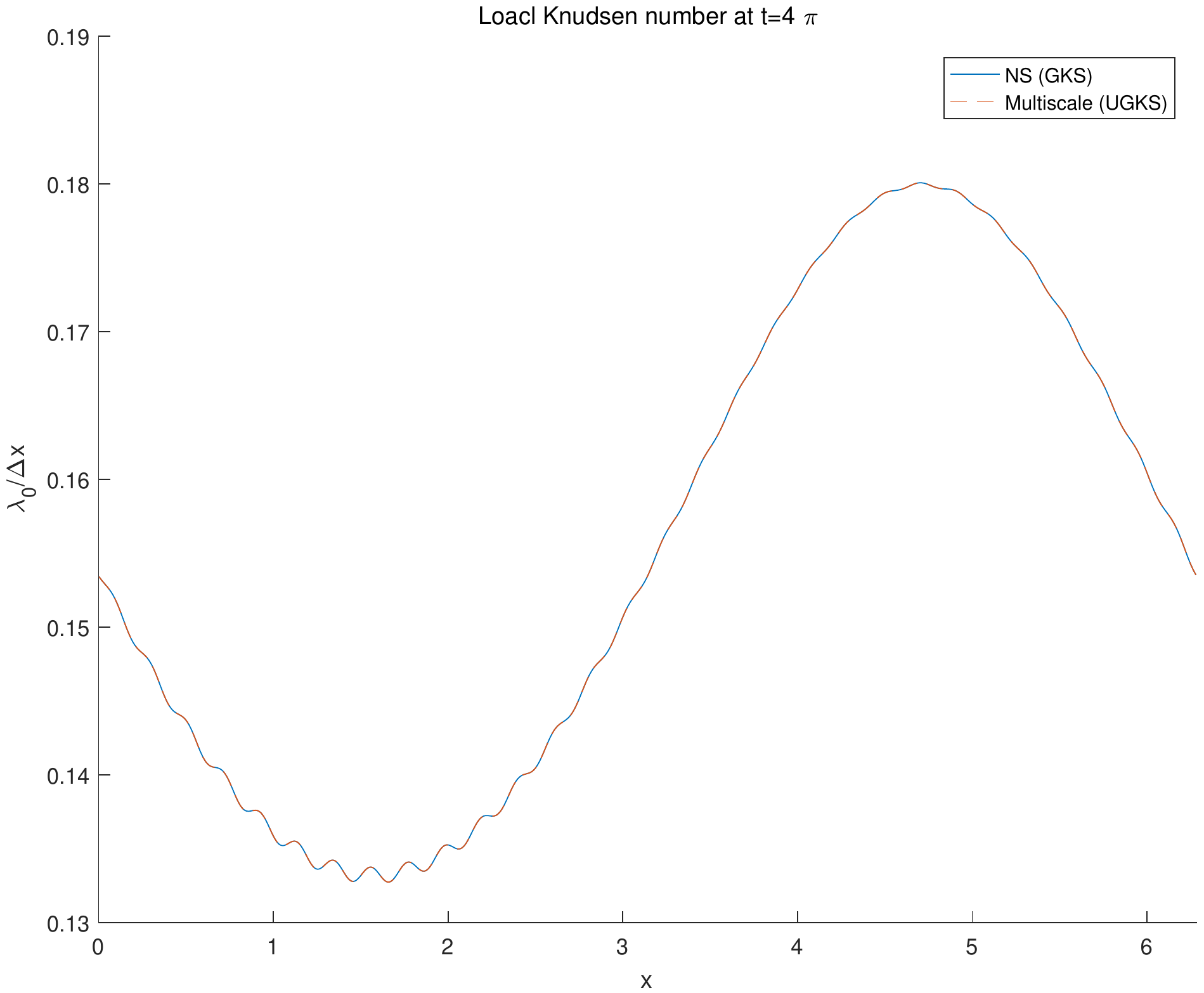}{b}
\caption{The density distribution at $t=4\pi$ on fine mesh (a),
and the cell Knudsen number at $t=4\pi$ on fine mesh (b).}
\label{sine-fine}
\end{figure}

In the next three test cases, we compare the NS and multiscale solutions under different cell Knudsen numbers.
First, we study the propagation of a density wave in argon gas.
The initial condition is set as
$$(\rho, U, p)=(5\sin(30x)+\sin(x)+10, 1.0, 0.5).$$
The computational domain is $[0,2\pi]$ with periodic boundary condition.
The variable hard sphere (VHS) model is used with the viscosity temperature dependency index $\omega=0.81$.
The Knudsen number is set to be $K\!n=5\times 10^{-2}$.
The velocity space is truncated from $-5$ to $5$ with $200$ velocity grids to minimize the velocity integration error.
A coarse mesh with $150$ cells and a fine mesh with $1000$ cells in the physical computational domain are adopted.
The UGKS is used to obtain the physically consistent cell averaged solutions.
The NS solutions are obtained by the gas kinetic scheme (GKS) \cite{xu-book}.
For any quantity $Q$, the deviation between the UGKS and NS solutions is calculated as
\begin{equation}\label{deviation}
  Dev=\frac{\|Q^{NS}-Q^{UGKS}\|_{L^2}}{\|Q^{UGKS}\|_{L^2}}.
\end{equation}
The initial density distribution is shown in Fig. \ref{sine-initial}, and the initial condition are numerically resolved on both meshes.
The density distribution as well as the cell Knudsen number at $t=4\pi$ are shown in Fig. \ref{sine-coarse}-\ref{sine-fine}, and the deviations are listed in Table \ref{sine-table}.

On a relative coarse mesh, the NS solution agrees well with the UGKS solution.
On such a numerical cell size scale, the cell Knudsen number is relatively small $K\!n_{c}\sim1\times10^{-2}$,
and the cell averaged velocity distribution is close to the local equilibrium.
Under the limitation principle, the numerical solution is the one corresponding to cell Knuden number.
When the grid points in physical space increase to $1000$, the local non-equilibrium effect in the high frequency {mode} appears.
Under such a resolution, the local flow physics is resolved by the mesh size, and the numerical Knudsen number for the solution is
the same as the physical Knudsen number.
The NS solution deviates from the physical one.
The results show that the physical solution is  more dissipative than the NS solution due to the inclusion of particle transport mechanism
in such a scale.
So, even with mesh converged solution, the NS solution still doesn't capture the physical solution in the corresponding resolution.
The UGKS can accurately recover the attenuation of high frequency wave in the experiments \cite{wang-xu}.

\begin{table}
\centering
    \begin{tabular}{| l | l | l | l |}
    \hline
    Density wave propagation & $K\!n_c$ & $\tau/\Delta t$ & $Dev$ \\ \hline
    Fine mesh & $\sim 2\times 10^{-1}$ & $\sim 3\times 10^{-2}$ & $ 1.29\times 10^{-6}$\\\hline
    Coarse mesh  & $\sim 2\times 10^{-2}$ & $\sim 3\times 10^{-3}$ & $ 7.15\times 10^{-8}$\\
    \hline
    \end{tabular}
    \caption{Parameters under different meshes, deviation between NS and multiscale solution for the density wave propagation.}
    \label{sine-table}
\end{table}

\subsection{Shock interface interaction}

In the shock interface interaction case, non-equilibrium effect beyond the NS modeling will emerge when the shock wave hits the interface with
large gradients in the flow variables.
The NS solutions will be  compared with the multiscale solutions under different cell resolution.
On the relative coarse mesh case, the cell Knudsen number is {small} and  the NS solver will have the same results
as the multiscale {ones}, because under limitation principle this is the best result a multiscale method can have under such a mesh resolution.
When the cell size decreases and cell Knudsen number increases, the NS solutions deviate from the multiscale ones, which is due to the limitation of NS modeling in capturing the non-equilibrium phenomena.

\begin{figure}
\centering
\includegraphics[width=0.45\textwidth]{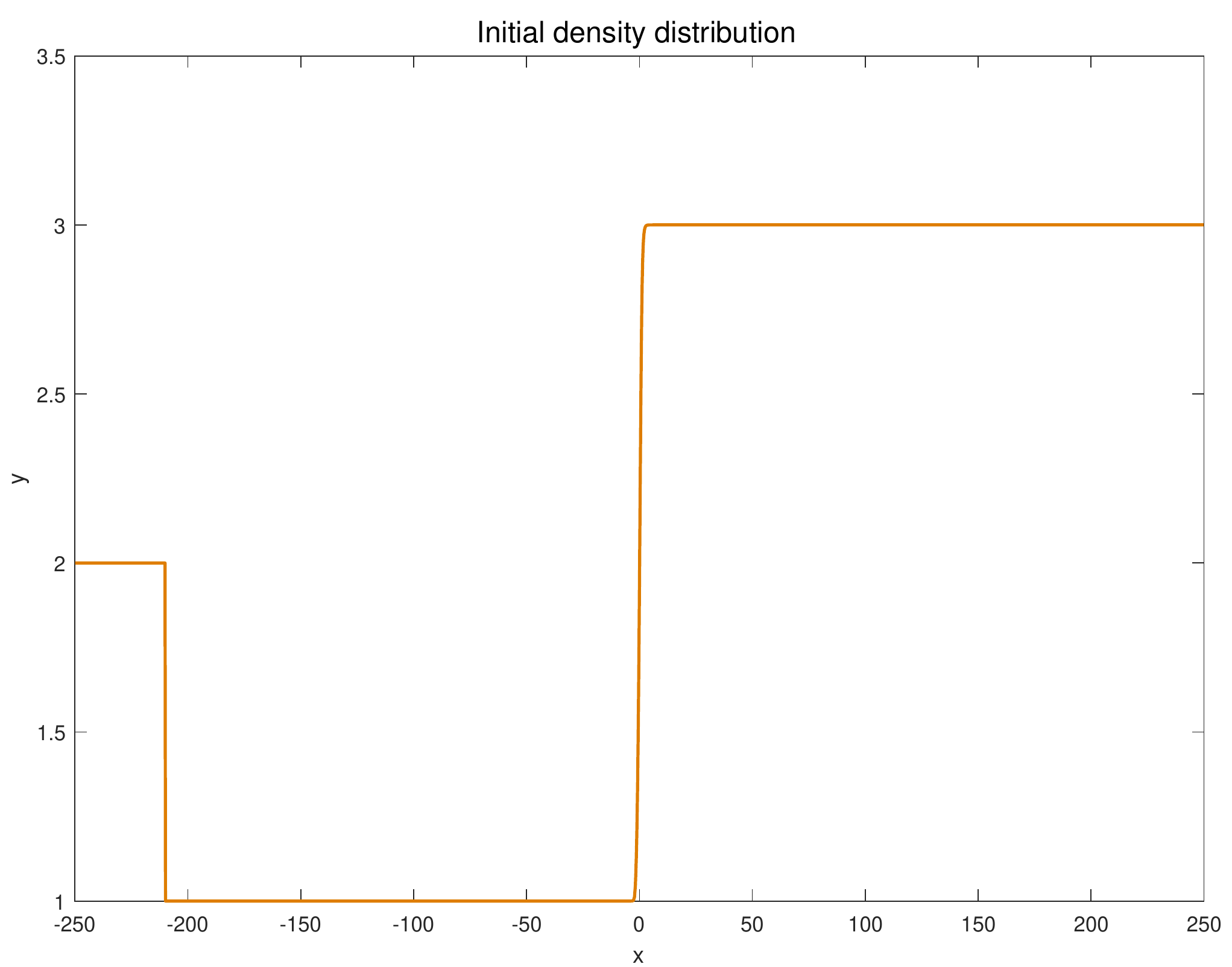}
\caption{Initial condition of the shock-contact interaction}
\label{sc-initial}
\end{figure}

\begin{figure}
\centering
\includegraphics[width=0.45\textwidth]{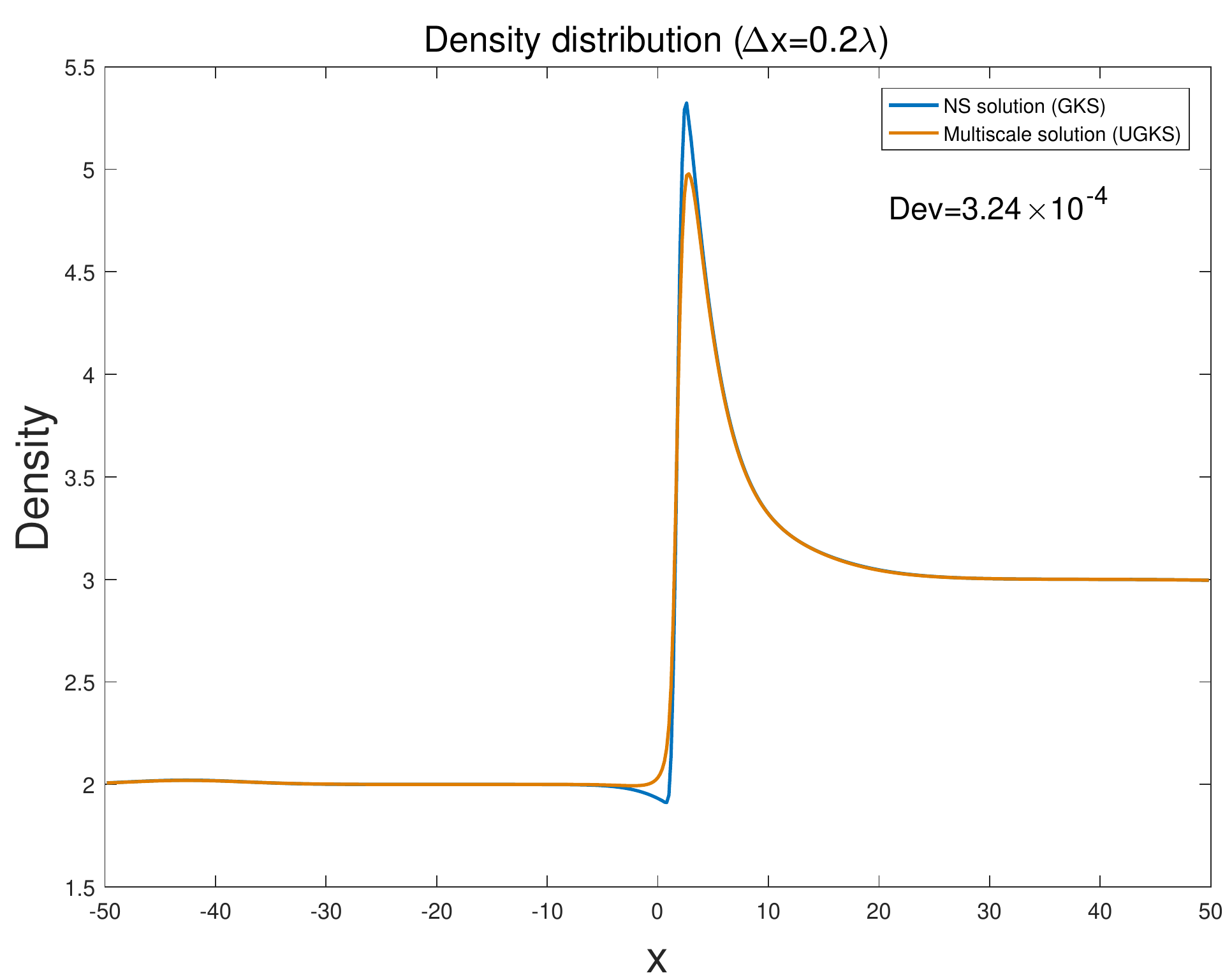}{a}
\includegraphics[width=0.45\textwidth]{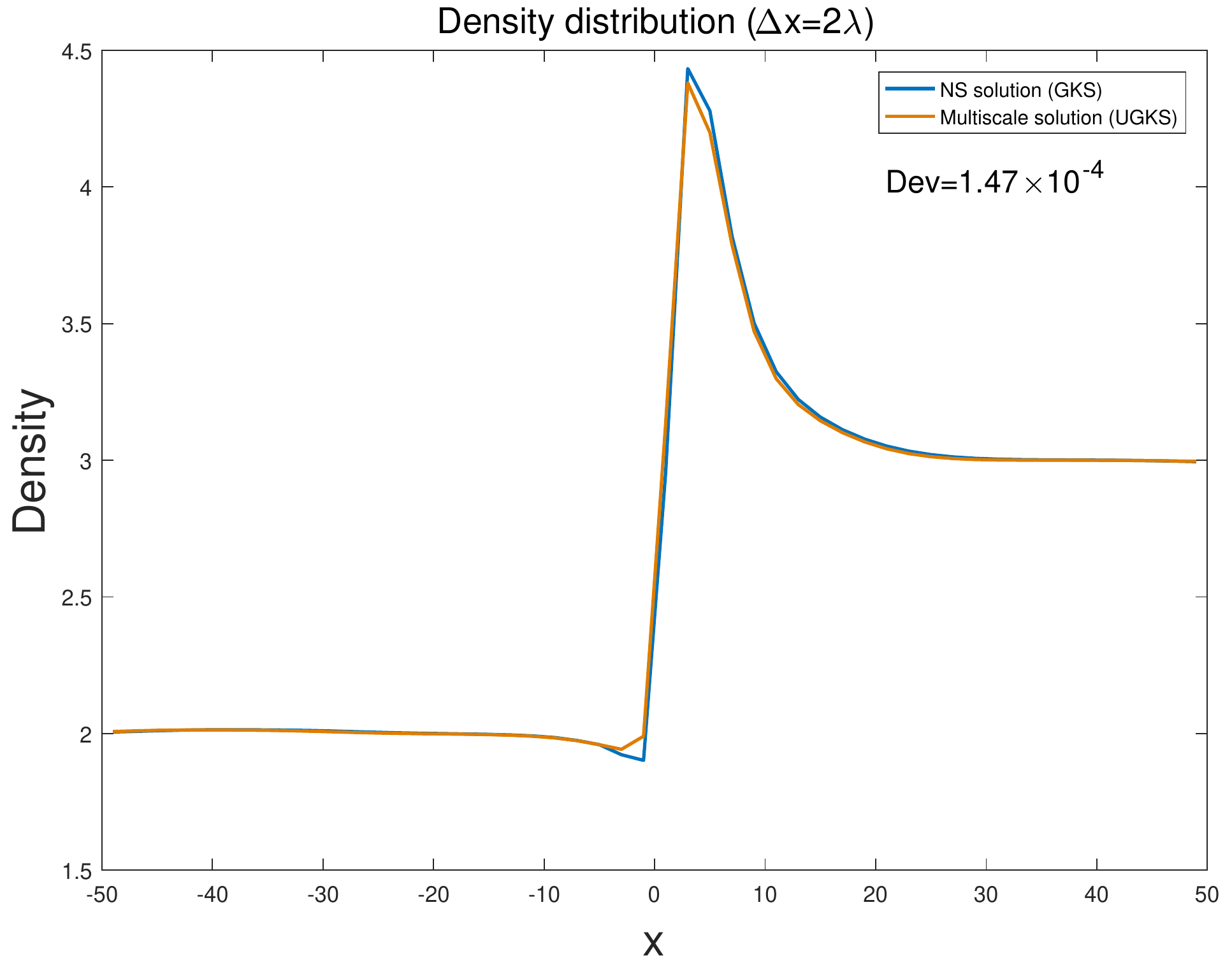}{b}
\caption{Density of the shock-contact interaction at $t=80$. The solutions of NS and multiscale method are compared under
a fine mesh (a) and a coarse mesh (b).}
\label{sc}
\end{figure}

For 1-D shock interface interaction, the working gas is argon modeled by VHS.
As shown in Fig. \ref{sc-initial}, initially a normal shock wave with $M\!a=3.0$ is situated with its density barycenter at $x=0$.
The upstream initial condition is set as
\begin{equation}\nonumber
(\rho, u, T)=\left\{
\begin{aligned}
  &(1.0,2.74,1.0) \quad x \geq -210,\\
  &(2.0,2.74,0.5) \quad x<-210.
\end{aligned}\right.
\end{equation}
The x-axis is normalized by the shock upstream mean free path, which means the Knudsen number is one.
The NS solutions are compared with the multiscale solutions under a fine mesh $\Delta x=0.2$ and a coarse mesh $\Delta x=2$.
For UGKS, the velocity domain is $[-8,8]$ covered by $100$ velocity points.
The solutions at $t=80$ are shown in Fig. \ref{sc} when the interface moves to the origin and interacts with the shock wave.
As shown in Table \ref{sc-table}, a larger deviation is observed under a fine mesh since the NS equations
are not able to recover the physical structure of shock wave, which is supposedly resolved in such a fine mesh case.
Under a coarse mesh, the shock wave structure is not resolved by the numerical resolution, and the physical shock wave is replaced by a numerical discontinuity, such as the shock capturing scheme.
Therefore, both the NS and the multiscale method give the same result under the limitation principle.

\begin{table}
\centering
    \begin{tabular}{| l | l | l | l |}
    \hline
    {Shock interface interaction} & $K\!n_c$ & $\tau/\Delta t$ &  $Dev$ \\ \hline
    Fine mesh  & $\sim 5$ & $\sim 50$ & $ 3.24 \times 10^{-4}$\\ \hline
    Coarse mesh & $\sim 0.5$ & $\sim 5$ & $ 1.47\times 10^{-4}$\\
    \hline
    \end{tabular}
    \caption{Parameters under different meshes, deviation between NS and multiscale solution for the shock interface interaction.}
    \label{sc-table}
\end{table}

\subsection{Richtmyer-Meshkov instability}

\begin{figure}
\begin{minipage}{0.4\linewidth}
  \centerline{\includegraphics[width=1.0\linewidth]{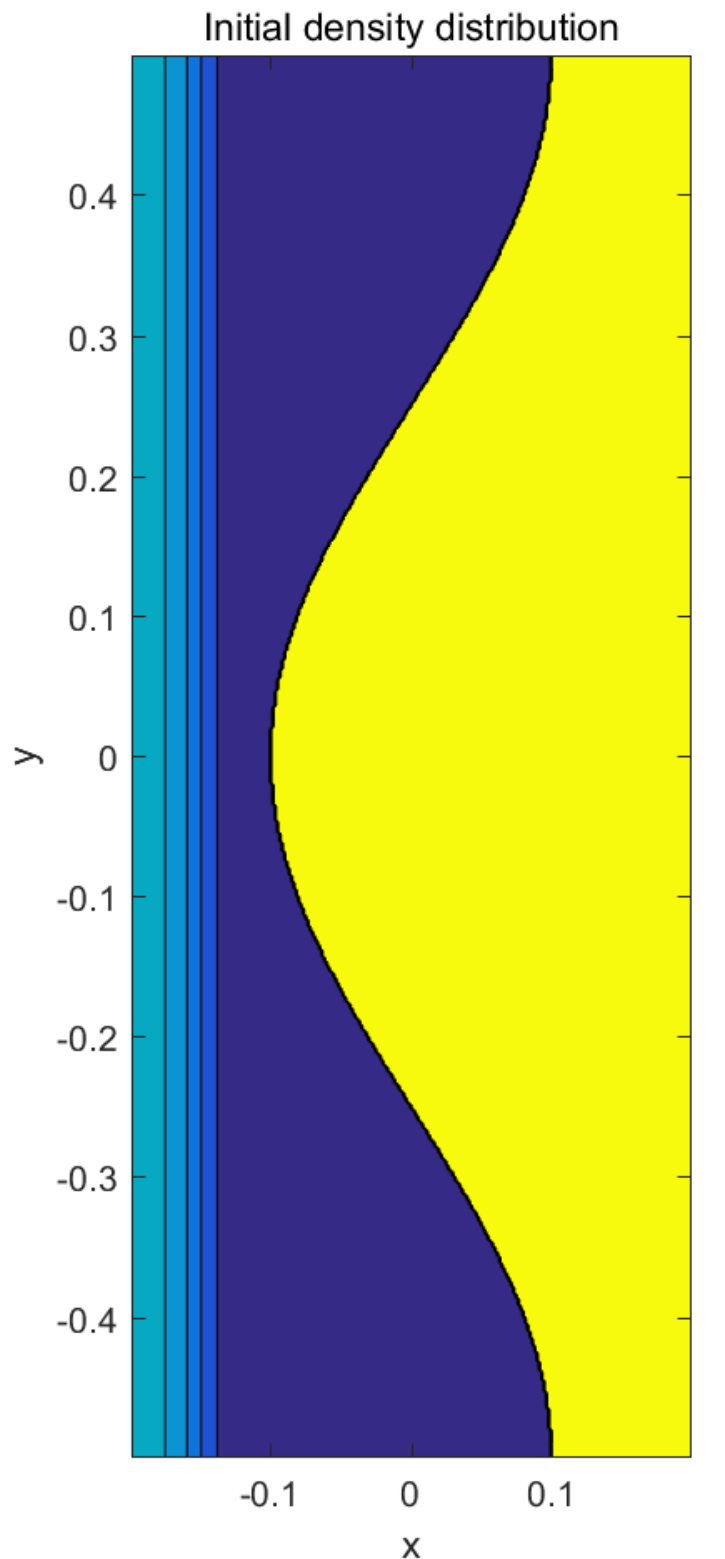}}{a}
\end{minipage}
\hfill
\begin{minipage}{0.5\linewidth}
    \begin{minipage}{1.0\linewidth}
      \centerline{\includegraphics[width=1.0\linewidth]{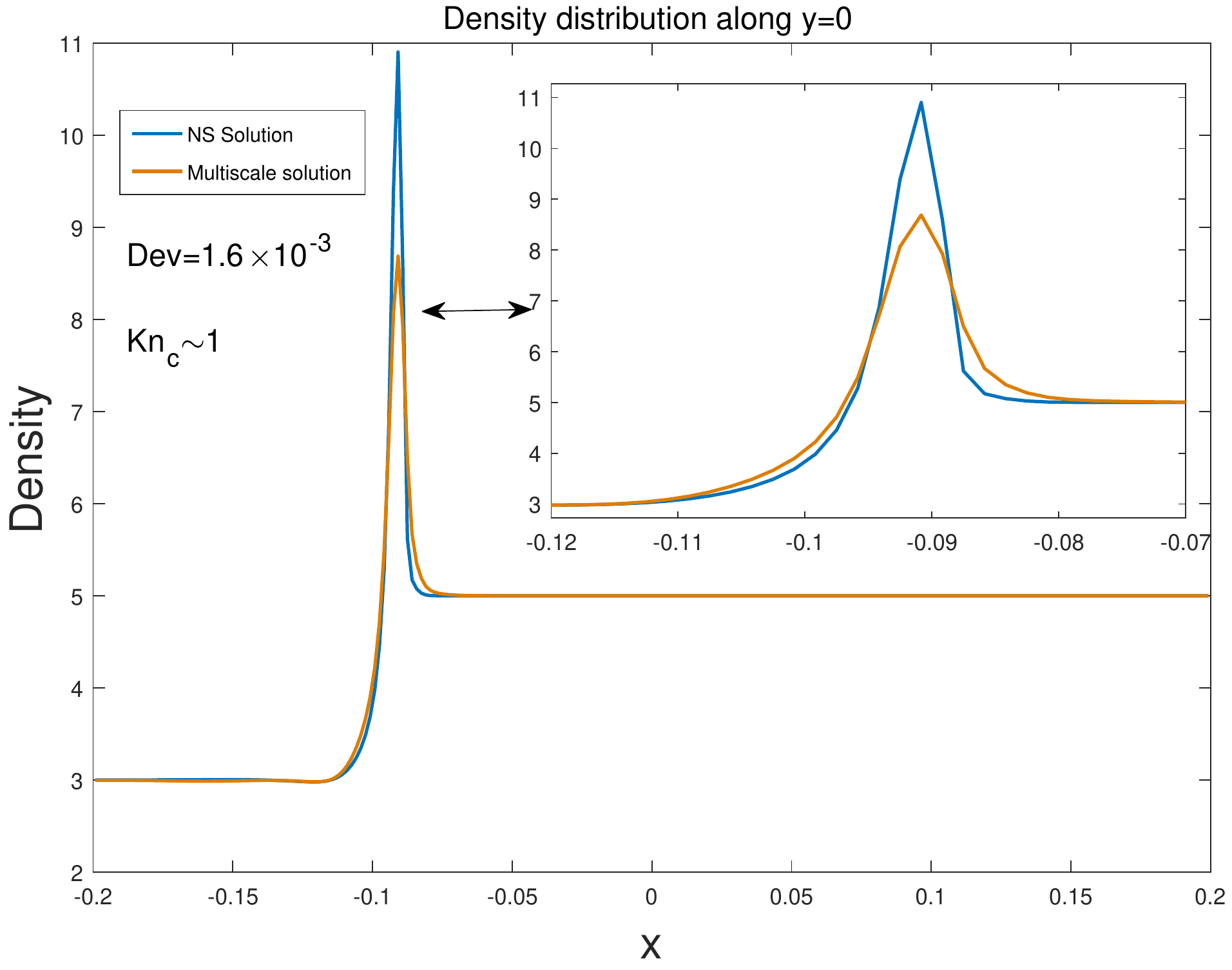}}{b}
     \end{minipage}
    \vfill
      \begin{minipage}{1.0\linewidth}
       \centerline{\includegraphics[width=1.0\linewidth]{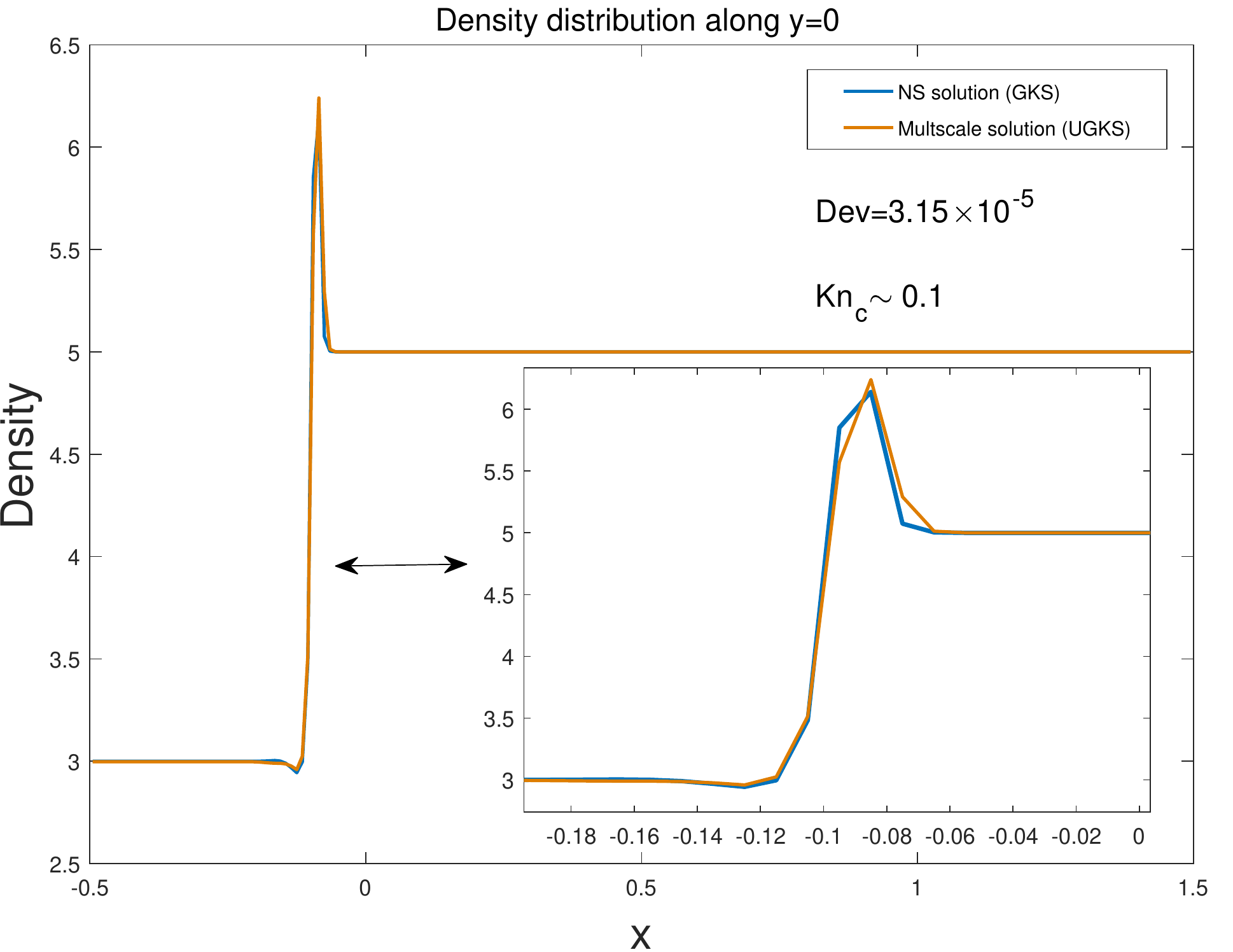}}{c}
       \end{minipage}
\end{minipage}
\caption{Left figure shows the initial condition of RM instability. Right figures show the density distribution along $y=0$ at $t=0.026$.
The NS solutions and multiscale solutions are compared under a fine mesh (b) and a coarse mesh (c).}
\label{rm1}
\end{figure}

The Richtmyer-Meshkov (RM) instability is caused by a shock wave passing though a contact interface.
Vortices will be generated during the passage of the shock wave and trigger interface instability.
In the following, the 2-D RM instability is numerically studied on different time scales.
The interaction between shock and interface is studied on the time scale $t\sim10\tau$,
and the development of the instability is studied on the time scale $t\sim10^{5}\tau$.
The NS solutions are calculated by GKS while the UGKS provides the multiscale solutions.
At the starting time, the initial condition is shown in Fig. \ref{rm1}(a).
The computational domain is $[-0.2,0.2]\times[-0.5,0.5]$, the top and bottom are imposed with periodic boundary,
and the left and right are imposed with inflow/outflow boundary condition.
The working gas is argon modeled by VHS, and the Knudsen number is $K\!n=5\times10^{-3}$.
Initially, a shock wave with $M\!a=3.0$ is located at $x=-0.15$, and a contact interface is located at $x=0$ as
\begin{equation}\nonumber
(\rho, u, T)=\left\{
\begin{aligned}
  &(1.0,0,1.0) \quad x<\sin(2\pi y+1.5\pi),\\
  &(5.0,0,0.2) \quad x>\sin(2\pi y+1.5\pi).
\end{aligned}\right.
\end{equation}
The shock wave is pre-calculated and the fully developed shock structure is used as the initial condition.
For UGKS, the velocity domain is $[-5,5] \times [-5,5]$, and $48\times32$ velocity grids are used.
The NS and multiscale solutions are calculated under a fine mesh $240\times600$ with $K\!n_c\sim1$,
and a coarse mesh $40\times100$ with $K\!n_c\sim0.1$.
The density distribution at $t=0.026\approx18\tau$ along $y=0$ is shown in Fig. \ref{rm1}(b-c).
Under the coarse mesh, the NS solutions agree well with the multiscale solutions with $Dev=3.15\times 10^{-5}$,
while large deviation can be observed under a fine mesh with $Dev=1.6\times 10^{-3}$. The data is presented in Table \ref{rm-table}.
The UGKS solution is limited by the limitation principle in the coarse mesh case.

\begin{figure}
\centering
\includegraphics[width=0.45\textwidth]{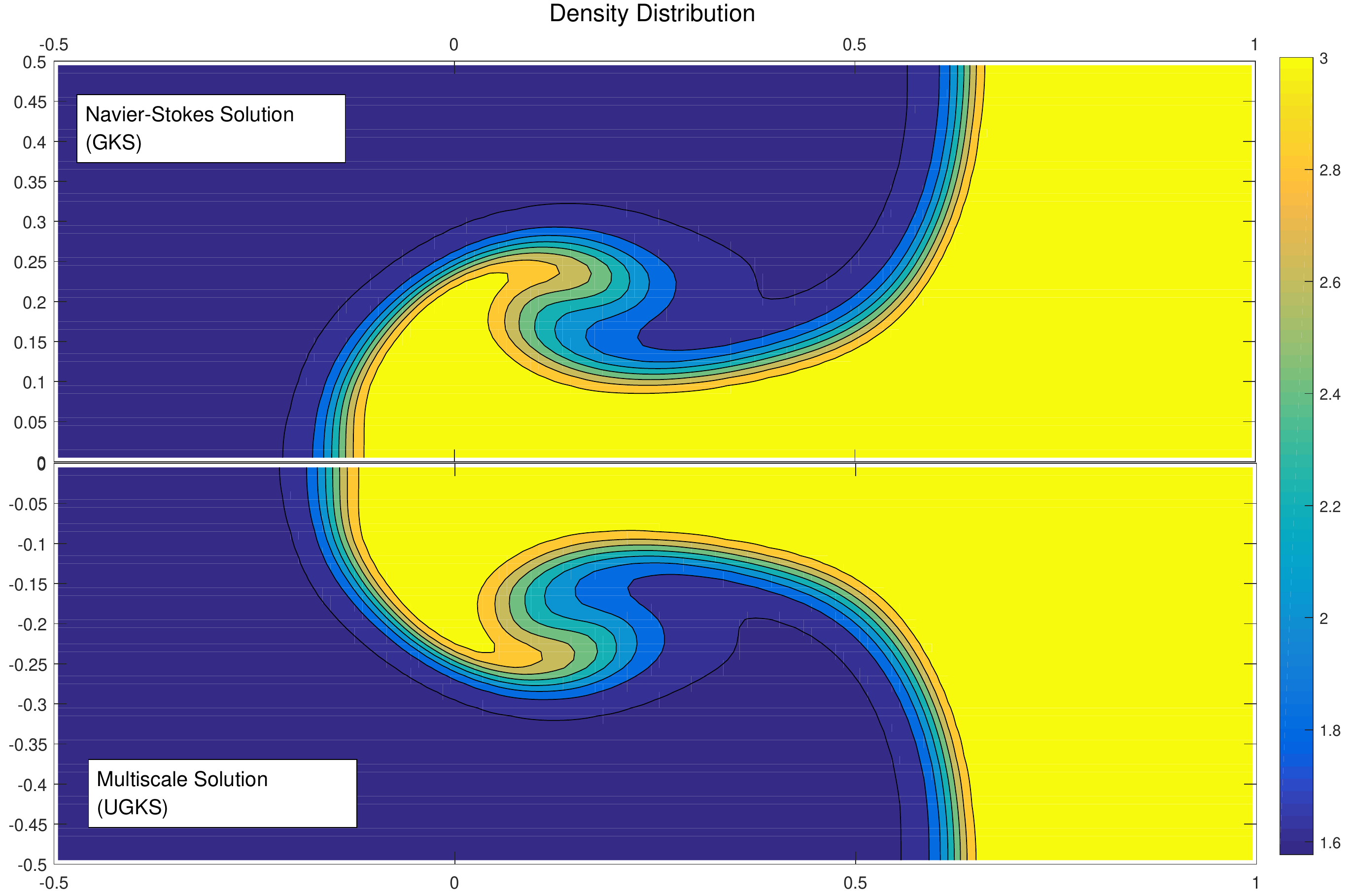}{a}
\includegraphics[width=0.45\textwidth]{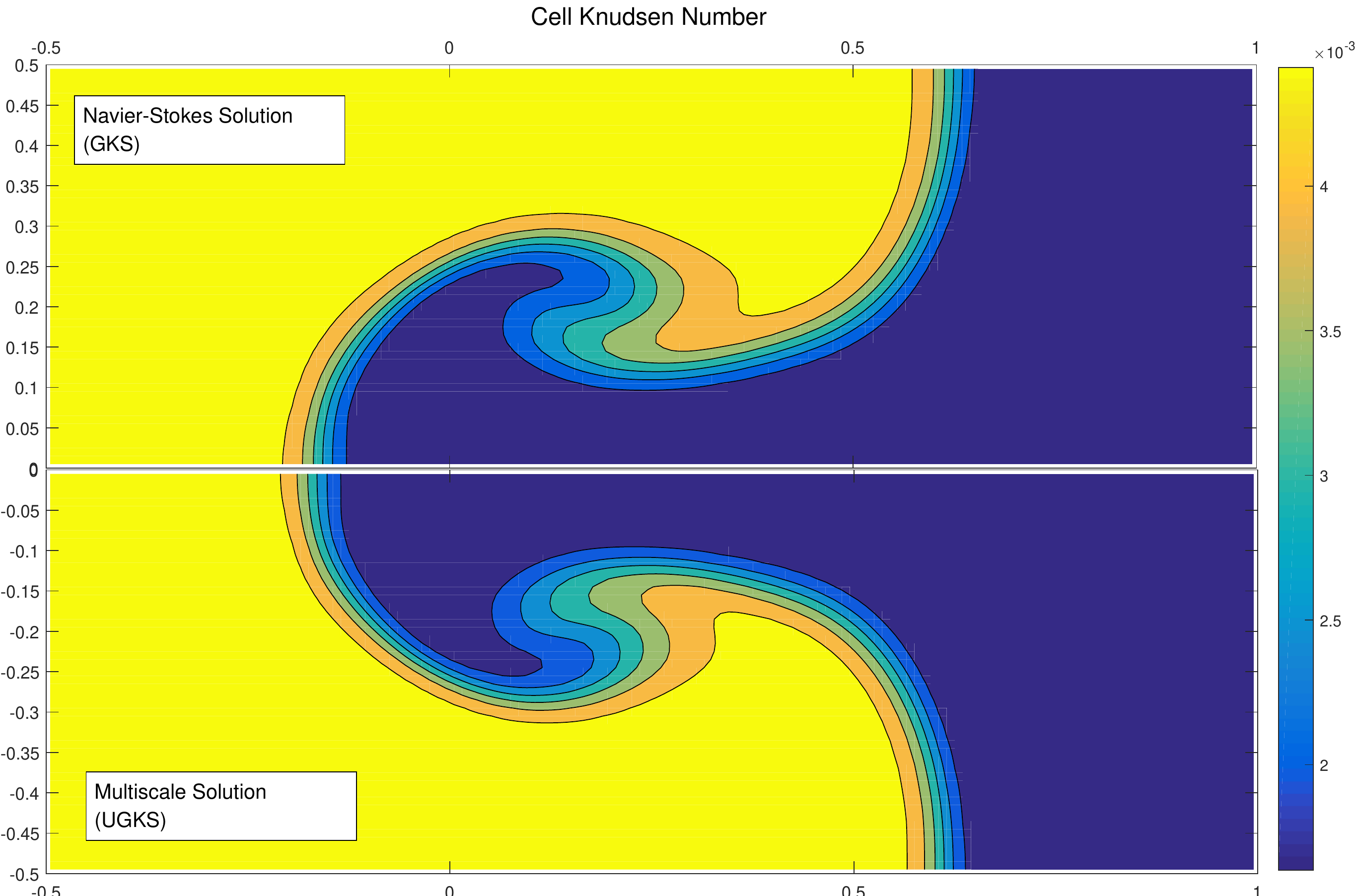}{b}
\caption{The density (a) and cell Knudsen (b) at t=18.6, with the NS solution (up) and the down multiscale solution (down).}
\label{rm2}
\end{figure}

\begin{figure}
\centering
\includegraphics[width=0.45\textwidth]{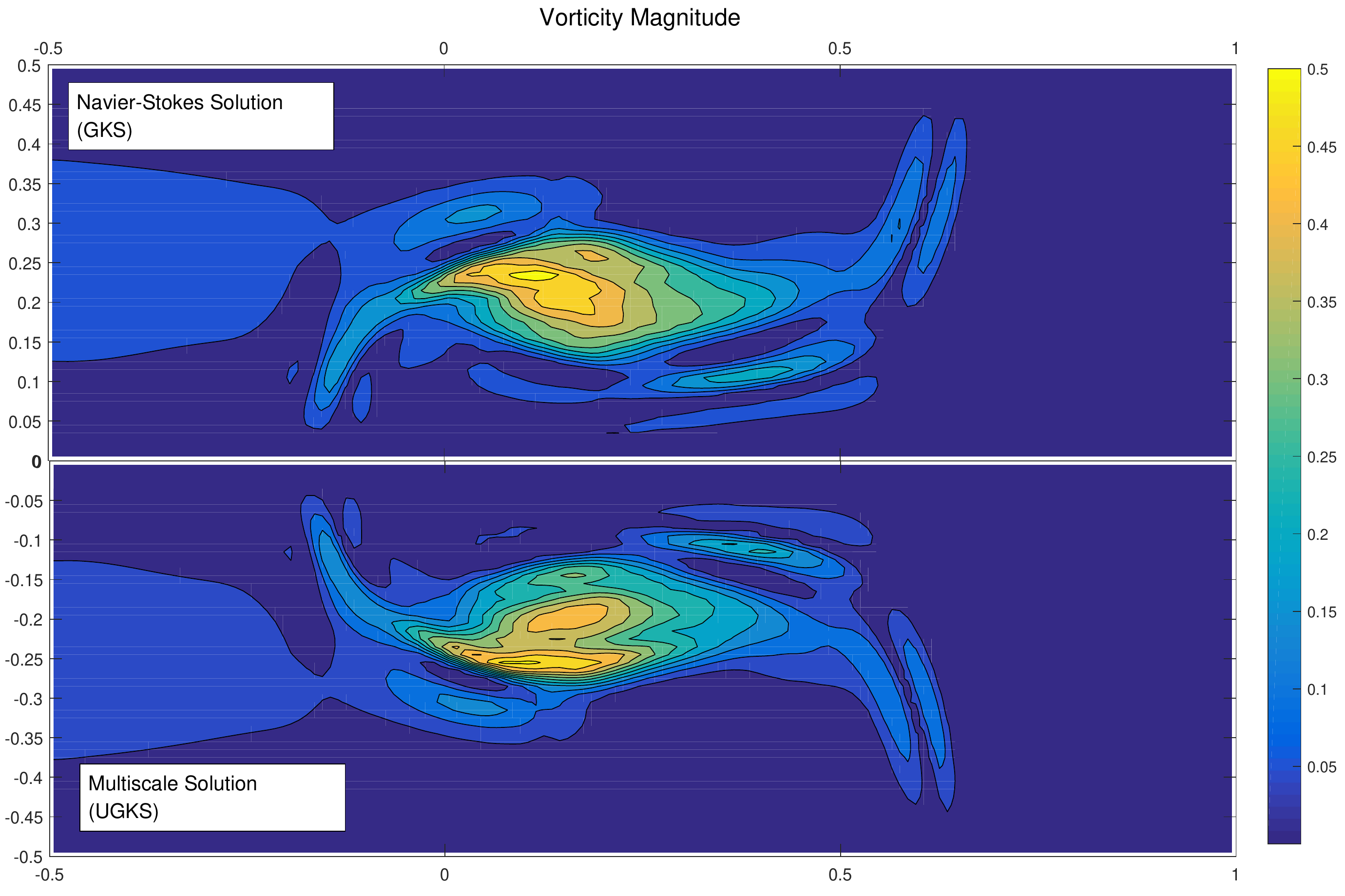}{a}
\includegraphics[width=0.42\textwidth]{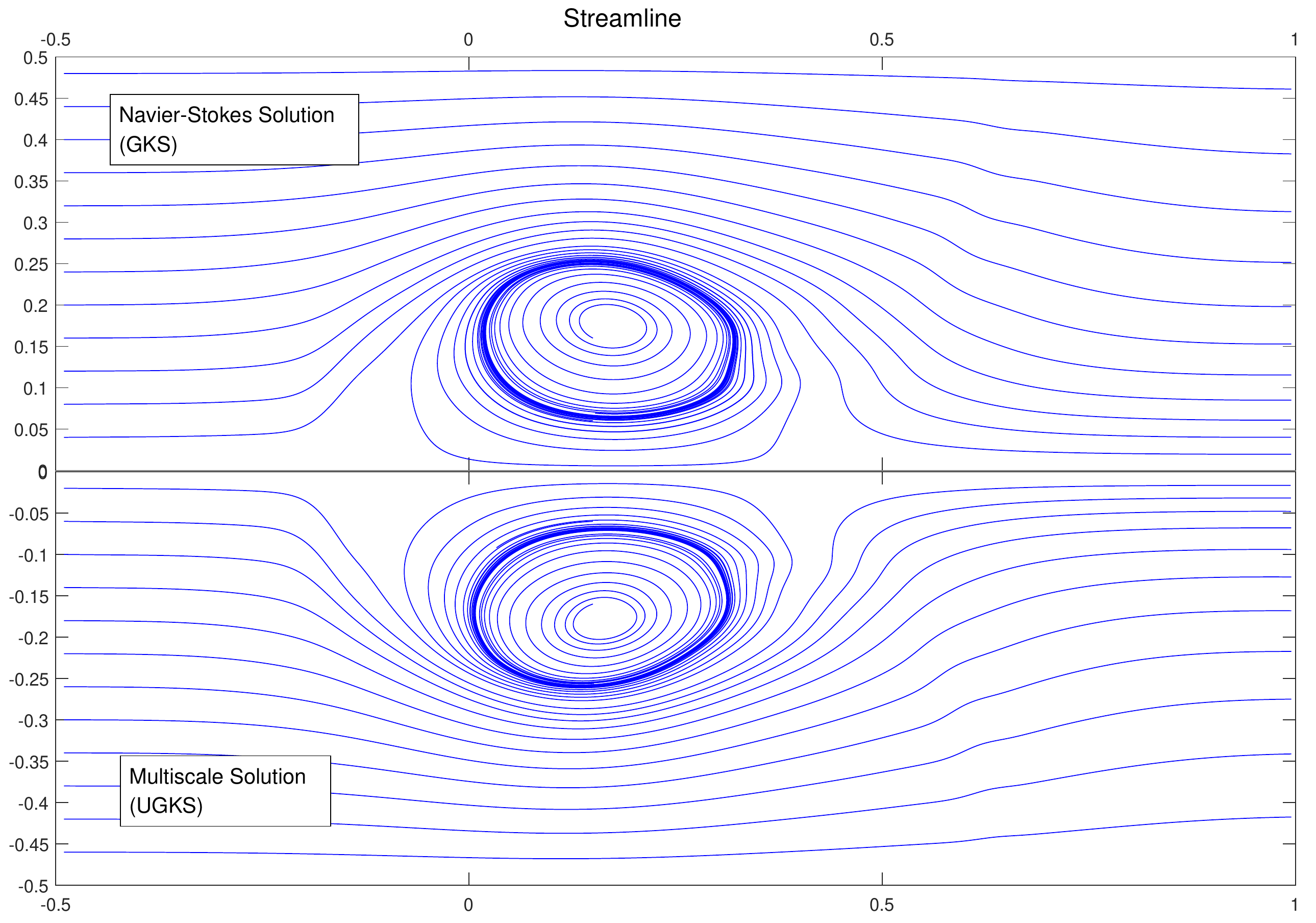}{b}
\caption{The vorticity (a) and streamline (b) at t=18.6, with the NS solution (up) and the multiscale solution (down).}
\label{rm3}
\end{figure}

Next, we calculate the development of the interface instability for a longer time scale.
The computational domain is $[-0.5,1.0]\times[-0.5,0.5]$ covered by $150\times100$ cells in the physical domain.
The Knudsen number is {$K\!n=1.0\times10^{-4}$}, and the Mach number of the shock wave is $M\!a=1.3$.
The contact discontinuity located at $x=0$ is
\begin{equation}\nonumber
(\rho, u, T)=\left\{
\begin{aligned}
  &(1.0,-0.255,1.0) \quad x<\sin(2\pi y+1.5\pi),\\
  &(2.0,-0.255,0.5) \quad x>\sin(2\pi y+1.5\pi).
\end{aligned}\right.
\end{equation}
The solutions at $t=18.6\approx2.4\times10^{5}\tau$ are shown in Fig.\ref{rm2}-\ref{rm3}, where the density, cell Knudsen number,
vorticity magnitude, and streamline are plotted together with the NS solution (up) and the  UGKS solution (down). On such large space and time scales, the NS solutions agree well with the multiscale solutions.

\begin{table}
\centering
    \begin{tabular}{| l | l | l | l |}
    \hline
    {Richtmyer-Meshkov instability} & $K\!n_c$ & $\tau/\Delta t$ &  $Dev$ \\ \hline
    Fine mesh  & $\sim 1$ & $\sim 50$ & $ 1.6 \times 10^{-3}$\\ \hline
    Coarse mesh & $\sim 0.1$ & $\sim 5$ & $ 3.15\times 10^{-5}$\\
    \hline
    \end{tabular}
    \caption{Parameters under different meshes, deviation between NS and multiscale solution for the Richtmyer-Meshkov instability.}
    \label{rm-table}
\end{table}

\subsection{Shock Bubble interaction}

\begin{figure}
\centering
\includegraphics[width=0.7\textwidth]{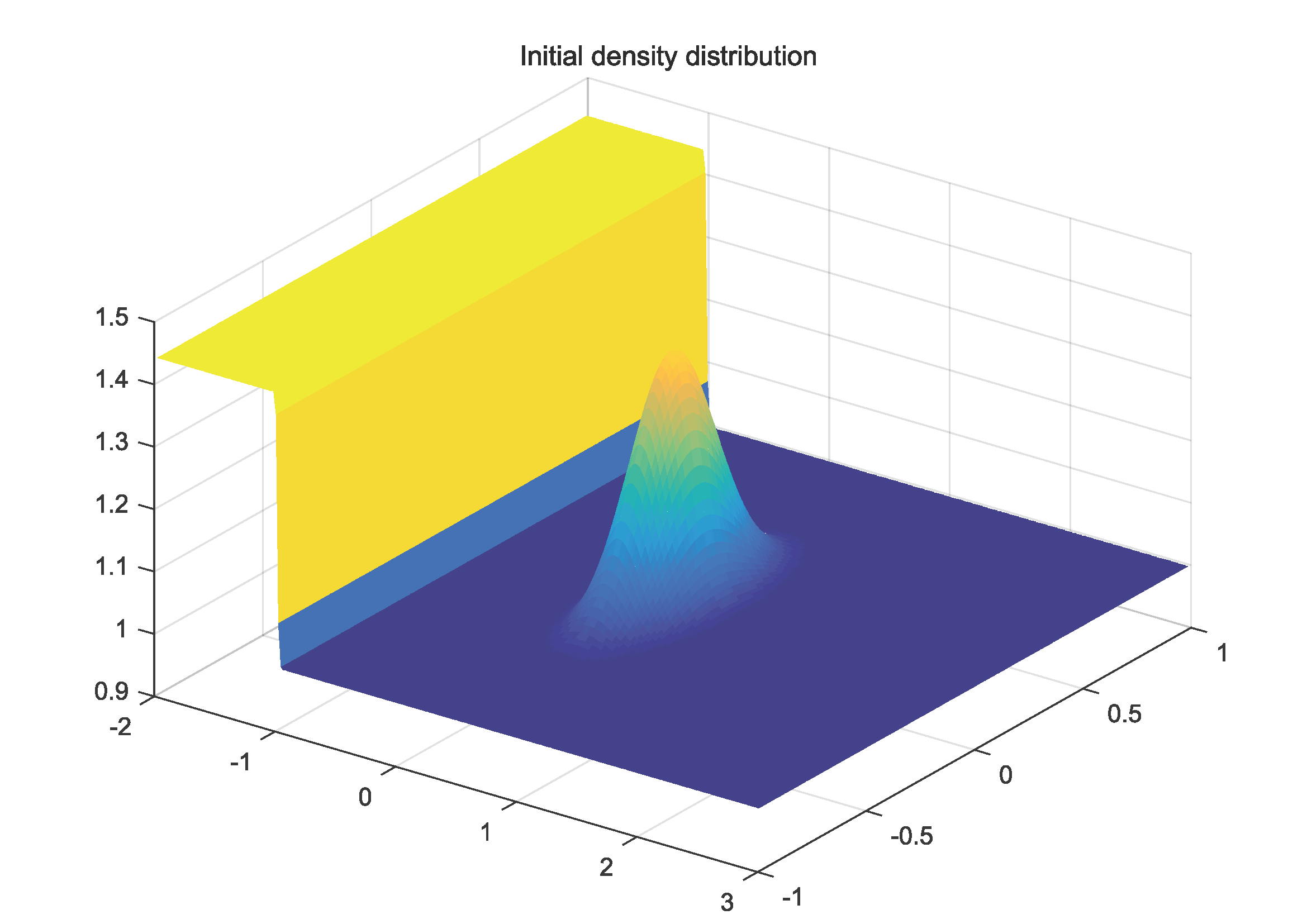}
\caption{Initial condition for the shock bubble interaction process.}
\label{bubble-initial}
\end{figure}

In this section, we study the process of a shock wave interacting with dense cold bubble to show the capability of NS equations in describing the flow with large gradients.
The initial condition is shown in Fig. \ref{bubble-initial}. A shock wave is situated with its density barycenter at $x=-1.0$,
travelling in the positive x-direction into a flow field at rest with $(\rho,U,p)=(1.0,0,0.5)$.
Around $(x,y)=(0.5,0)$, a dense circular bubble is placed with constant pressure $p=0.5$.
The computational domain is $[-2,3]\times[-1,1]$, and the inflow/outflow boundary is imposed.
The velocity space has a domain $[-8,8] \times [-8,8]$ covered by $100\times100$ velocity grid points.

\begin{figure}
\centering
\includegraphics[width=1.0\textwidth]{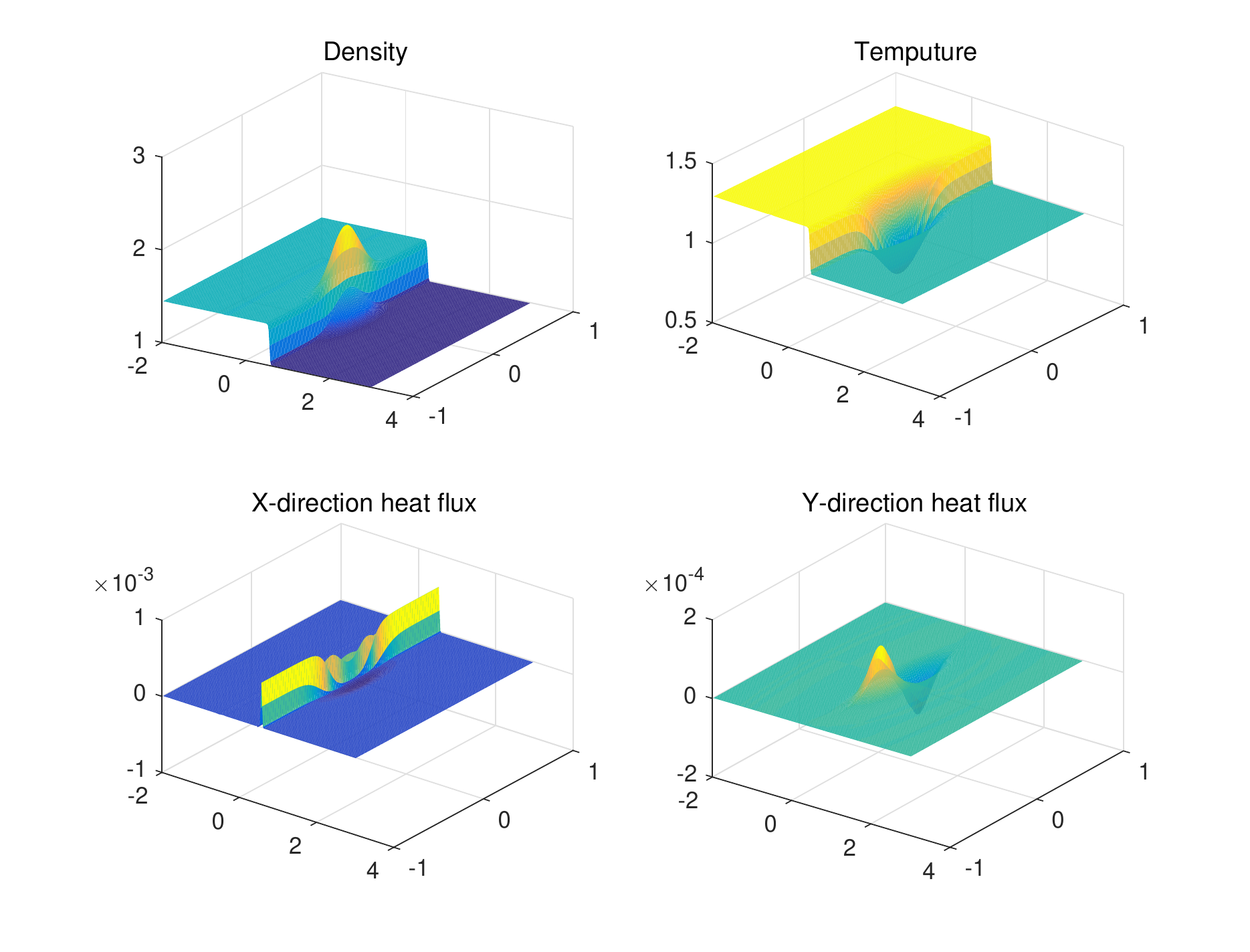}
\caption{Two-dimensional plots of the fields of density, temperature and heat fluxes at $t=1.3$.}
\label{bubble-ma13}
\end{figure}

\begin{figure}
\centering
\includegraphics[width=0.42\textwidth]{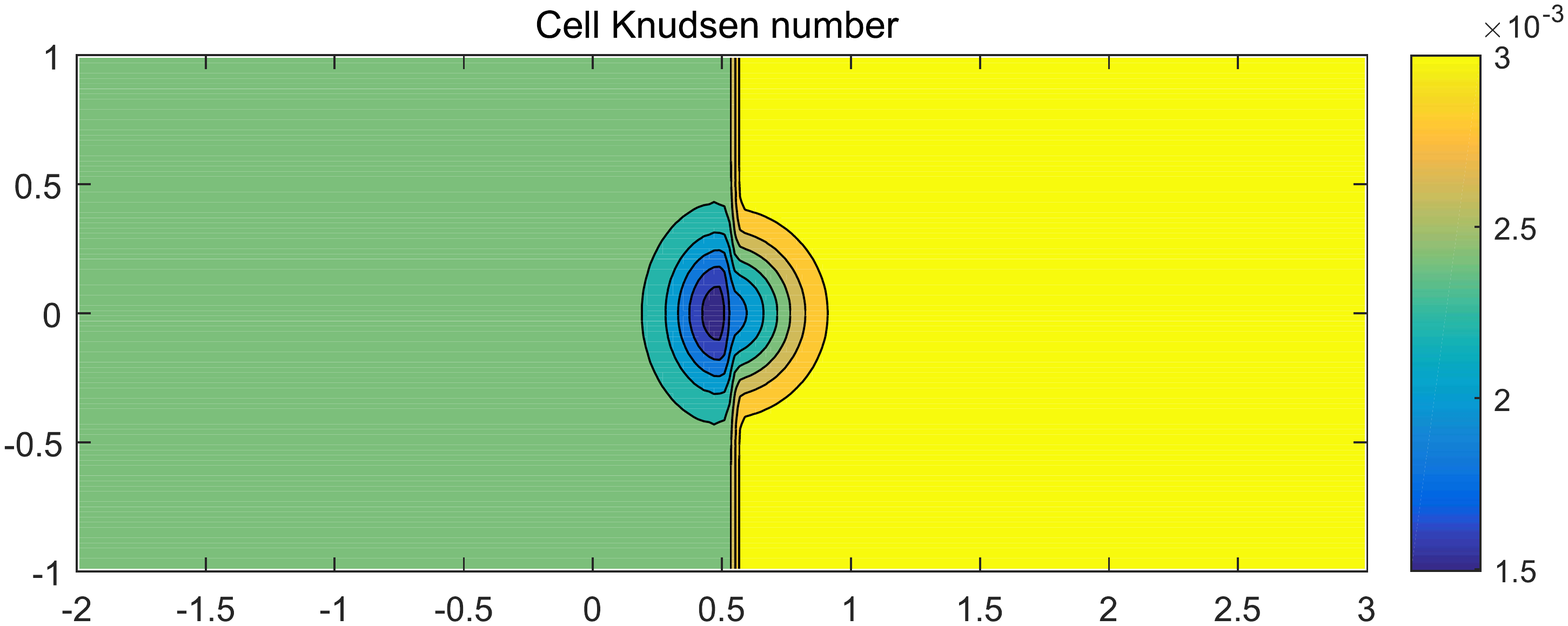}{a}
\includegraphics[width=0.42\textwidth]{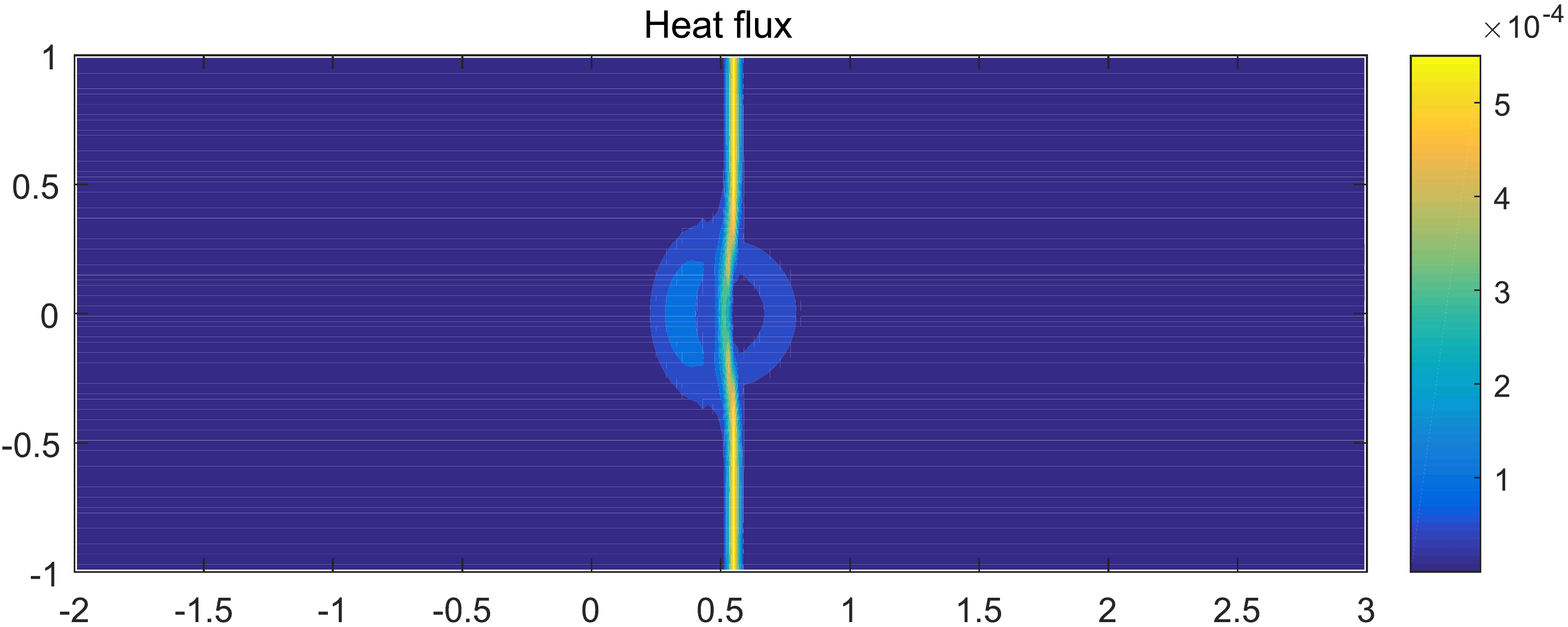}{b}\\
\includegraphics[width=0.4\textwidth]{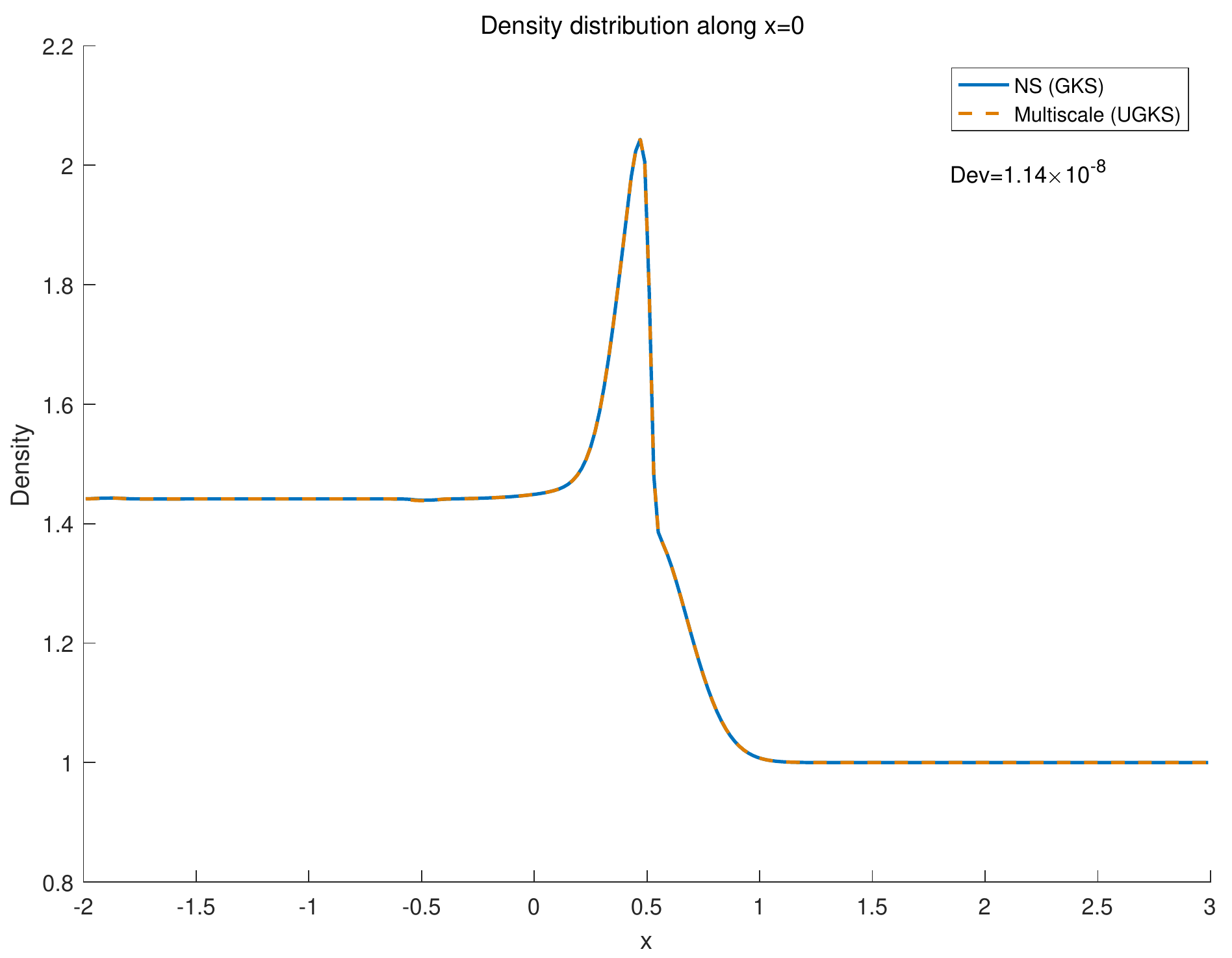}{c}
\includegraphics[width=0.4\textwidth]{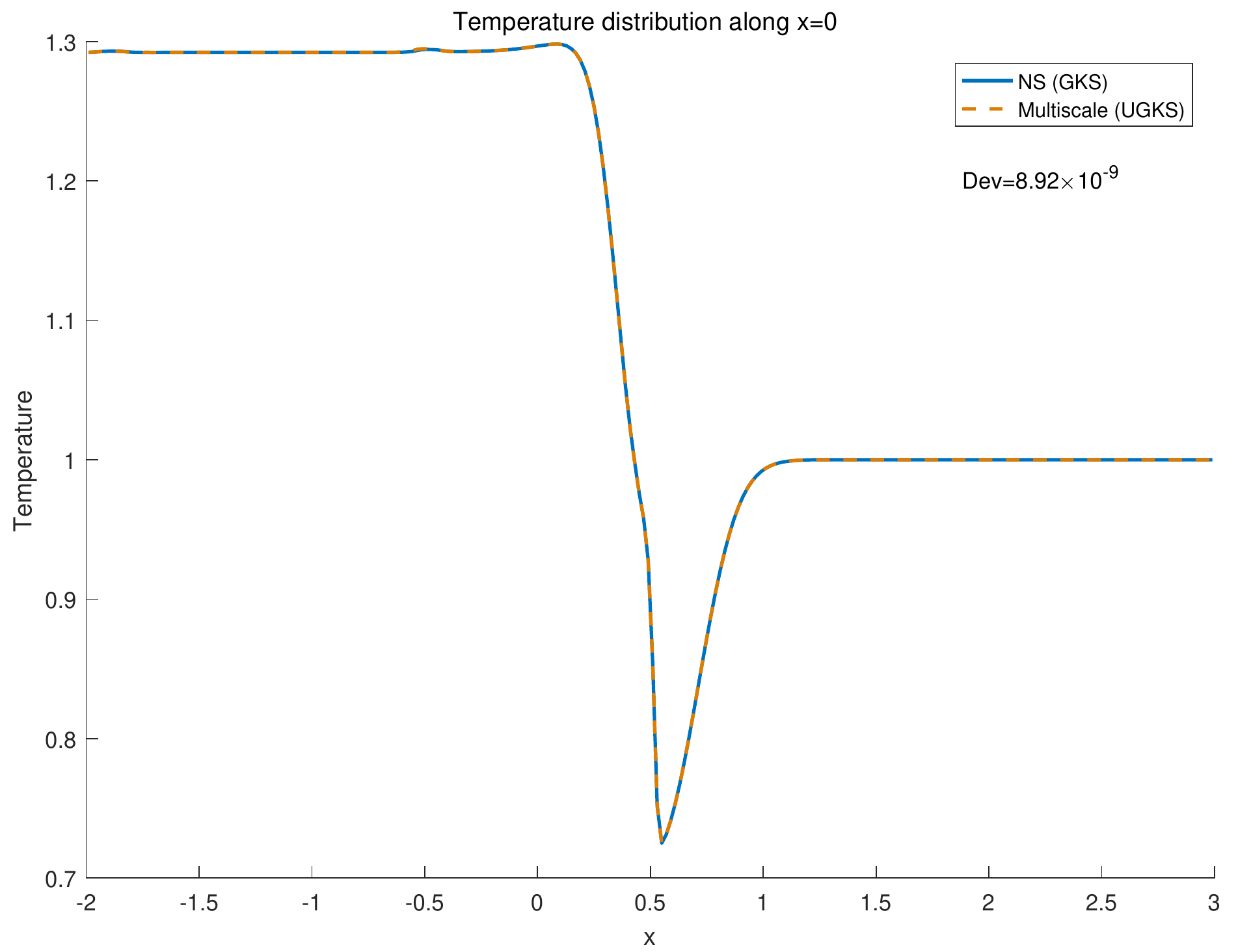}{d}\\
\includegraphics[width=0.4\textwidth]{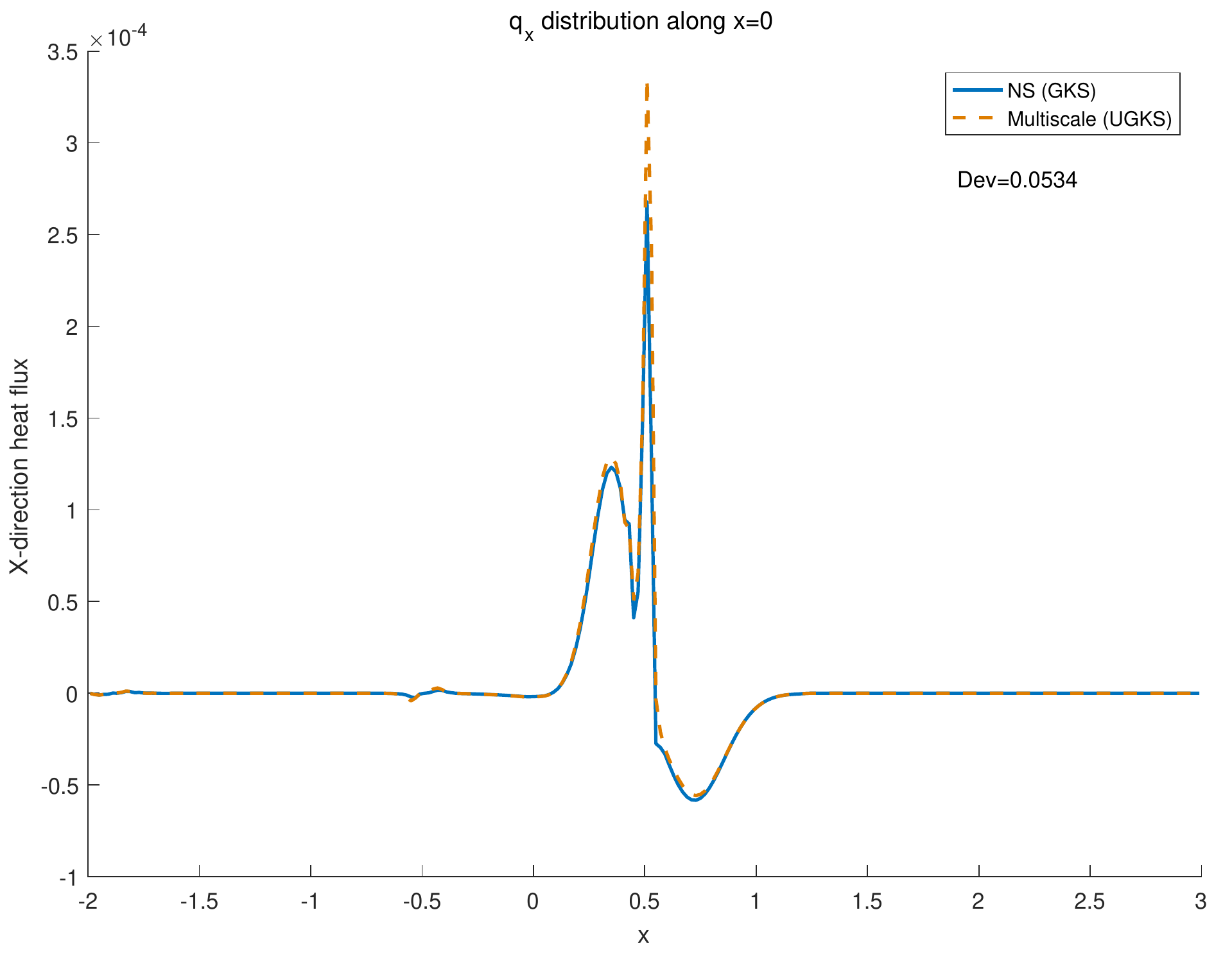}{e}
\includegraphics[width=0.4\textwidth]{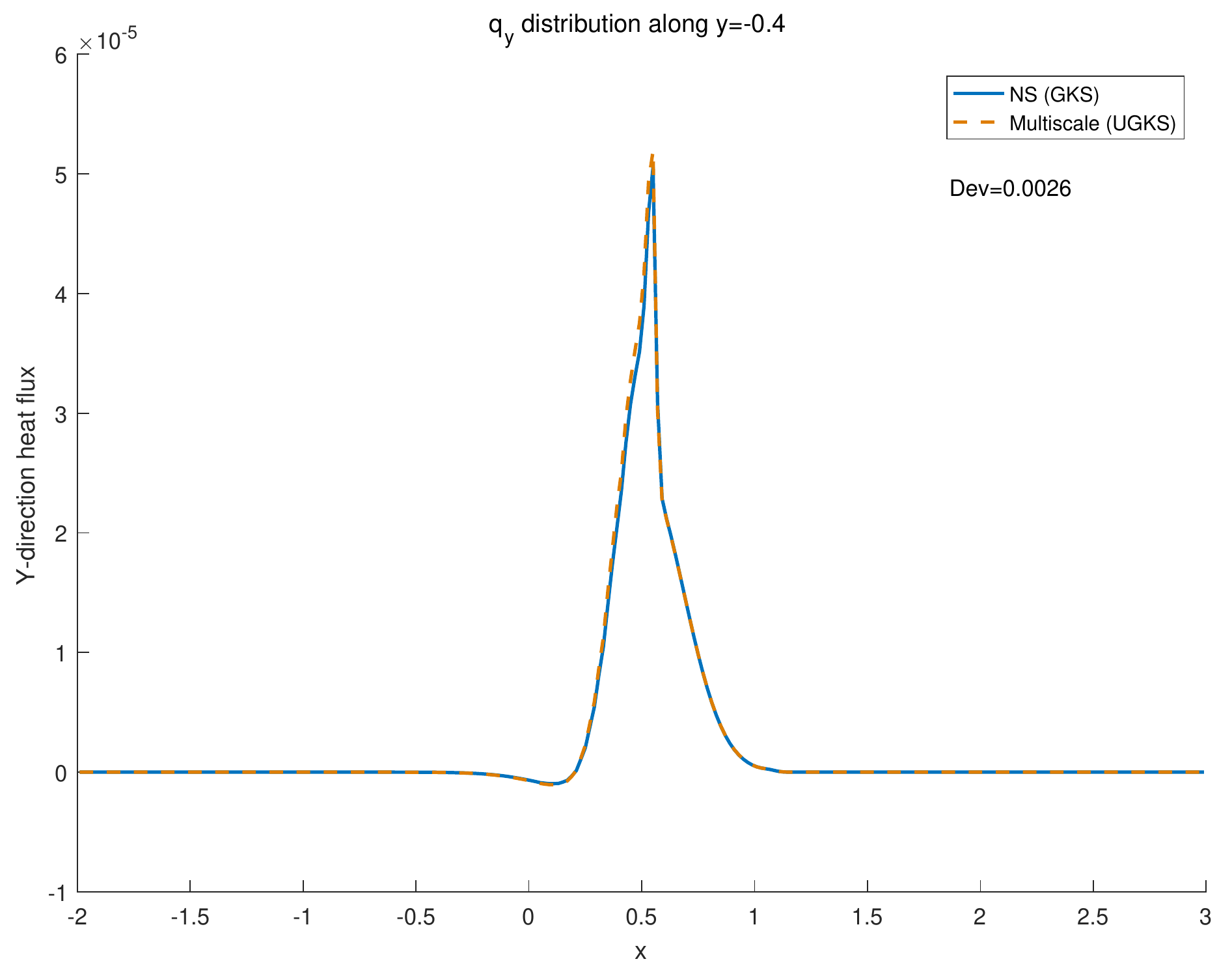}{f}
\caption{Results of the shock bubble interaction with $M\!a=1.3$ and $K\!n=1.0\times10^{-4}$ at $t=1.3$. Top two figures shows the cell Knudsen number (a) and total heat flux (b). (c)-(f) show the comparison between NS solution profiles (solid line) and UGKS solution profiles (dash line).}
\label{ma13kn-4}
\end{figure}

\begin{figure}
\centering
\includegraphics[width=0.42\textwidth]{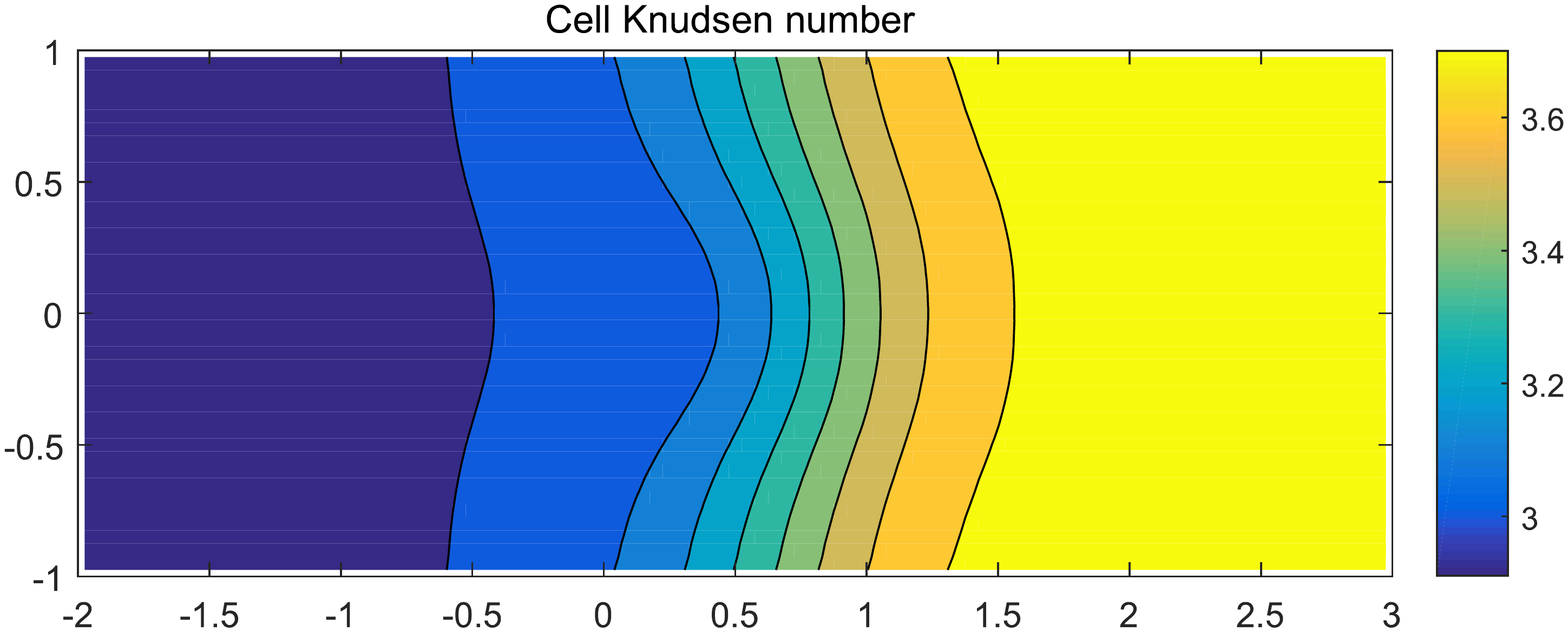}{a}
\includegraphics[width=0.42\textwidth]{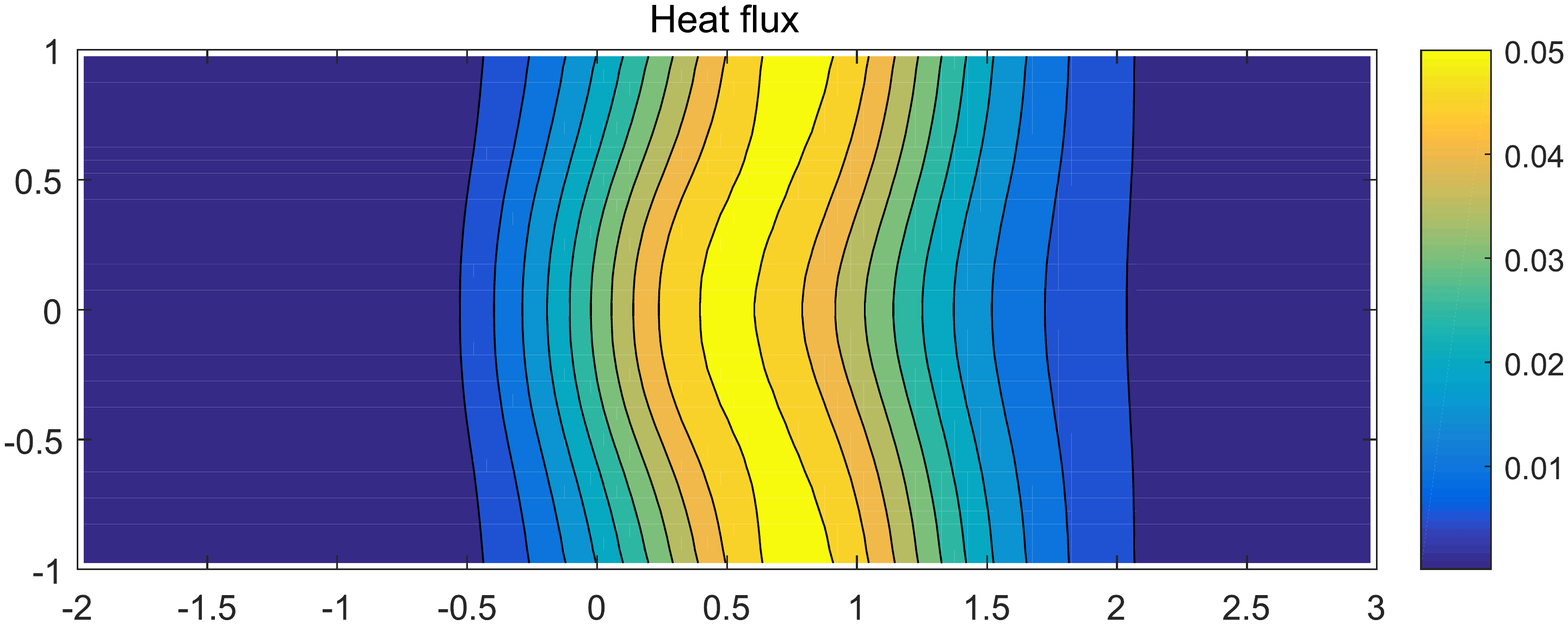}{b}\\
\includegraphics[width=0.4\textwidth]{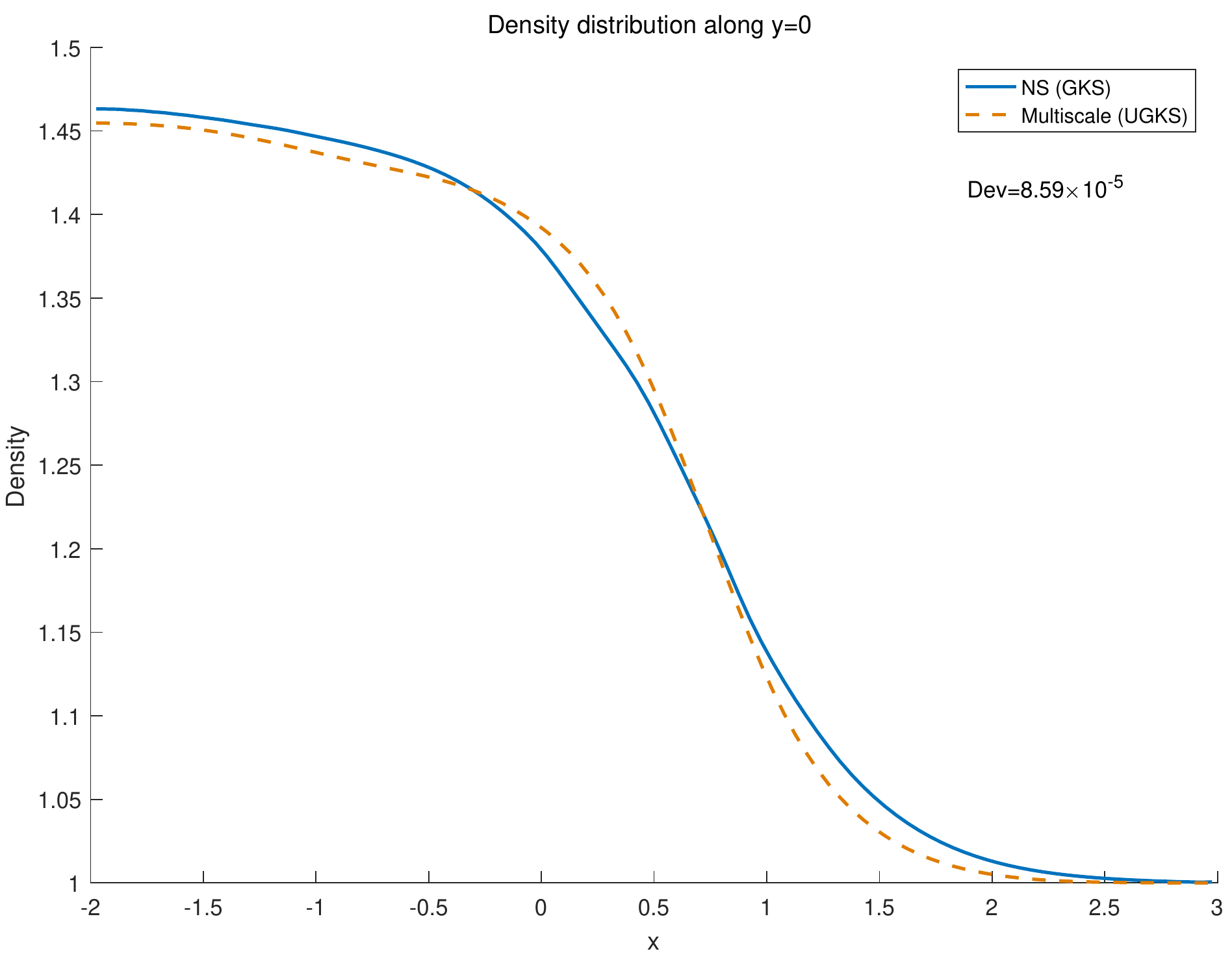}{c}
\includegraphics[width=0.4\textwidth]{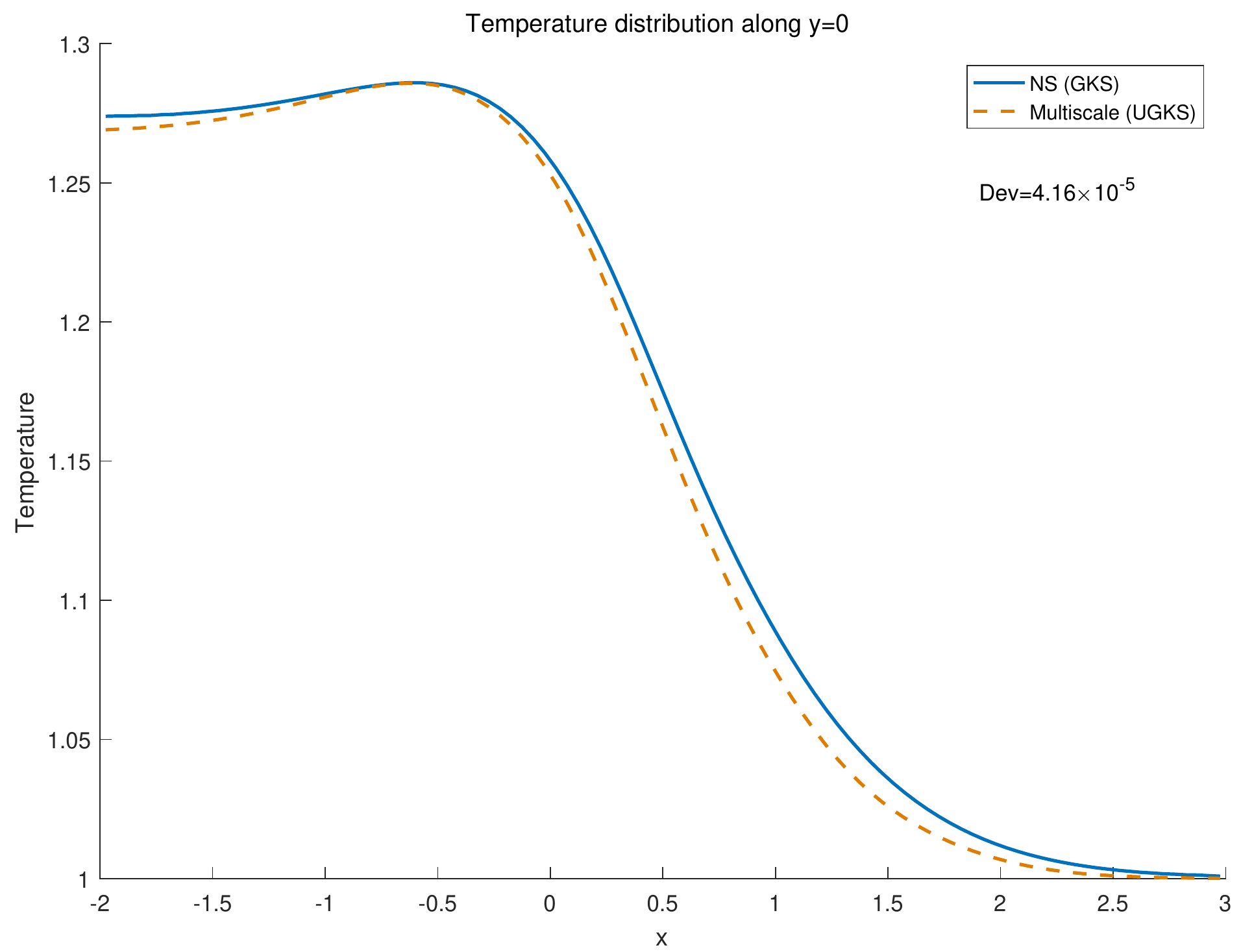}{d}\\
\includegraphics[width=0.4\textwidth]{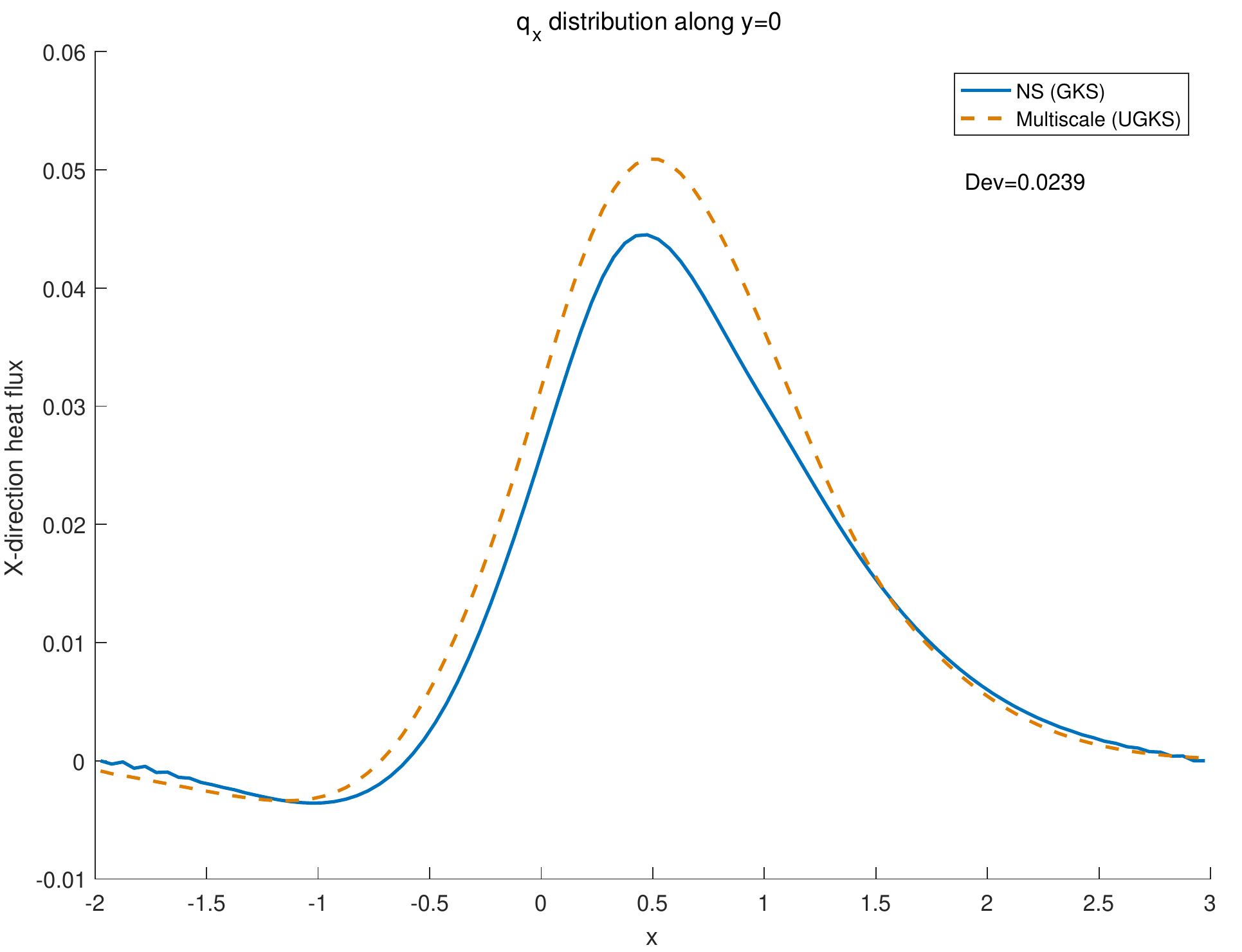}{e}
\includegraphics[width=0.4\textwidth]{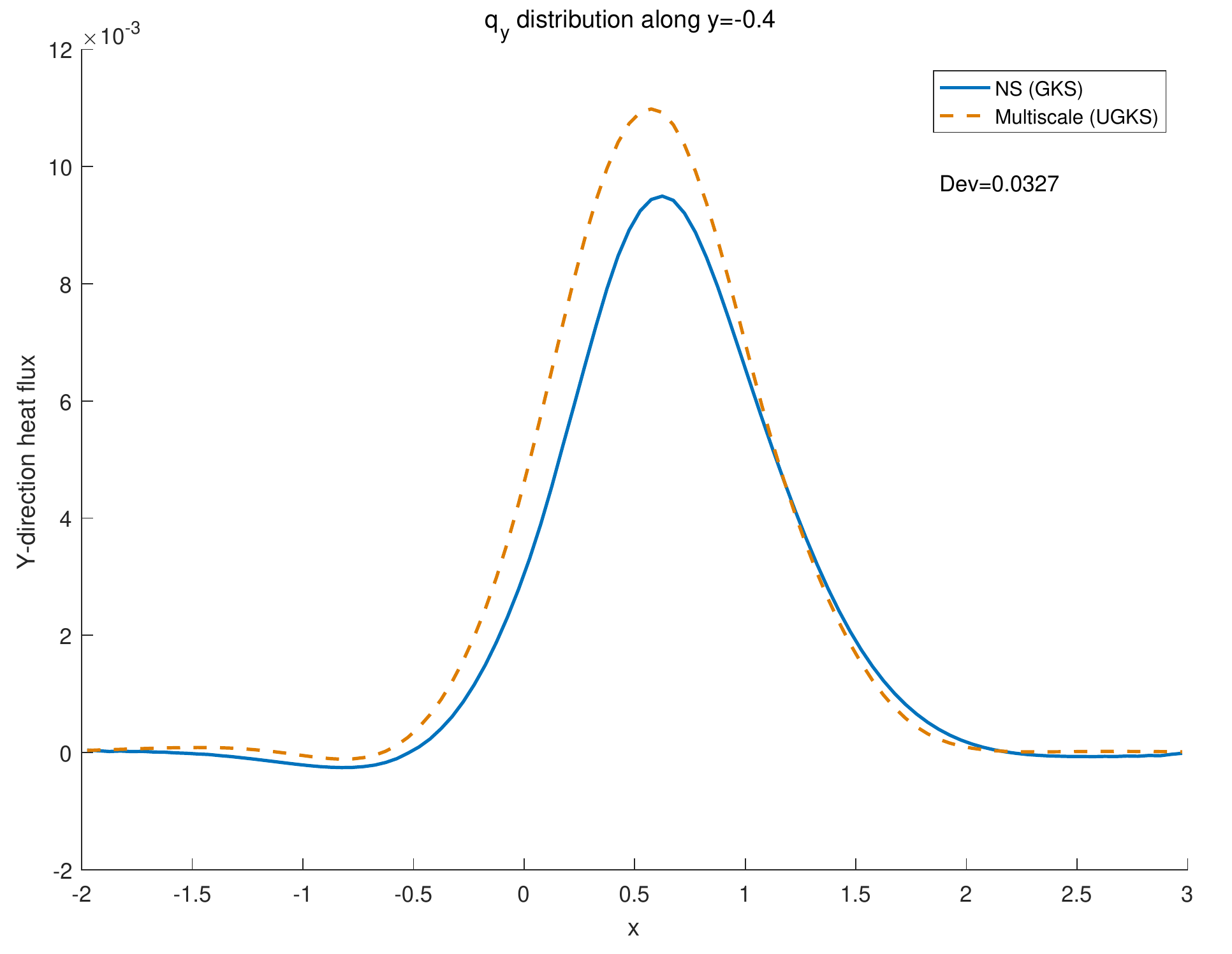}{f}
\caption{Results of the shock bubble interaction with $M\!a=1.3$ and $K\!n=0.3$ at $t=1.3$. Top two figures shows the cell Knudsen number (a) and total heat flux (b). (c)-(f) show the comparison between NS solution profiles (solid line) and UGKS solution profiles (dash line).}
\label{ma13kn03}
\end{figure}

First, we set the Mach number of the shock wave to be $M\!a=1.3$, and set the density of the bubble to be
\begin{equation}
  \rho(x,y)=1+0.4\exp(-16(x^2+y^2)).
\end{equation}
Two different Knudsen numbers are considered relative to the bubble radius, namely $K\!n=1.0\times10^{-4}$ and $K\!n=0.3$. For $K\!n=1.0\times10^{-4}$ case, $250\times100$ cells are used in the physical domain.
The solutions of UGKS at $t=1.3$ is shown in Fig. \ref{bubble-ma13} when the shock wave is about to pass the bubble.
This situation shows large density and temperature gradients in the shock layer which can lead to strong non-equilibrium.
Fig. \ref{ma13kn-4} shows the comparison between the NS and UGKS solutions, and
the agreements in the density and temperature profiles have been confirmed  at a small cell Knudsen number.
For high order moments, such as the x-direction heat flux, a slight deviation still can be observed inside the shock bubble interaction region.
For $K\!n=0.3$, the physical domain is divided into $100\times 40$ mesh points.
Solutions at $t=1.3$ are shown in Fig. \ref{ma13kn03}. For this case, the cell Knudsen number is relatively large and the non-equilibrium flow
can be resolved.
The NS and the multiscale solutions in density and temperature deviate from each other in this case.

\begin{figure}
\centering
\includegraphics[width=0.42\textwidth]{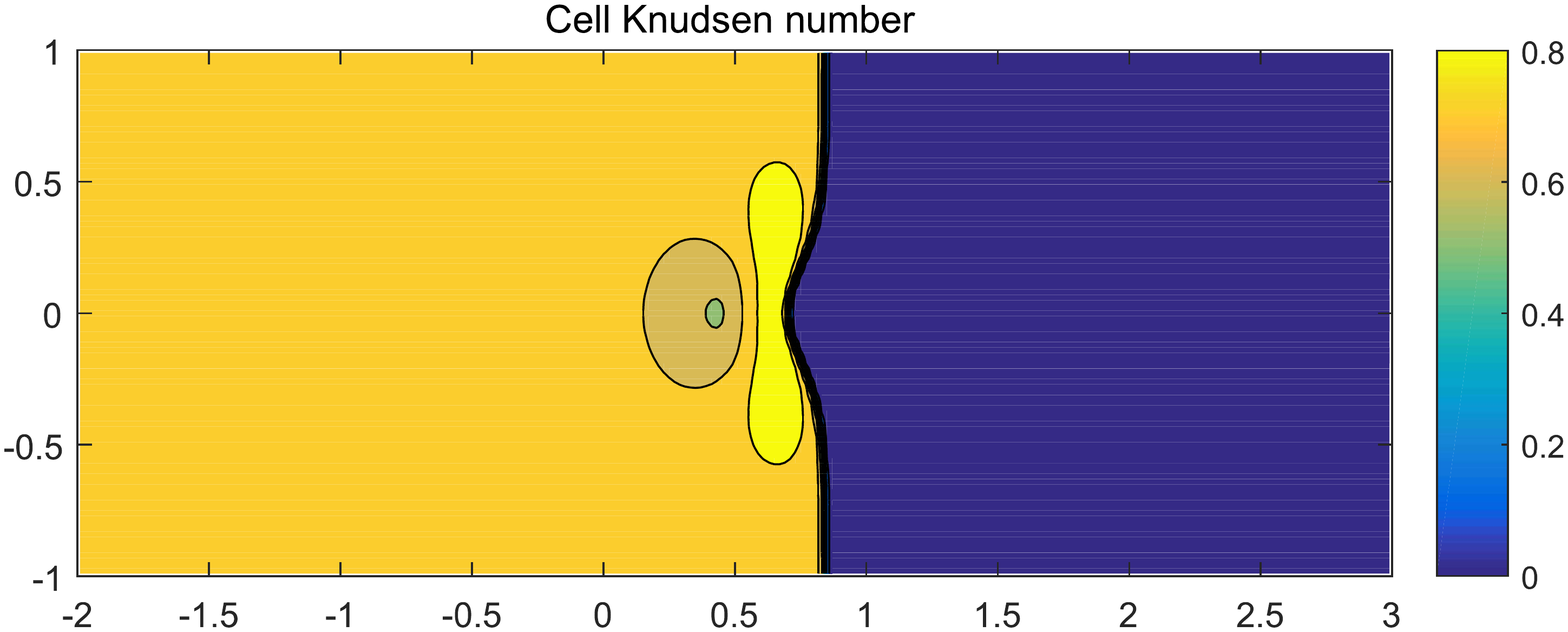}{a}
\includegraphics[width=0.42\textwidth]{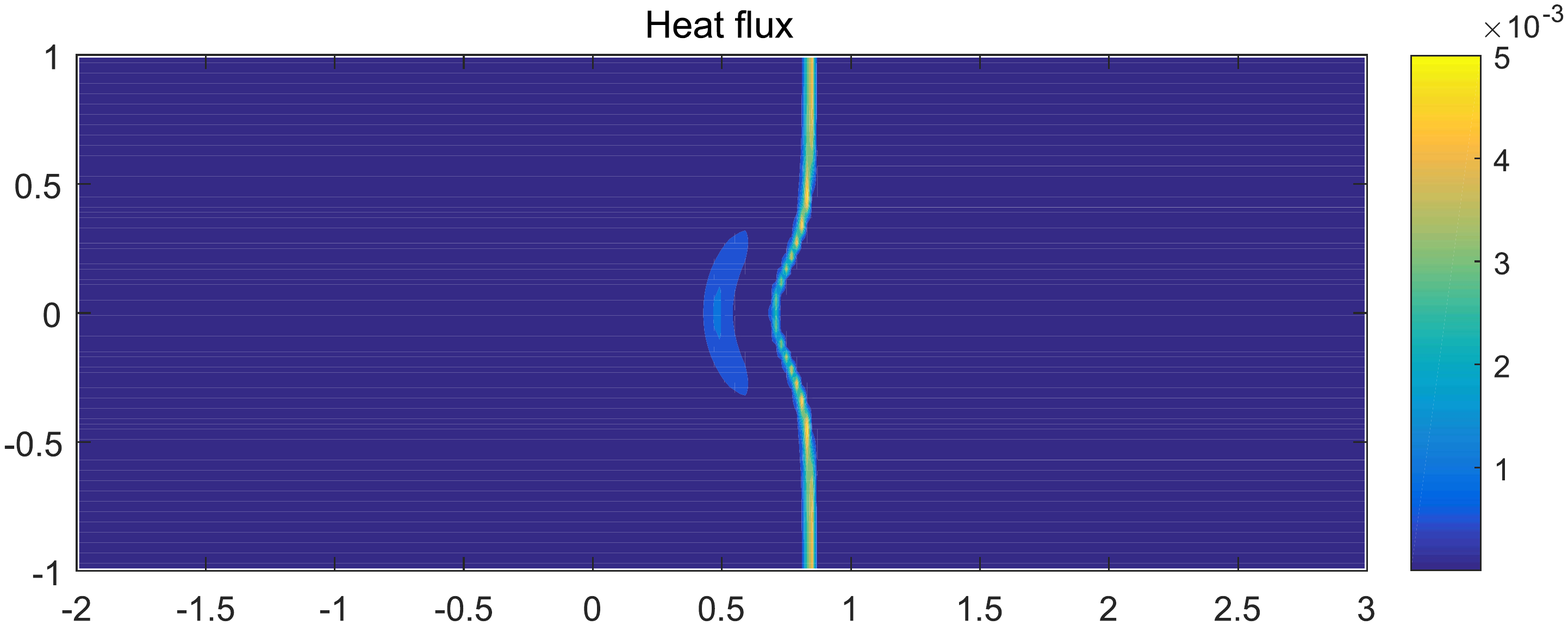}{b}\\
\includegraphics[width=0.4\textwidth]{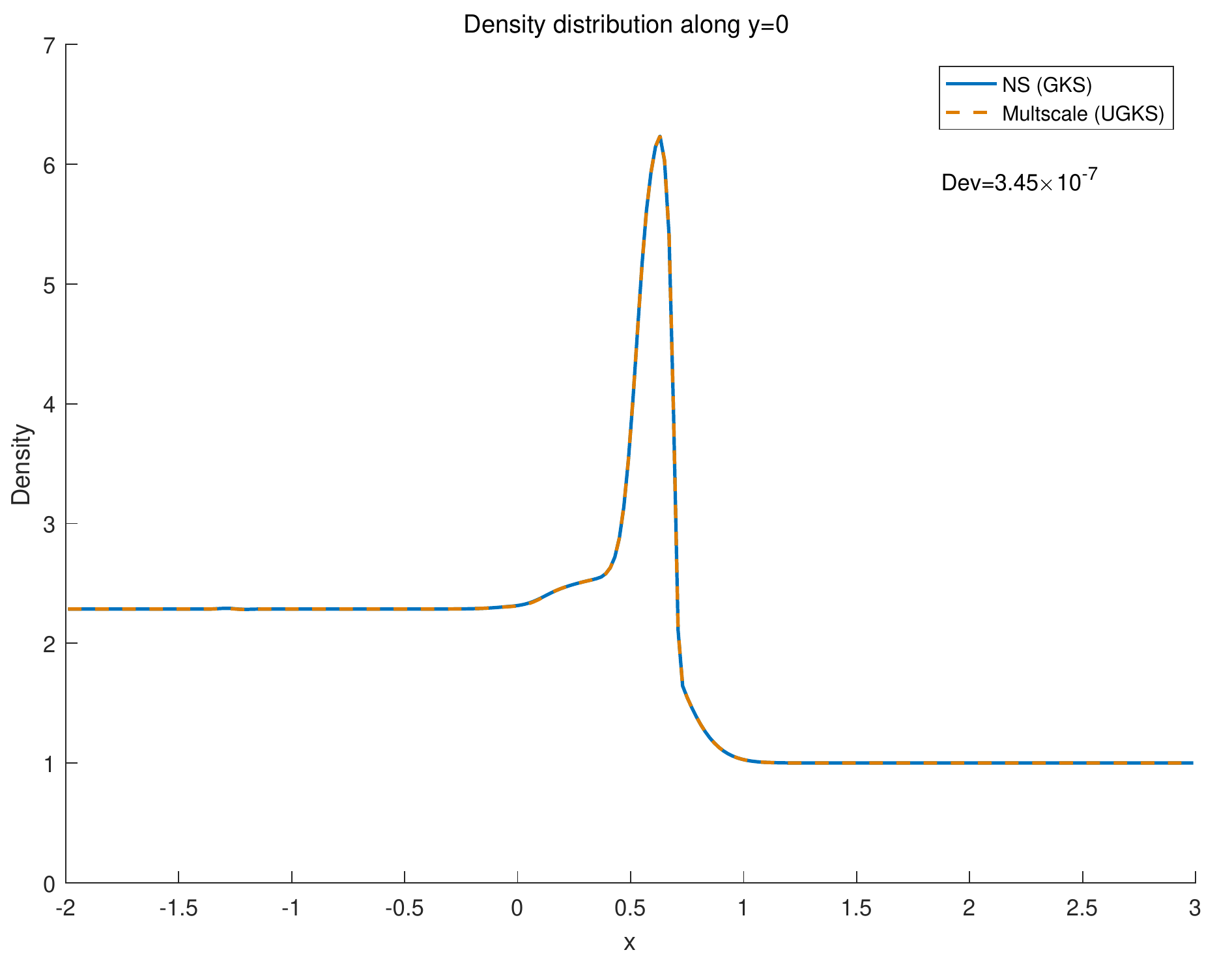}{c}
\includegraphics[width=0.4\textwidth]{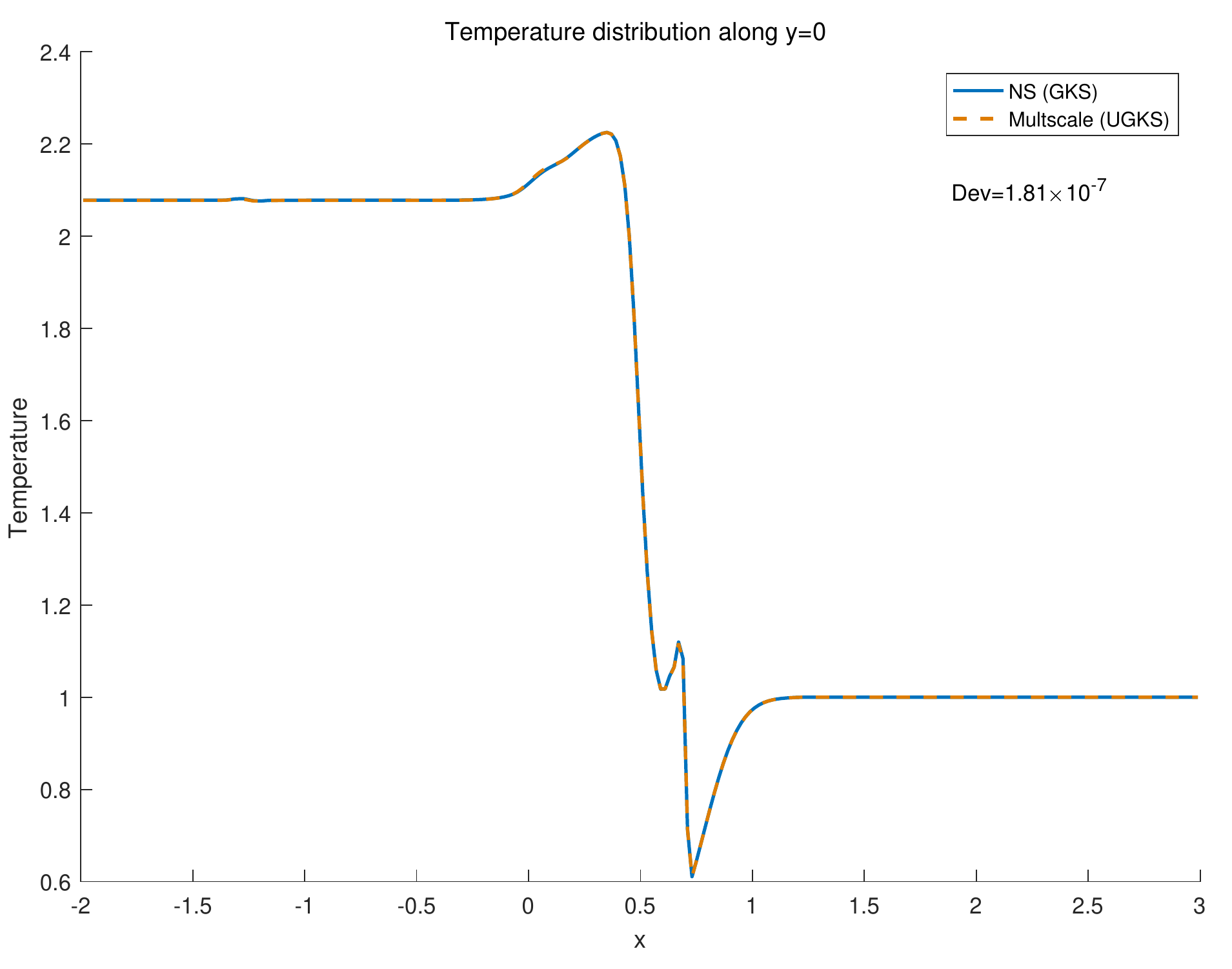}{d}\\
\includegraphics[width=0.4\textwidth]{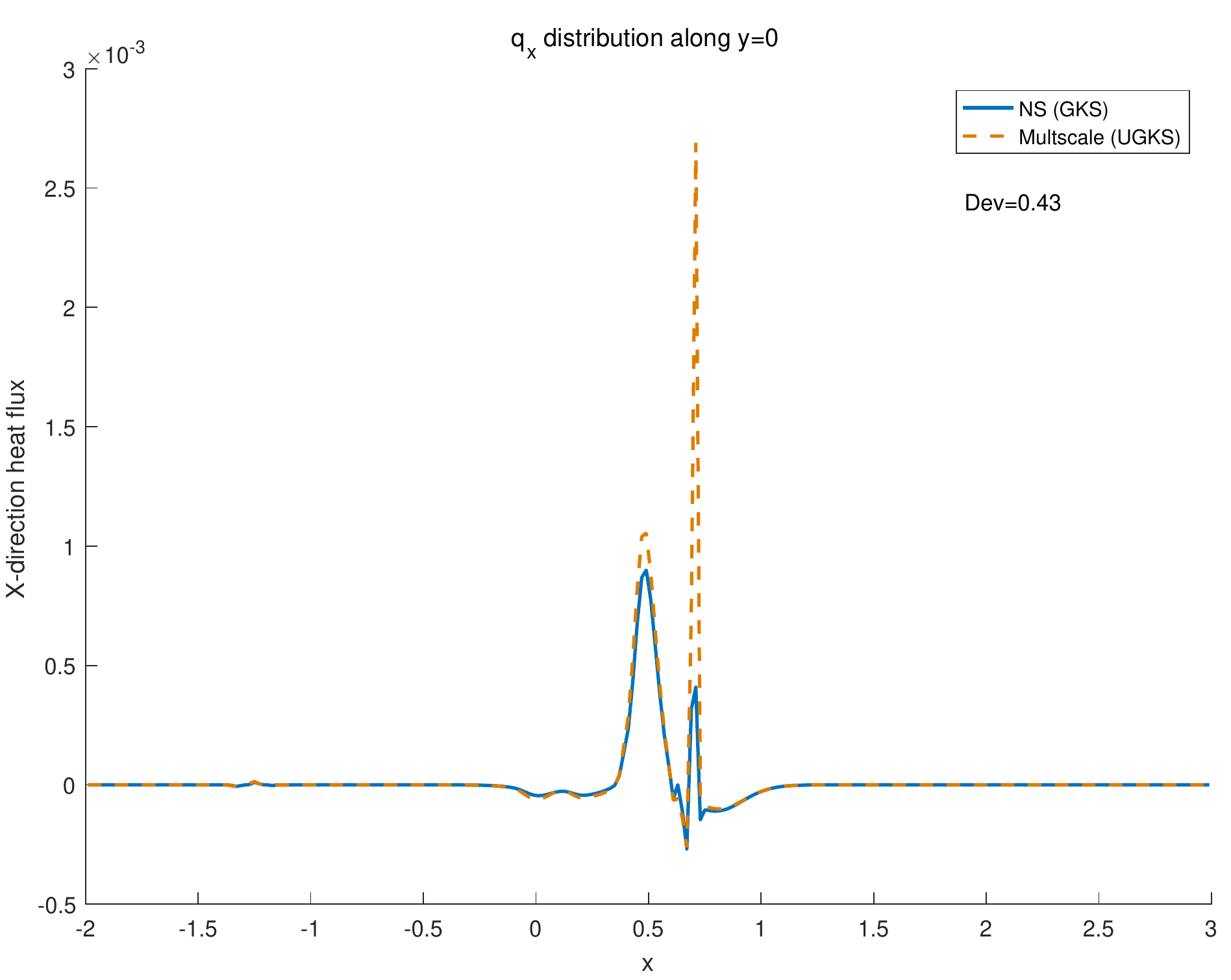}{e}
\includegraphics[width=0.4\textwidth]{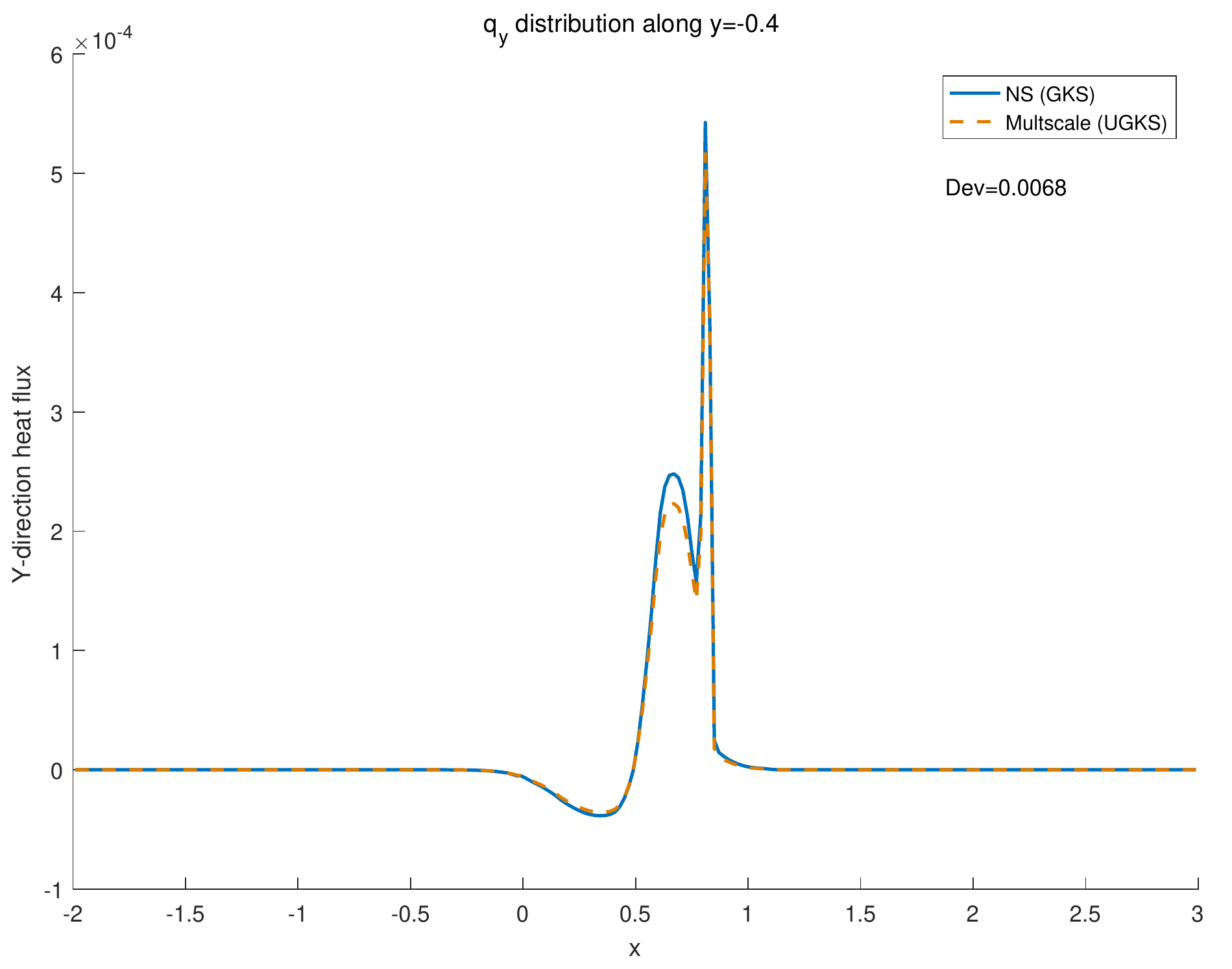}{f}
\caption{Results of the shock bubble interaction with $M\!a=2$ and $K\!n=1.0\times10^{-4}$ at $t=1.0$. Top two figures shows the cell Knudsen number (a) and total heat flux (b). (c)-(f) show the comparison between NS solution profiles (solid line) and UGKS solution profiles (dash line).}
\label{ma2kn-4}
\end{figure}

\begin{figure}
\centering
\includegraphics[width=0.42\textwidth]{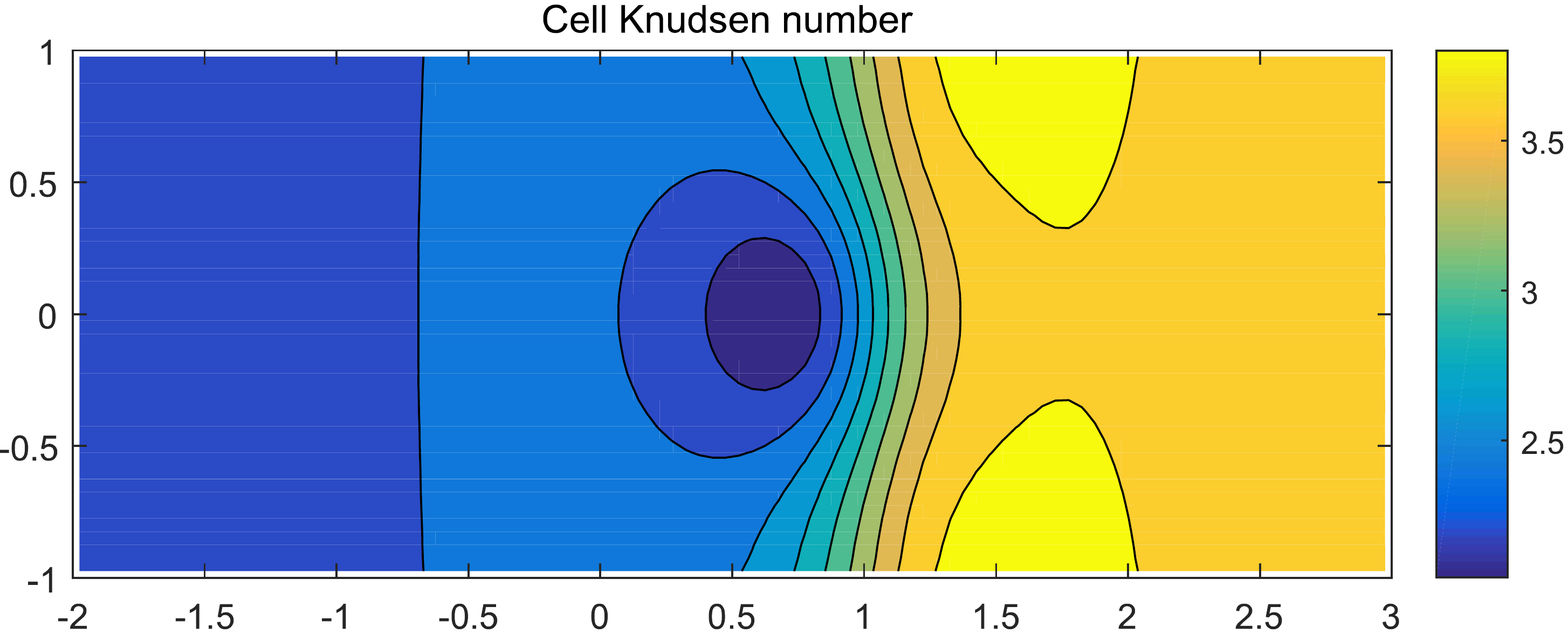}{a}
\includegraphics[width=0.42\textwidth]{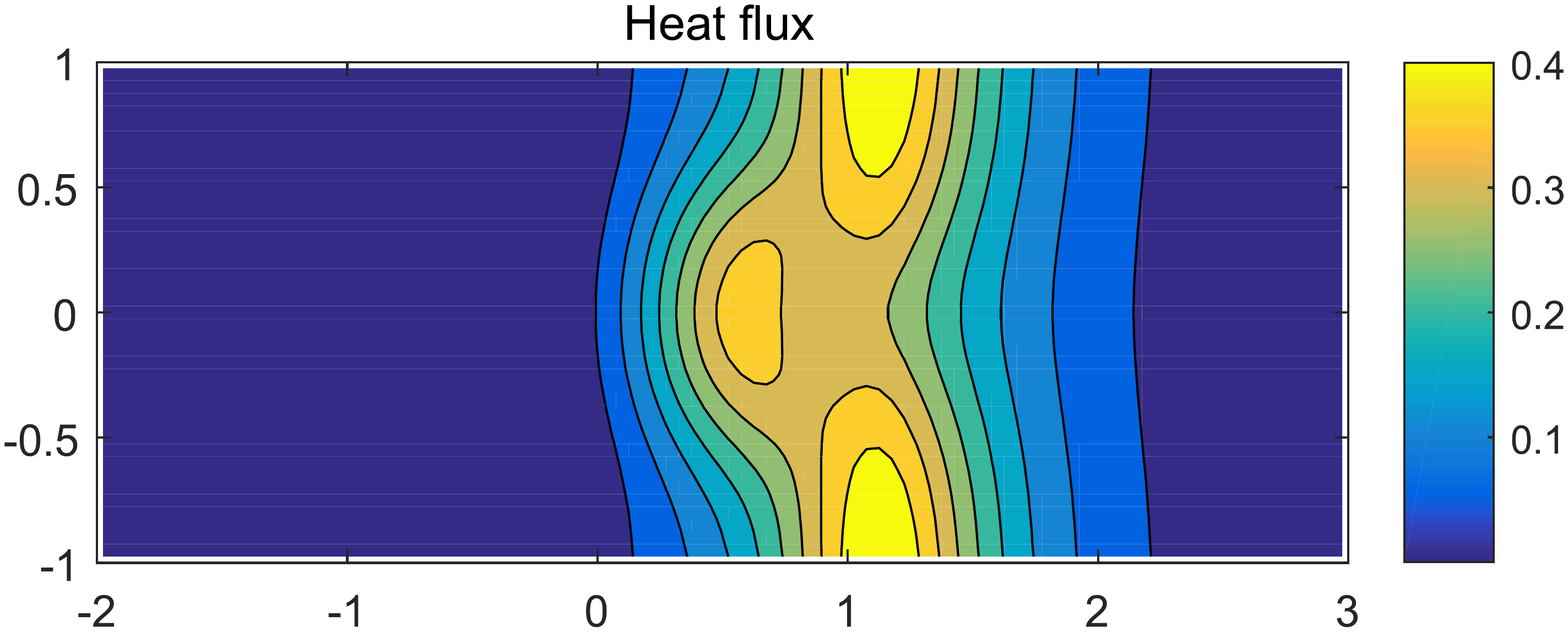}{b}\\
\includegraphics[width=0.4\textwidth]{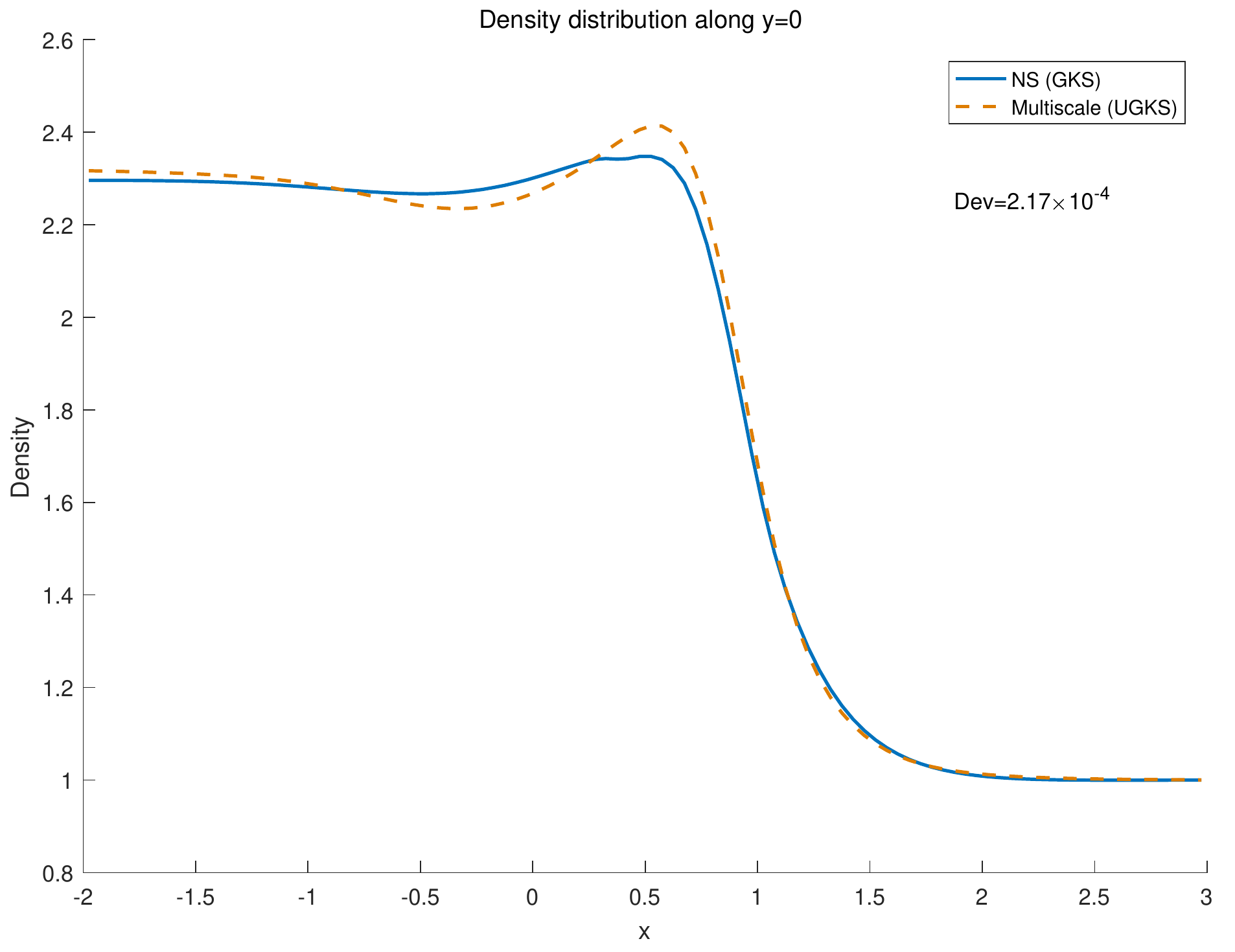}{c}
\includegraphics[width=0.4\textwidth]{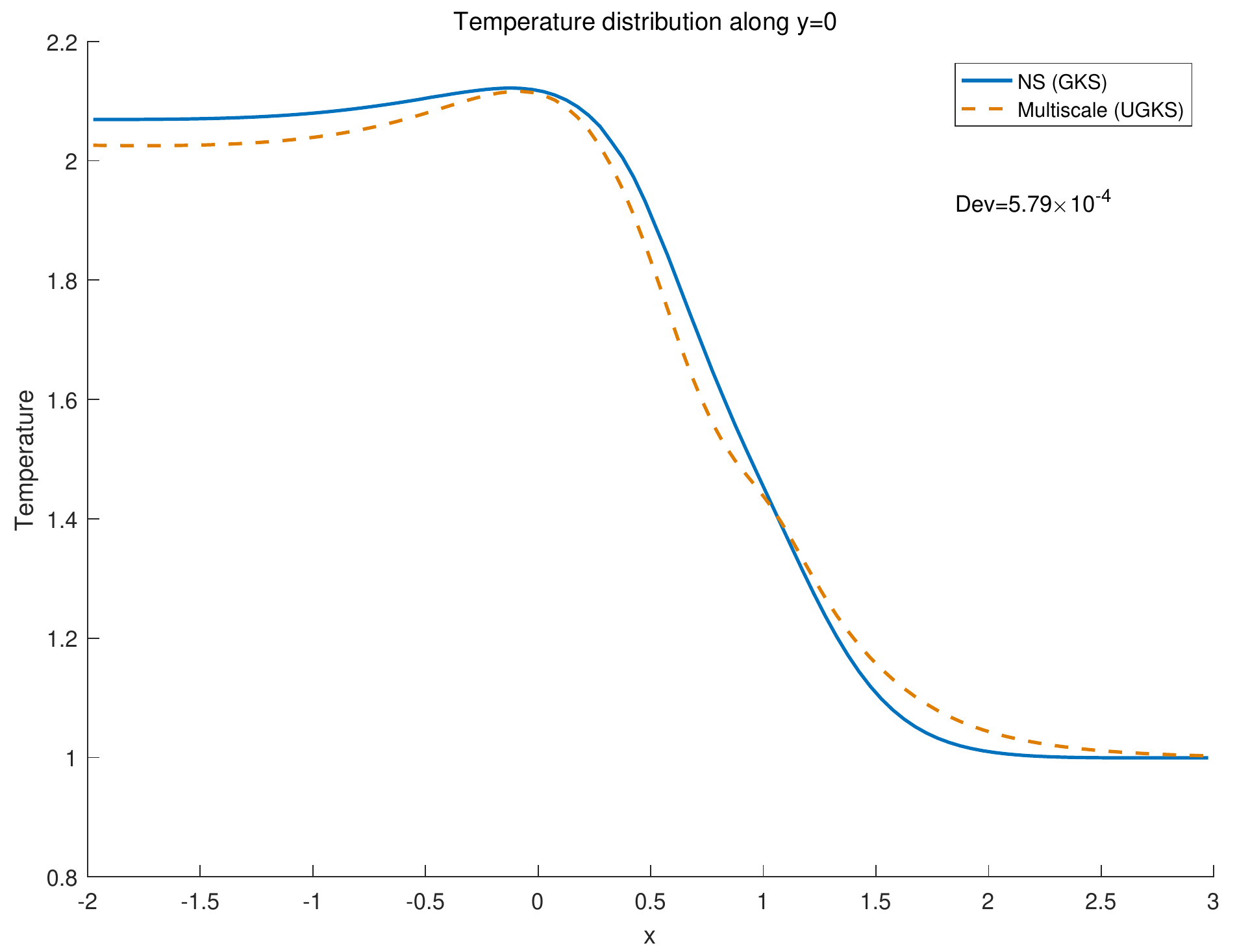}{d}\\
\includegraphics[width=0.4\textwidth]{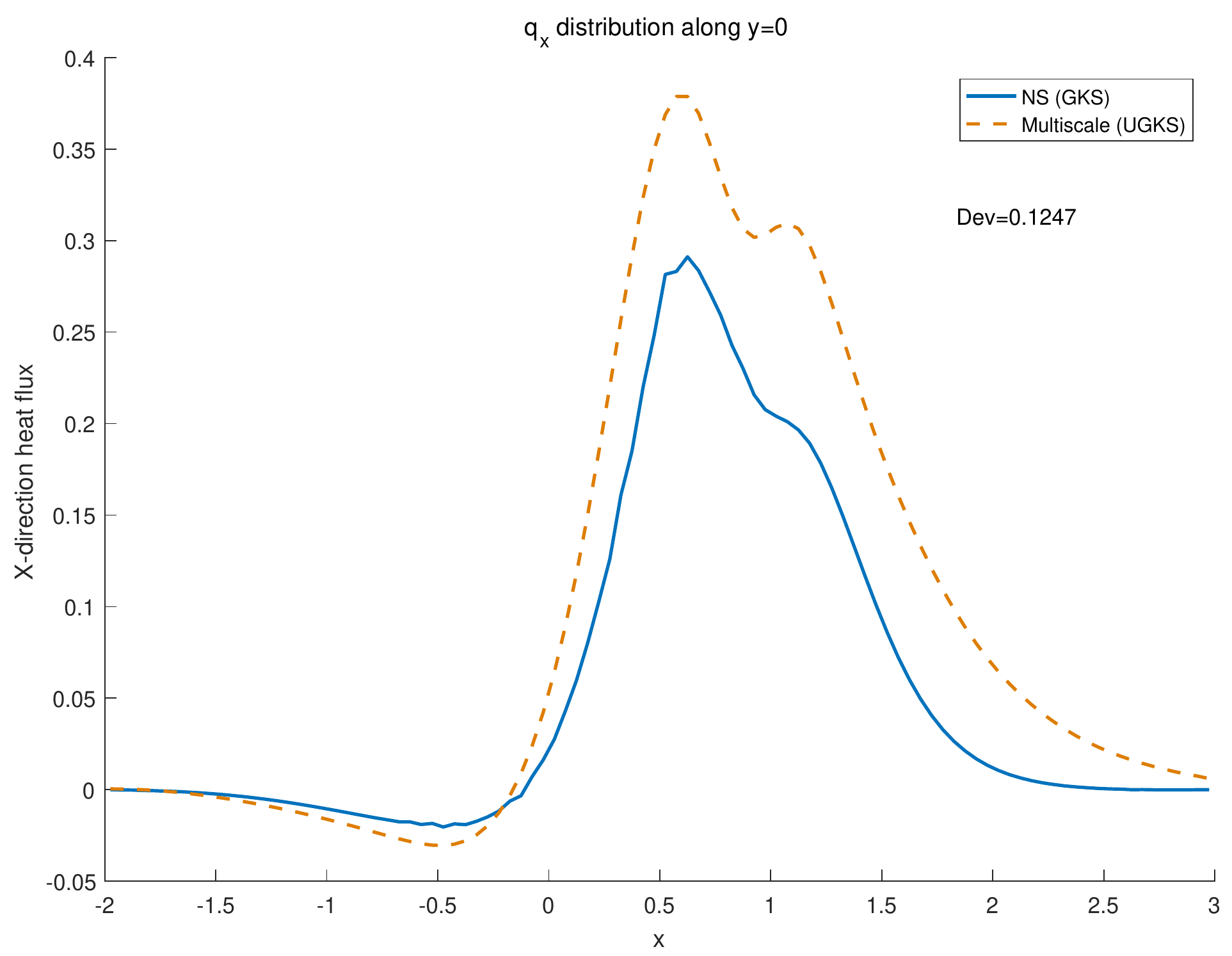}{e}
\includegraphics[width=0.4\textwidth]{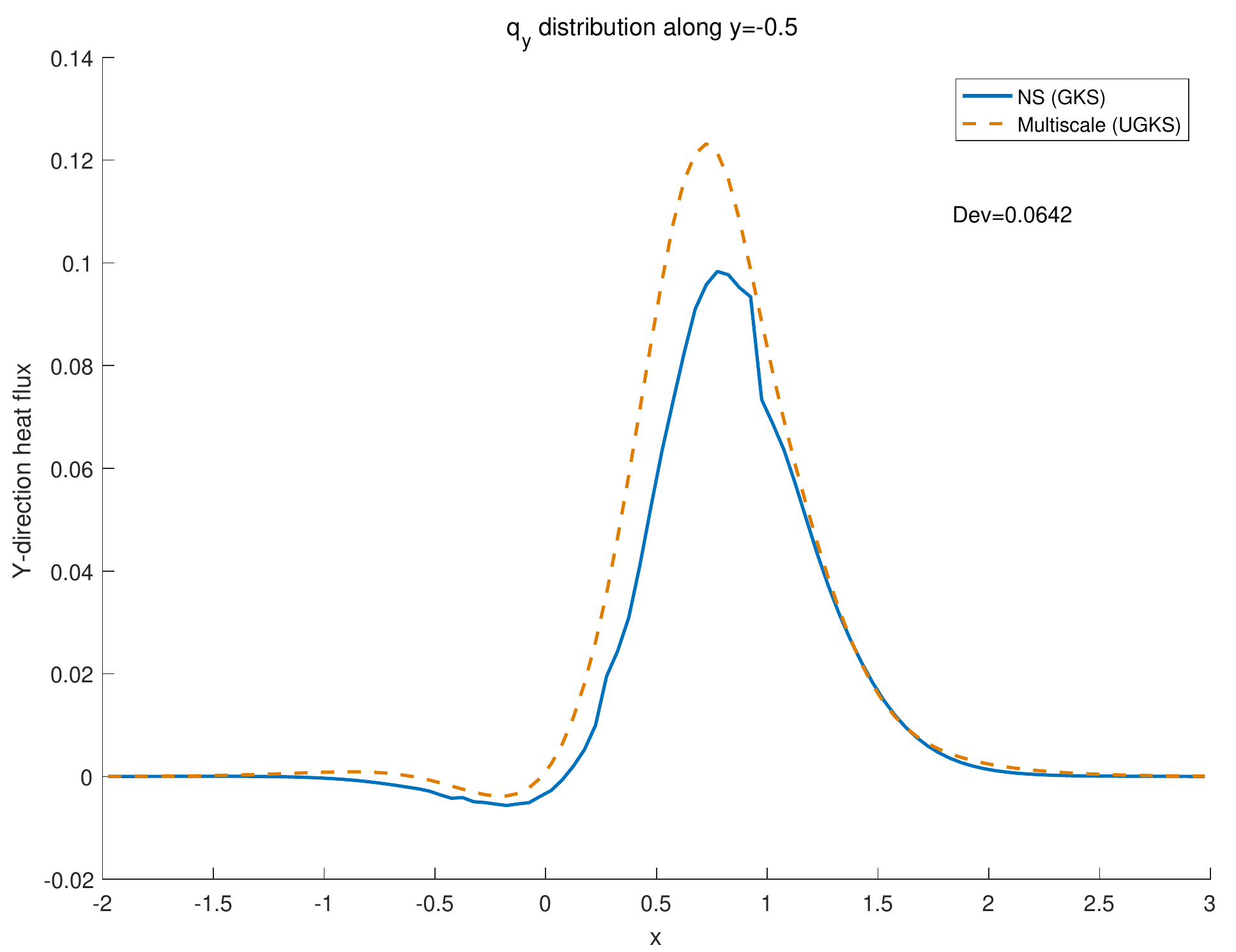}{f}
\caption{Results of the shock bubble interaction with $M\!a=2$ and $K\!n=0.3$ at $t=1.0$. Top two figures shows the cell Knudsen number (a) and total heat flux (b). (c)-(f) show the comparison between NS solution profiles (solid line) and UGKS solution profiles (dash line).}
\label{ma2kn03}
\end{figure}

Next, we increase the gradients of the flow variables by increasing the Mach number of the shock wave to $M\!a=2.0$
and a bubble density distribution
\begin{equation}
  \rho(x,y)=1+1.5\exp(-16(x^2+y^2)).
\end{equation}
The solutions at small Knudsen number $K\!n=1.0\times10^{-4}$ and at $t=1.0$ are shown in Fig. \ref{ma2kn-4}.
For this case, the Reynolds number relative to the bubble radius is larger than $10^3$.
The NS density and temperature profiles agree well with the multiscale solutions,
which have a small increment in the deviation in comparison with $M\!a=1.3$ and $K\!n=1.0\times10^{-4}$ case.
For the heat flux, large deviation can be clearly observed, especially for the x-directional heat flux with $43\%$ deviation.
When the Knudsen number is increased to $K\!n=0.3$, relative large deviation in all flow variables can be observed in Fig. \ref{ma2kn03}.

For the shock bubble interaction at a relative large cell Knudsen number,
the flow physics is not in NS regime and the NS equations don't provide accurate solutions.
At a small cell Knudsen number, the low order moments from the NS, such as density and temperature, agree with multiscale solution due to the
cell averaged effect under the limitation principle.
However, even at a small cell Knudsen number the high order moments, such as the heat flux, will not be well predicted by NS equations
at the region with large flow gradient.
Therefore, even in the near continuum flow regime, the multiscale numerical method can provide more accurate
physical solutions under the same cell resolution as NS solver.

\subsection{Force-driven Poiseuille flow}
In the following, through force-driven Poiseuille flow
we study the capability of NS equations in the description of non-equilibrium flow physics, even in the continuum flow regime.
This example clearly indicates the multiscale nature of the flow physics and the NS modeling is not fully complete for the physical solution.

The Poiseuille flow is about the flow confined between two parallel isothermal plates.
A constant body force is applied in the x-direction along the channel.
Periodic boundary condition is taken in the flow direction.
The simulated gas is a hard sphere monatomic gas.
The initial condition is set as $\rho_0=1.0$, $u_x=0$, $T=1$. In the following, the solutions at a variety of Knudsen numbers
and external force will be obtained.
The Knudsen number is defined by the ratio between the mean free path and the wall distance, $K\!n=\lambda/L_y$.
The results from DSMC and NS have been previously presented \cite{DSMC}.

\begin{figure}
\centering
\includegraphics[width=0.45\textwidth]{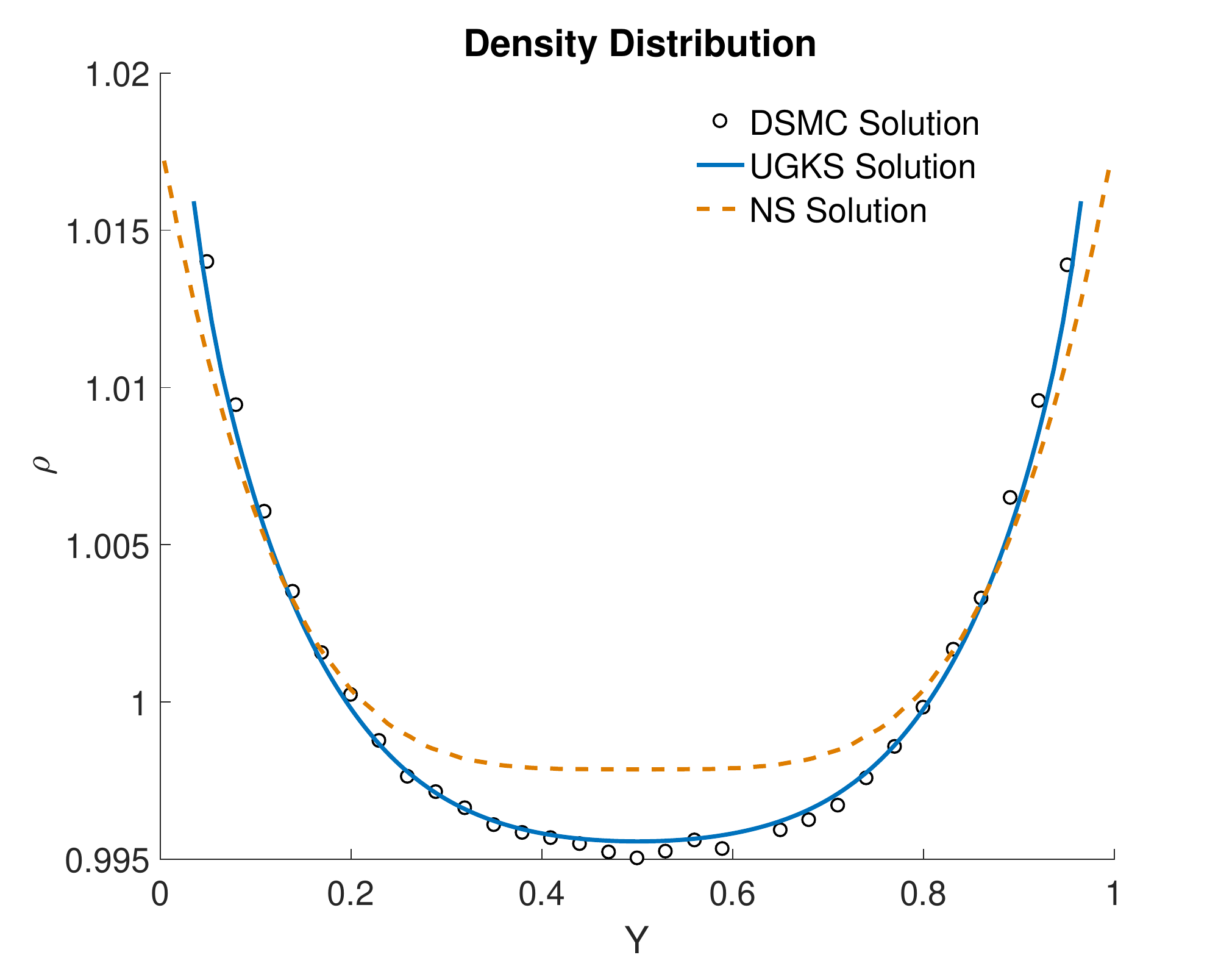}{a}
\includegraphics[width=0.45\textwidth]{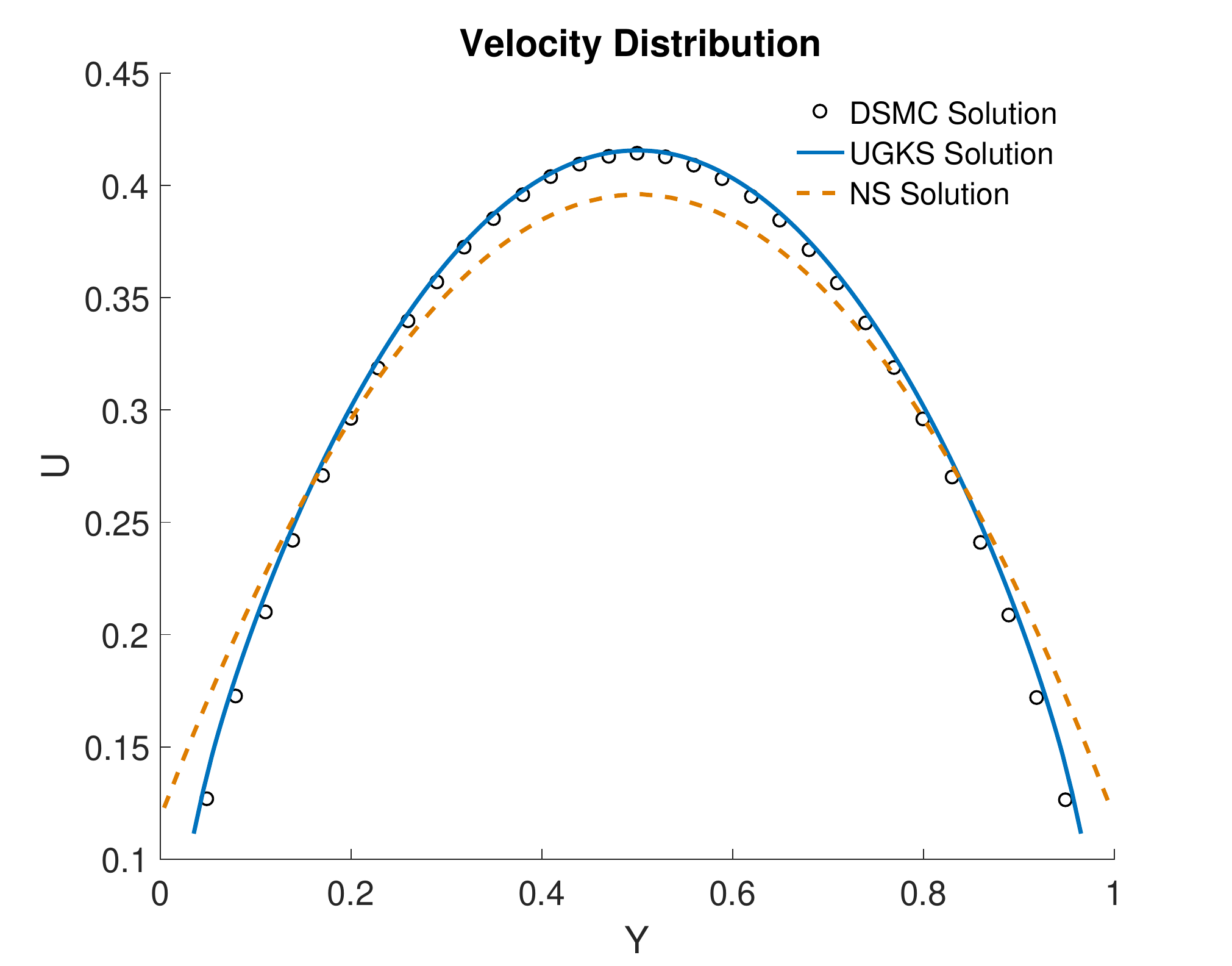}{b}\\
\includegraphics[width=0.45\textwidth]{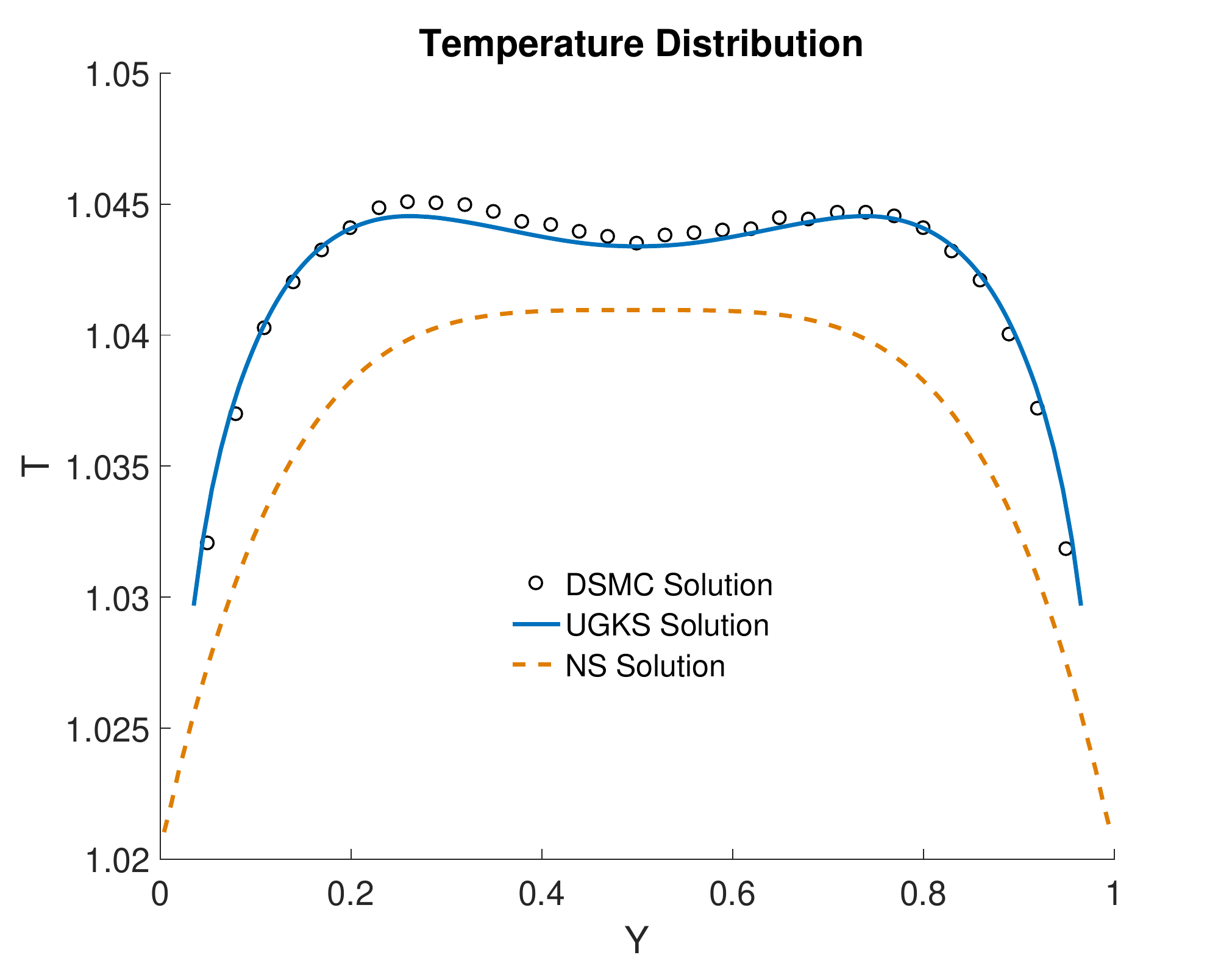}{c}
\includegraphics[width=0.45\textwidth]{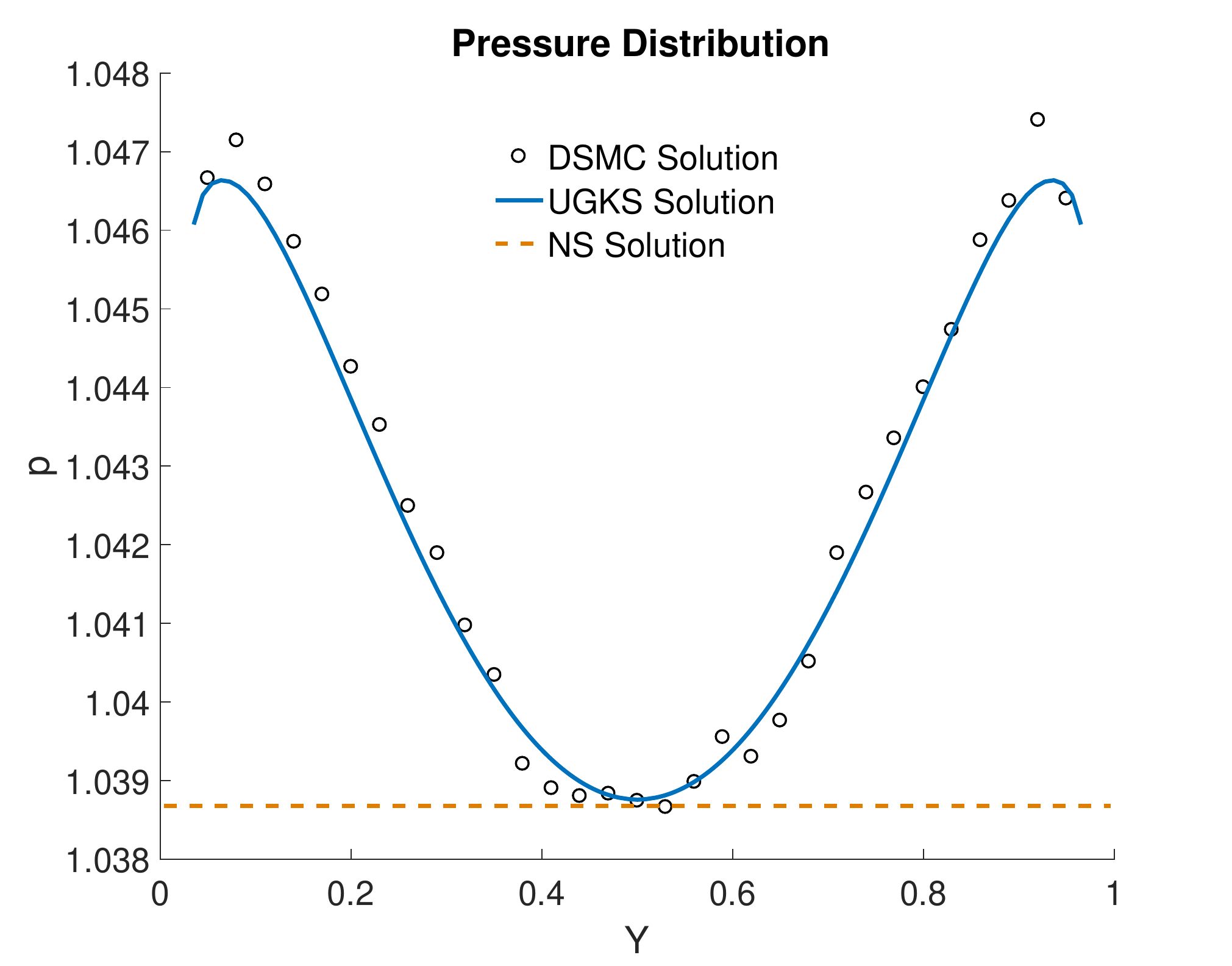}{d}
\caption{The (a) density, (b) x velocity, (c) temperature, (d) pressure distribution for the force-driven Poiseuille flow at steady state {with Knudsen number $K\!n=0.1$ and acceleration $F_x=0.126$}. The solid lines are the UGKS solutions, the dashed lines are the NS solutions and symbols stand for DSMC solutions.}
\label{poi1}
\end{figure}

\begin{figure}
\centering
\includegraphics[width=0.45\textwidth]{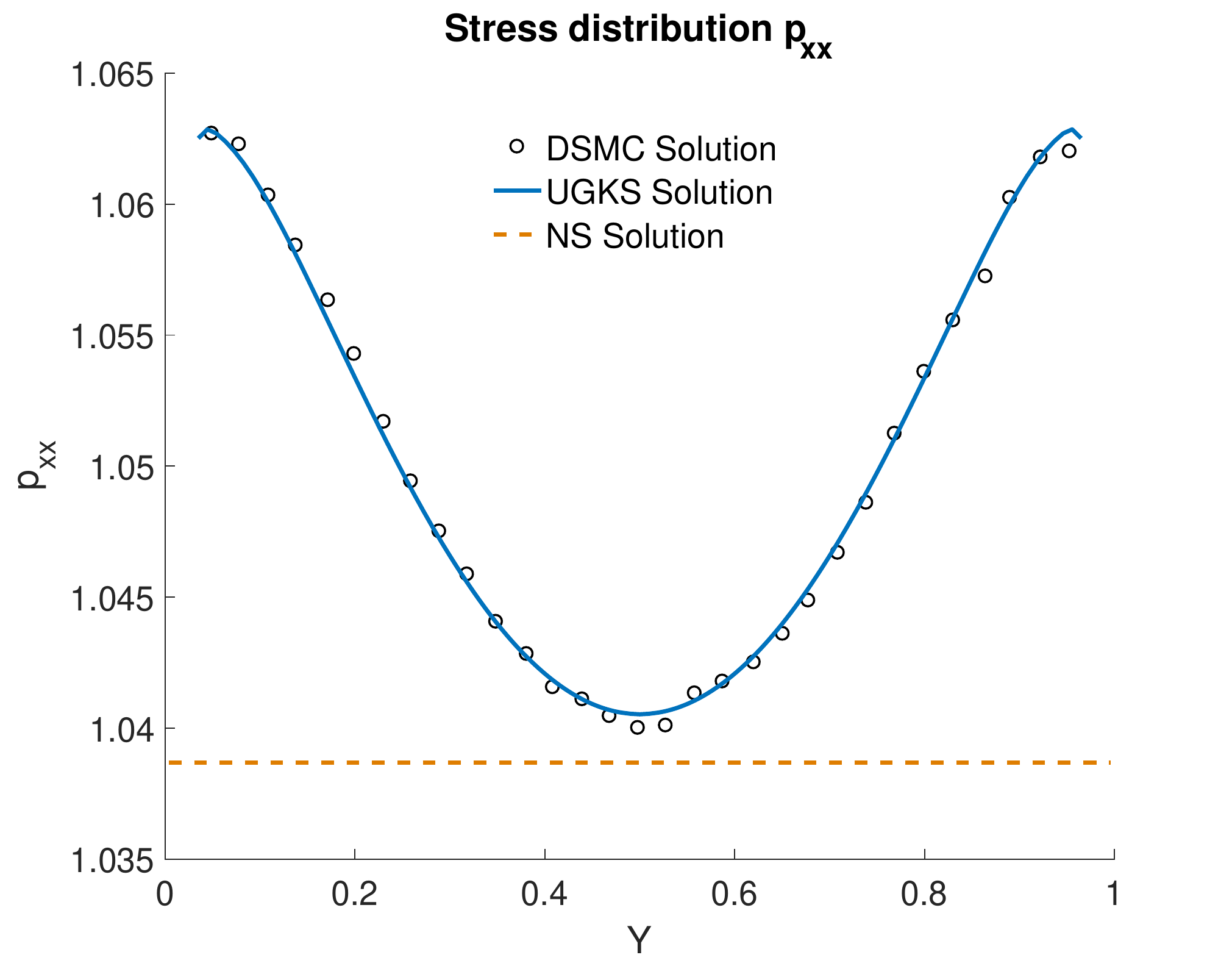}{a}
\includegraphics[width=0.45\textwidth]{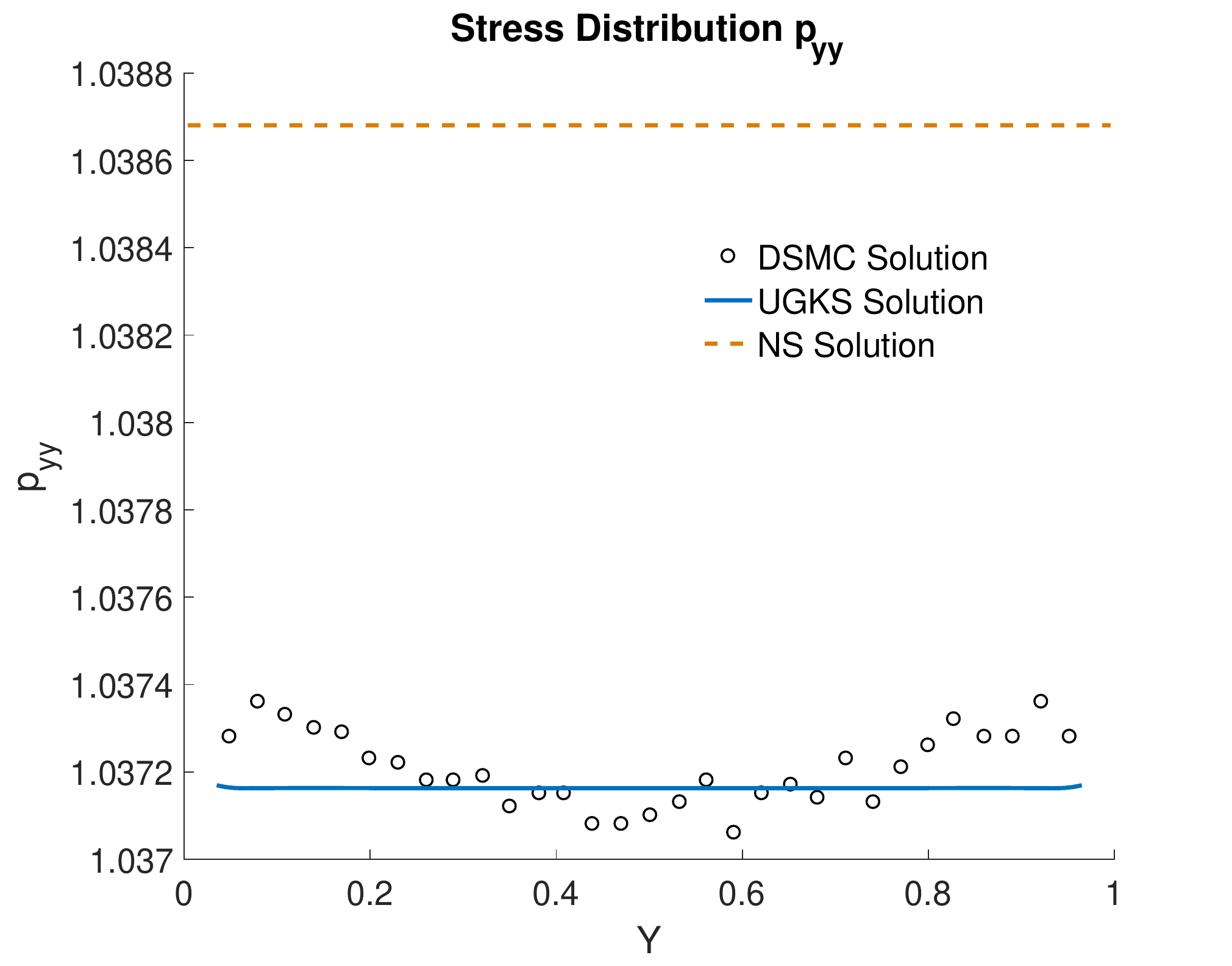}{b}\\
\includegraphics[width=0.45\textwidth]{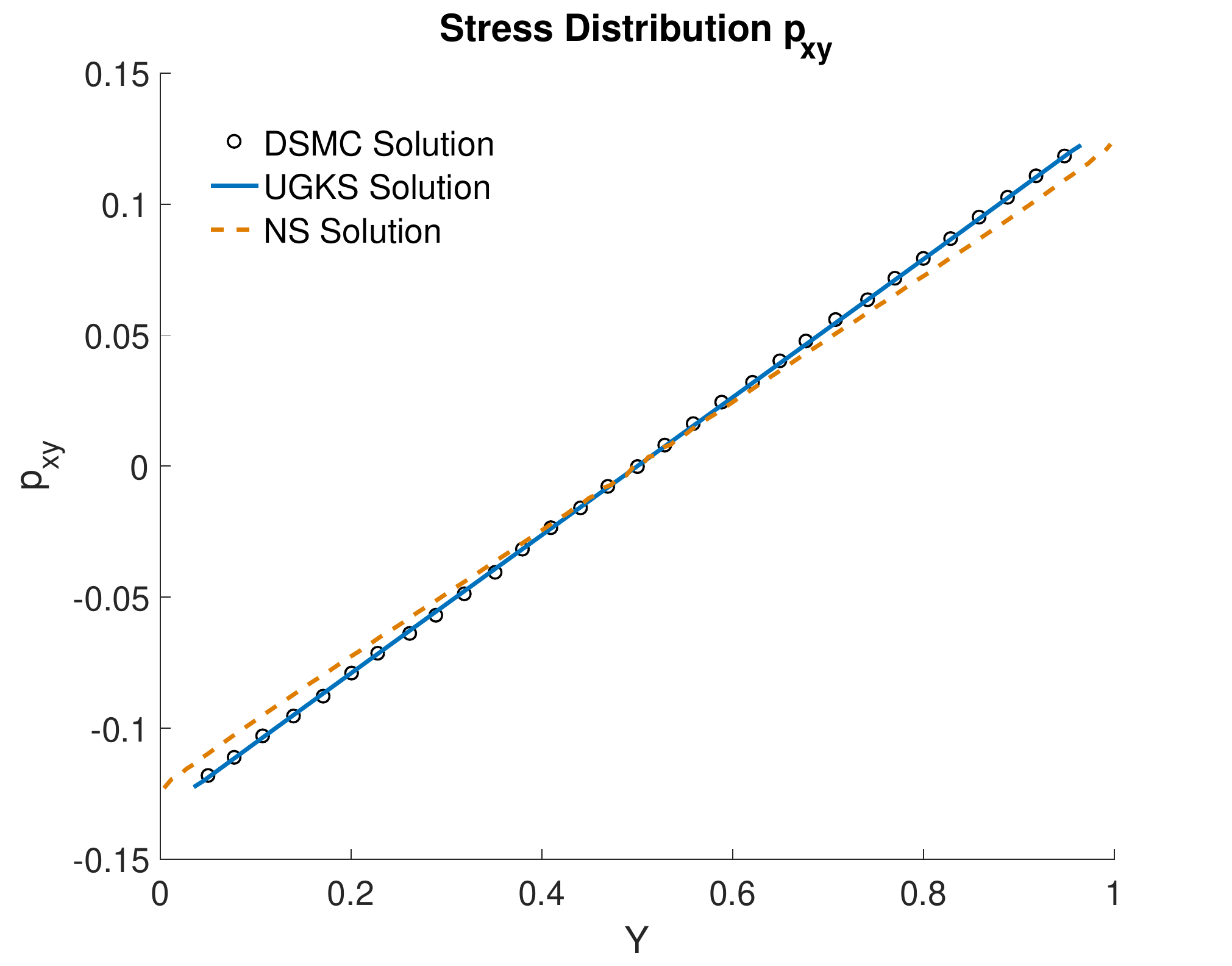}{c}
\includegraphics[width=0.45\textwidth]{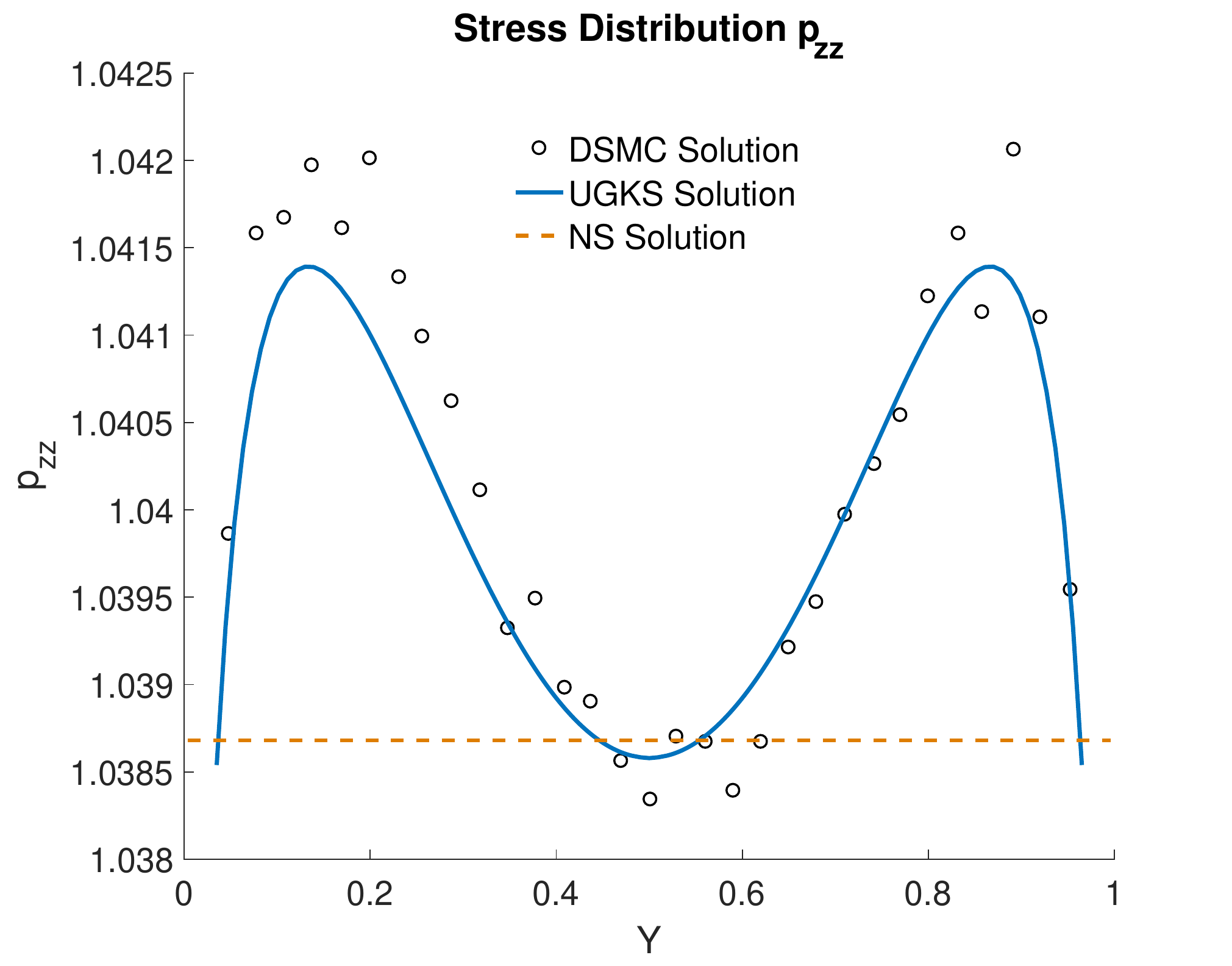}{d}
\caption{The stress distribution for the force-driven Poiseuille flow at steady state: (a) $p_{xx}$, (b)$p_{yy}$, (c) $p_{xy}$, (d)$p_{zz}$. The solid lines are the UGKS solutions, the dashed lines are the NS solutions and symbols stand for DSMC solutions.}
\label{poi2}
\end{figure}

\begin{figure}
\centering
\includegraphics[width=0.45\textwidth]{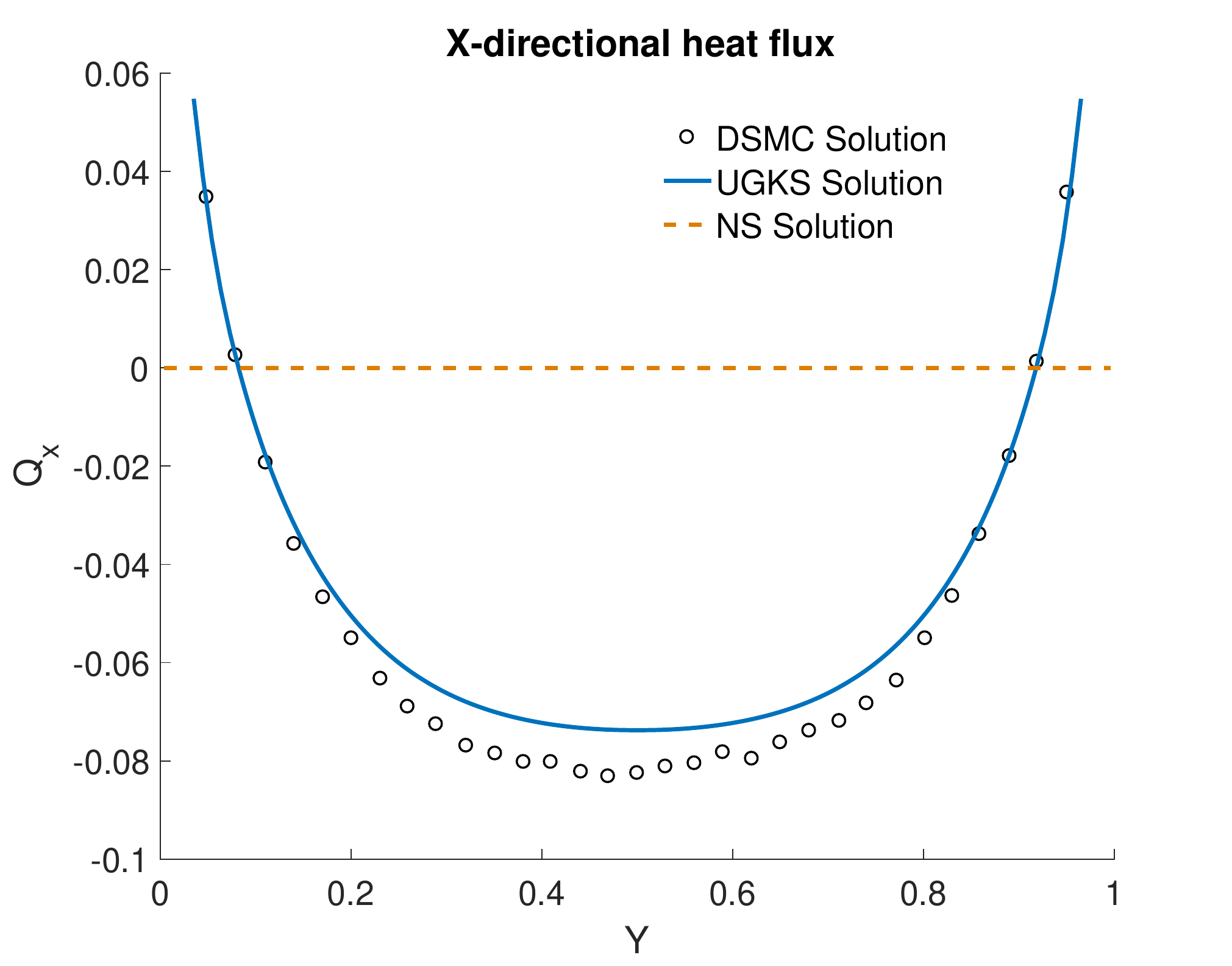}{a}
\includegraphics[width=0.45\textwidth]{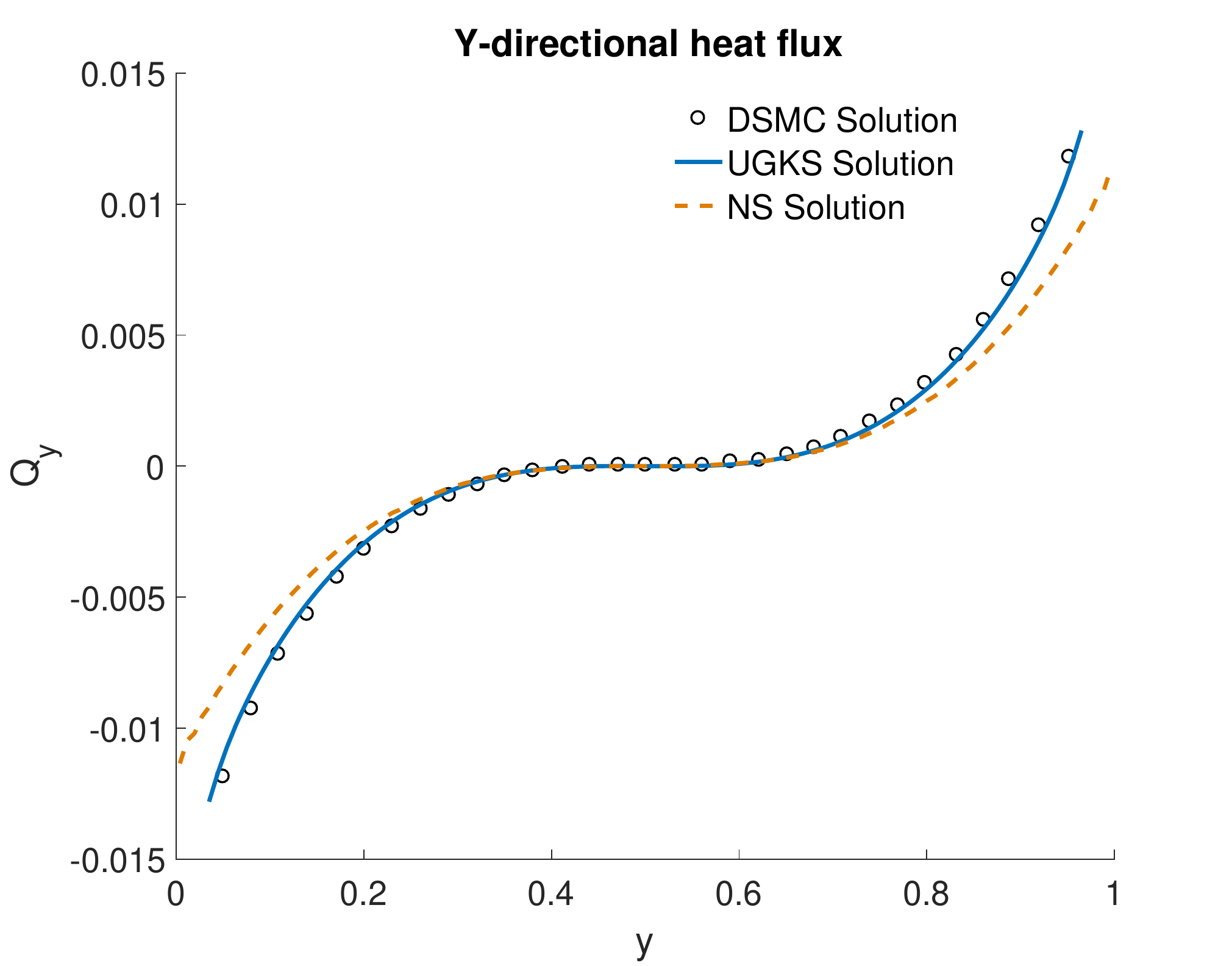}{b}
\caption{The heat flux distribution for the force-driven Poiseuille flow at steady state: (a) x-directional heat flux $Q_x$, (b) y-directional heat flux $Q_{y}$. The solid lines are the UGKS solutions, the dashed lines are the NS solutions and symbols stand for DSMC solutions.}
\label{poi3}
\end{figure}

The first test has a Knudsen number $K\!n=0.1$ and acceleration $F_x=1.26\times10^{-1}$.
The physical domain is divided into $51$ cells in the y-direction,
and velocity domain $[-6,6]^3$ is divided by $32$ velocity points in x,z-direction and $64$ velocity points in y-direction.
For this set of parameters, the time step is about $100$ times smaller than the local collision time, therefore the Boltzmann collision term takes effect in Eq.\eqref{update-f}. It can be observed in Fig. \ref{poi1}-\ref{poi3} that the UGKS solutions agree well with DSMC results
in the prediction of density, velocity, temperature, stress, and heat flux.
The NS solution is calculated by GKS with the first order slip boundary condition
$$u_{\text{slip}}=\alpha\lambda\frac{\partial u}{\partial y}, \quad T_{\text{slip}}=\beta\frac{2\gamma}{\gamma+1}\frac{\lambda}{\text{Pr}}\frac{\partial T}{\partial y},$$
with $\alpha=1.11$ and $\beta=1.13$. In the flow resolved calculation, the non-equilibrium effect is not well captured by NS equations at all,
especially the stress and heat flux.

\begin{figure}
\centering
\includegraphics[width=0.45\textwidth]{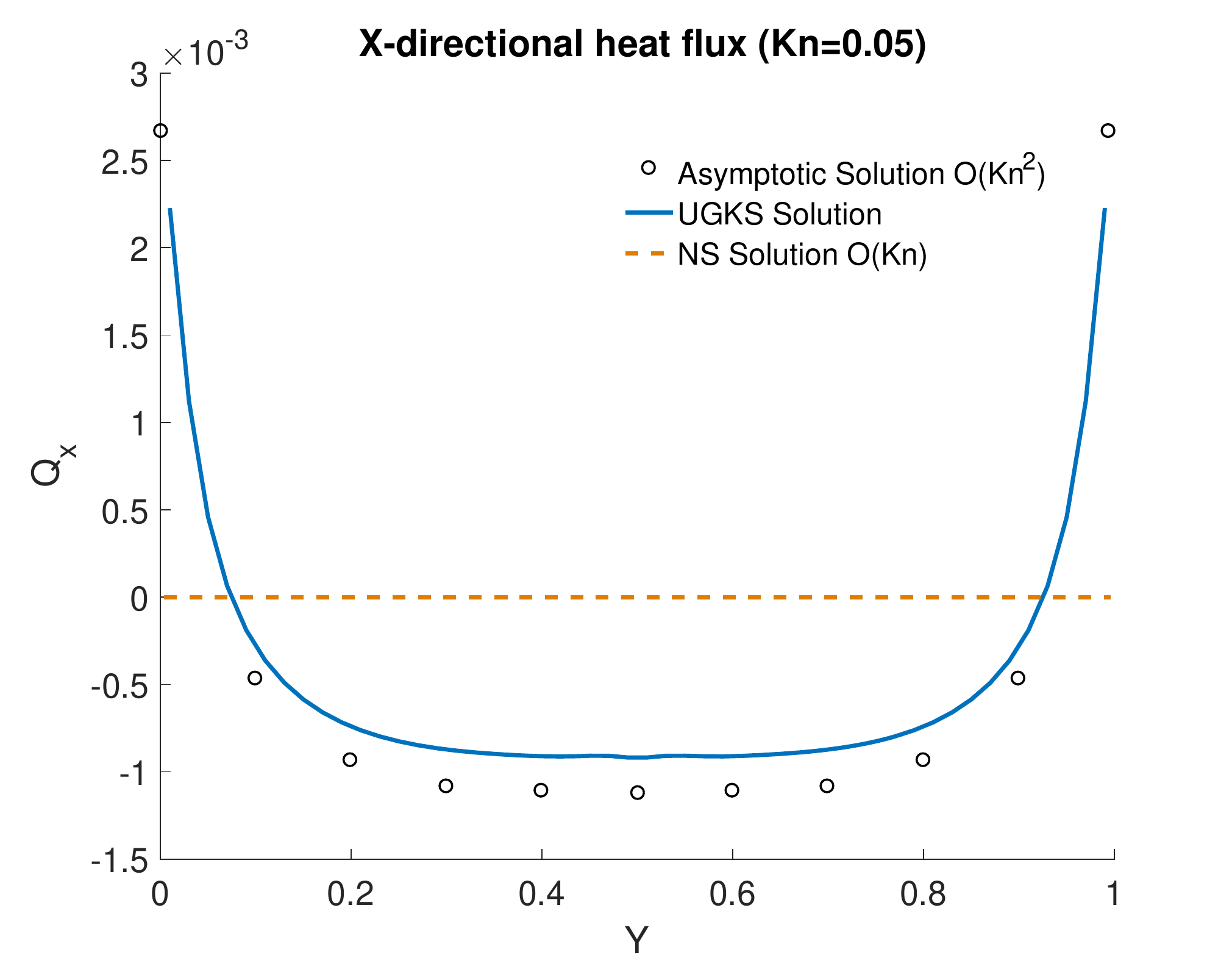}{a}
\includegraphics[width=0.45\textwidth]{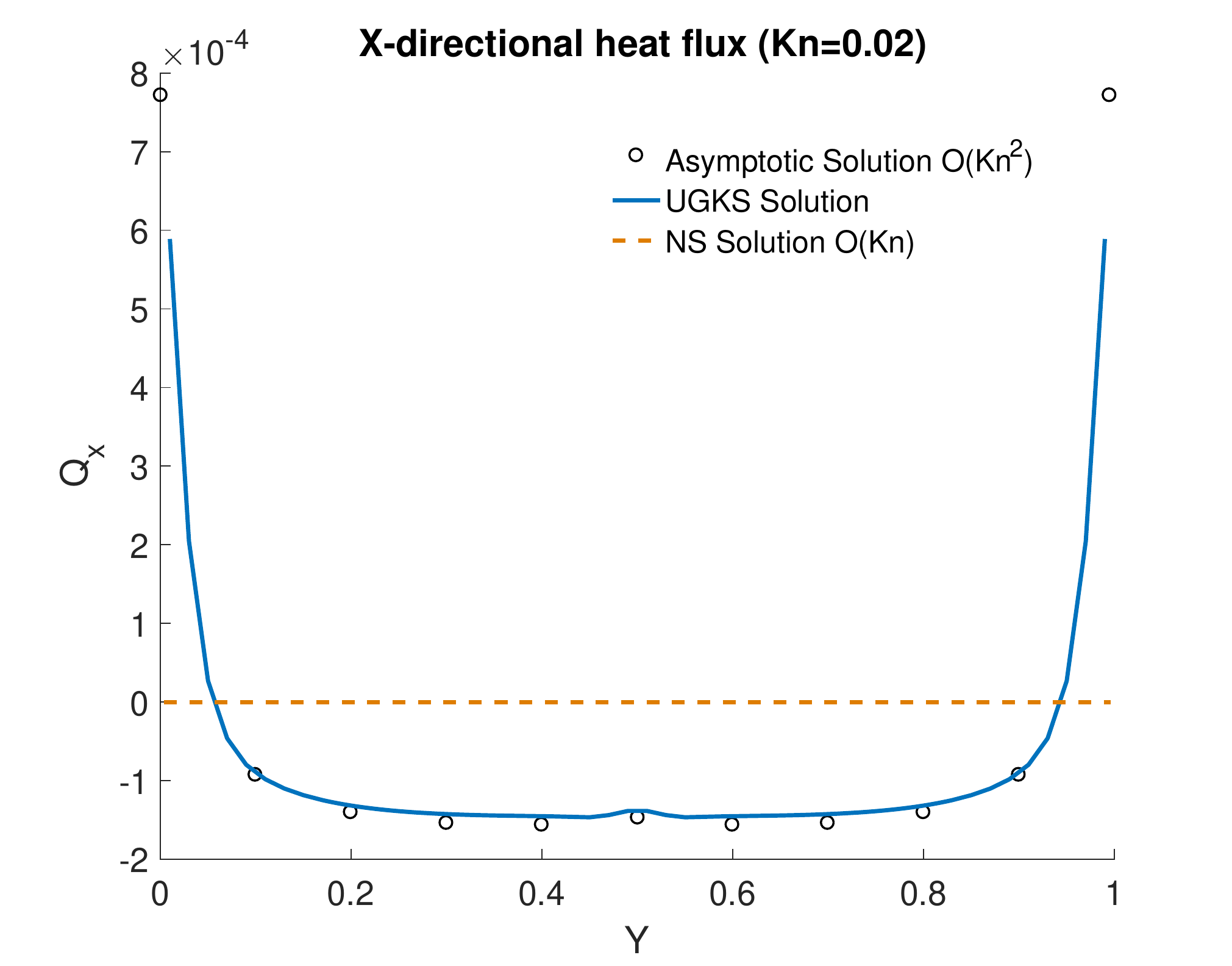}{b}
\caption{X-directional heat flux with {(a) Knudsen number $K\!n=0.05$ and acceleration $F_x=0.05$, (b) Knudsen number $K\!n=0.02$ and acceleration $F_x=0.02$}. The UGKS solutions (solid lines) are compared with the asymptotic solutions (symbols) and NS solutions (dashed lines).}
\label{poi4}
\end{figure}

Next, we reduce the Knudsen number and external force to $F_x=K\!n=5\times10^{-2}$ and $F_x=K\!n=2\times10^{-2}$.
For small Knudsen number and small acceleration, an asymptotic solution from the BGK equation
is derived by Aoki et al. \cite{aoki}. The UGKS solutions are compared with the second order asymptotic solutions and NS solutions.
As shown in Fig. \ref{poi4}, asymptotic solution agrees well with DSMC solution especially for $K\!n=2\times 10^{-2}$,
while the NS solutions always predict zero heat flux in the flow direction.

\begin{figure}
\centering
\includegraphics[width=0.45\textwidth]{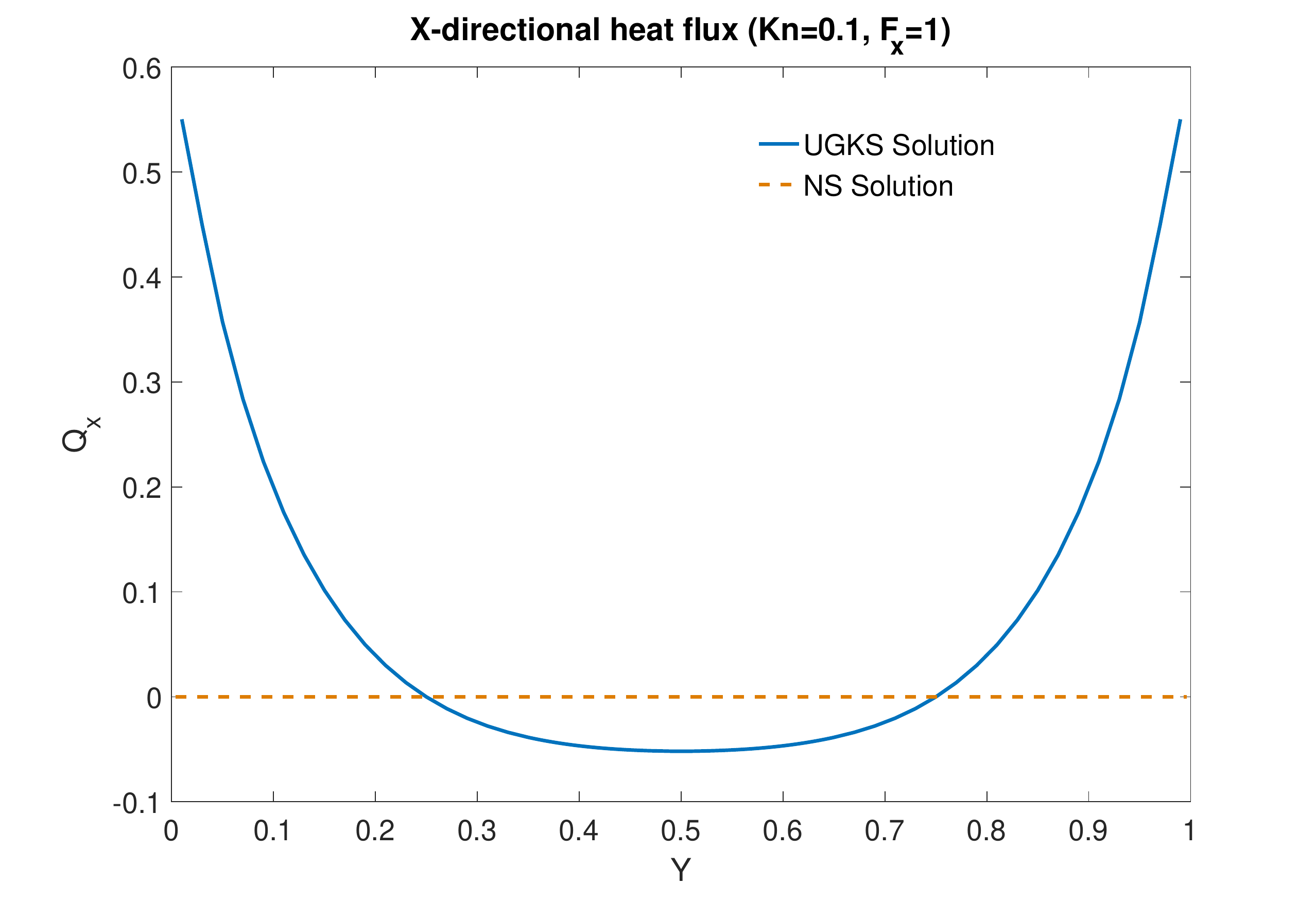}{a}\\
\includegraphics[width=0.45\textwidth]{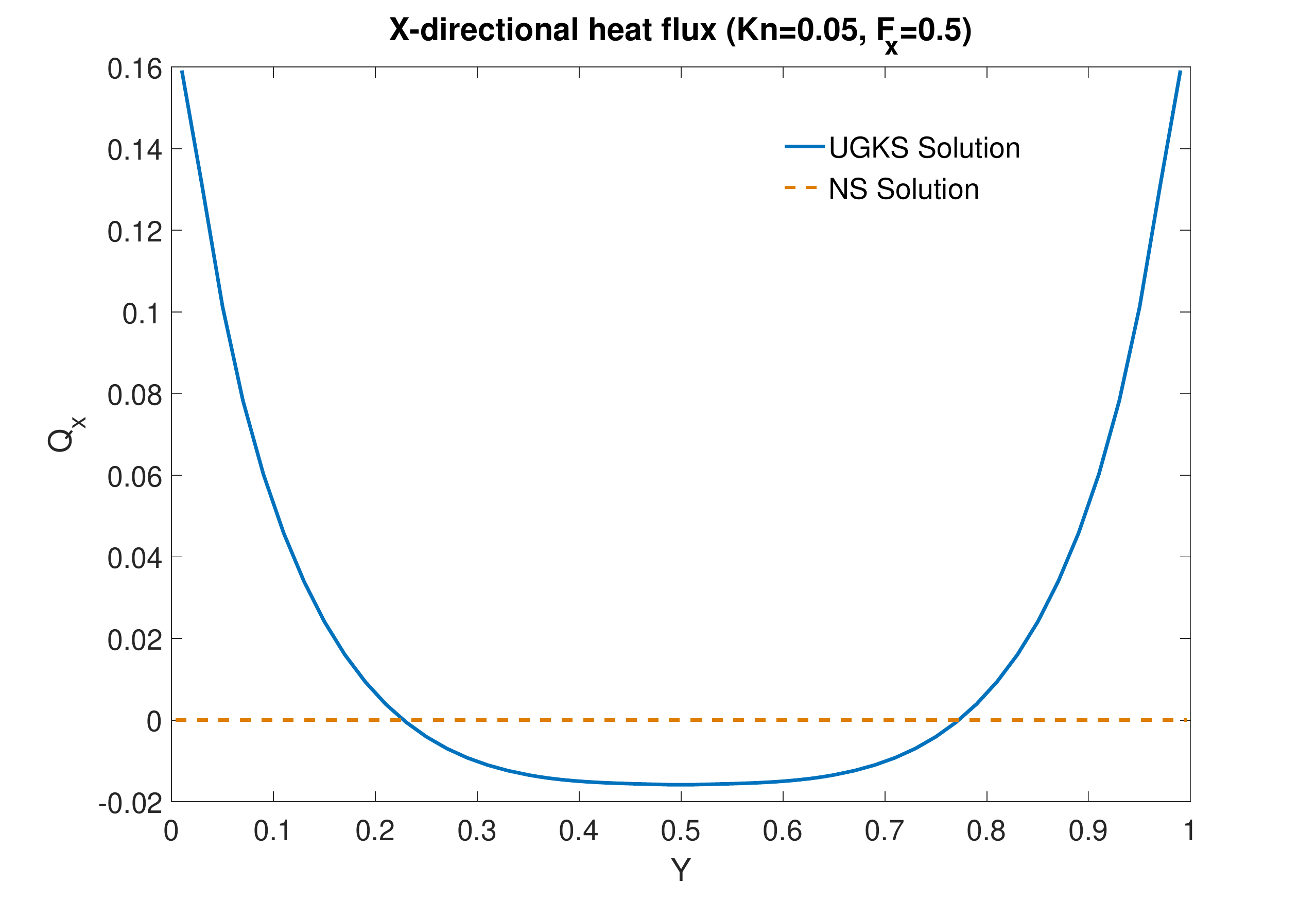}{b}
\includegraphics[width=0.45\textwidth]{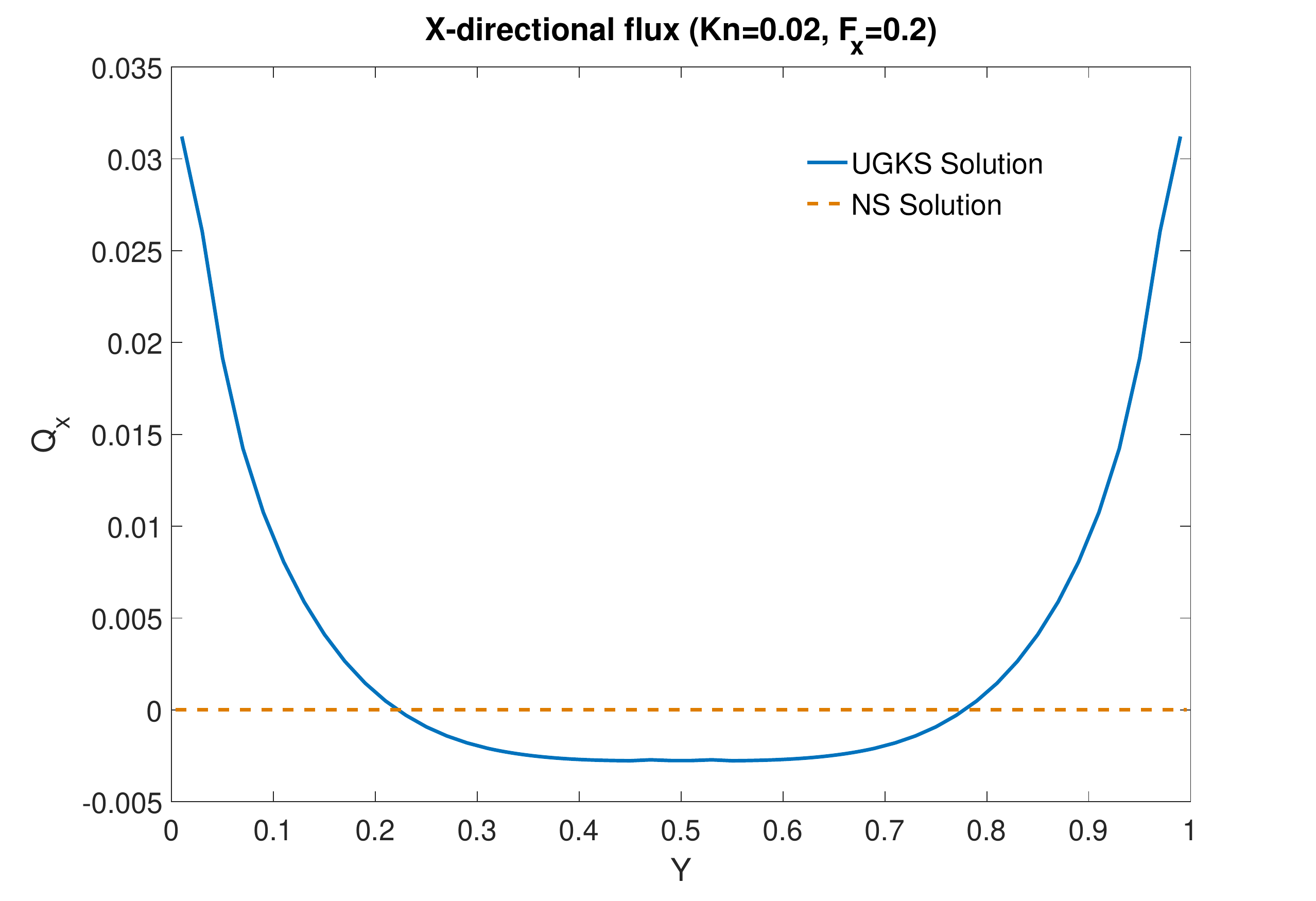}{c}
\caption{X-directional heat flux with relative large external force {(a) Knudsen number $K\!n=0.1$ and acceleration $F_x=1.0$, (b) Knudsen number $K\!n=0.05$ and acceleration $F_x=0.5$, (c) Knudsen number $K\!n=0.02$ and acceleration $F_x=0.2$.} The solid lines represent UGKS solutions and dashed lines stand for the NS solutions.}
\label{poi5}
\end{figure}

Next, we study the Poiseuille flow under a relatively large external force.
The calculation is performed with three set of parameters: $K\!n=0.1, F_x=1$, $K\!n=5\times10^{-2}, F_x=0.5$,
and $K\!n=2\times 10^{-2}, F_x=0.2$.
The x-directional flux results are shown in Fig. \ref{poi5}.
It is shown that the heat flux induced by the external force is proportional to the first order of Knudsen number,
which is on the same order as the Fourier's heat flux induced by the temperature gradient.
Therefore, with the external force field, the Fourier heat flux in the NS equations is not complete in the
description of energy transport, even in the continuum flow regime \cite{xiao}.
This test case indicates the importance of a multiscale method for the capturing of intrinsically multiple scale flow dynamics.
The limitation principle presented in this paper is for the multiscale method. For the NS solver, the quality of the solution is far away from
the limiting solution constrained by the limitation principle.

\section{Conclusion}

Gas dynamics is associated with multiscale nature. The NS equations describe a flow physics for very small variation of
flow variables, and this variation is measured by the physical Knudsen number.
At the same time, the NS uses the fluid element for modeling the dynamics,
and the scale of the fluid element is not explicitly defined, which introduces inaccuracy in the prediction of physical solution
even in the analytical level.
For  CFD computation, the situation becomes  more complicated in a discretized space.
Here we have to have a clear picture about the modeling scale of the governing equation and the numerical scale,
i.e., the mesh size and time step, for the evaluation of the solution.
With the variation of the mesh size, the coarse-grained flow variables have a numerical dynamics with the corresponding
numerical Knudsen number, and only a limited solution can be recovered by a numerical scheme.
With the definitions of cell Knudsen number $K\!n_c$ and the physical solution Knudsen number $K\!n_s$, the representable numerical solution
in a mesh size scale is controlled by the numerical Knudsen number $K\!n_n$, which is a function of $K\!n_c$ and $K\!n_s$ with the relationship
$K\!n_n = \text{softmin} (K\!n_s , K\!n_c )$.
 This is the limitation principle which states the  best numerical solution can be obtained by a multiscale CFD method, such as the UGKS.
In order to understand such a limitation principle, in this paper we first demonstrate the physical modeling deficiency in the NS equations
in the mesh refinement process. As the mesh size goes to the particle mean free path in the high speed flow simulation,
the wave interaction modeling, such as Riemann solver for the NS solution, doesn't hold anymore.
Instead, particle transport and collision will be the correct physics here.
Therefore, corresponding to different mesh size scale relative to the particle mean free path,
we need a consistent gas dynamics in a numerical scheme.
The flow transport in UGKS depends on the ratio of $\Delta t / \tau$.
With a variation of $\tau /\Delta t$,
the multiscale gas dynamics is covered from the particle transport as $\tau \simeq \Delta t$ to the wave interaction as $\tau \ll \Delta t$,
 with a continuous variation between them \cite{xu-liu}.
Even with a multiscale method, due to possible mismatch between the mesh size and physical structure thickness,
the final numerical solution is constrained by the limitation principle with a newly defined
the numerical Knudsen number instead of the physical one.
The numerical gas dynamic solution is related to
the flow variable variation (physical Knudsen number) and
mesh resolution {(cell Knudsen number)}.
In order to illustrate such a limitation principle, many numerical examples are presented,
such as shock interface interaction and Richtmyer-Meshkov instability,
where different solutions can be identified under different mesh resolution.
The force-driven Poiseuille flow clearly shows the multiscale nature of gas dynamics.
Based on the UGKS solution, even in the continuum flow regime it shows that
the single scale NS equations are not adequate to present a  physical solution.
In a CFD simulation, we need to consider both limitations from the governing equations and the cell size resolution.
The limitation principle is about the best result  a best mutiscale method can simulate
in the mesh size scale.
Any CFD simulation is associated with uncertainty principle due to the variation of mesh resolution.
There is no faithful cell converged physical solution if the flow solver is targeting on the NS equations alone, even in the
continuum flow regime.

\section*{Acknowledgement}
The current research was supported by Hong Kong research grant council (16206617,16207715,16211014),
and  National Science Foundation of China (91530319,11772281).

\section*{References}

\bibliography{mybibfile}

\newpage

\clearpage

\end{document}